\documentclass[reprint,amsmath,amssymb,aps]{revtex4-2}

\usepackage{graphicx}
\usepackage{subcaption}
\usepackage[justification=Justified]{caption}
\usepackage{dcolumn}
\usepackage{bm}
\usepackage{xcolor}
\usepackage{tabularx}
\usepackage[export]{adjustbox} 
\usepackage{booktabs}  
\usepackage{tabularx}  
\usepackage{makecell}  
\usepackage{array}
\usepackage{caption}
\usepackage{float}
\usepackage{lipsum}
\usepackage{soul}
\usepackage{multirow}
\usepackage{hyperref}
\usepackage{tikz}
\usepackage[dvipsnames]{xcolor}
\usepackage{appendix}
\hypersetup{
    pdfborder={0 0 0}, 
    colorlinks=true,   
    linkcolor=black,   
    urlcolor=blue,     
    citecolor=blue    
}
\usepackage{stmaryrd}
\usepackage[none]{hyphenat}

\newcommand{\cm}{\text{\;cm}^{-1}}
\newcommand{\fs}{\text{\;fs}^{-1}}
\newcommand{\Q}{\mathcal{Q}}
\newcommand{\Qset}{[\, E_2 \;\Omega_{\text{2P}} \; \gamma_2 \; \Gamma_{12}\,]}
\newcommand{\QEset}{[\,\Omega_{\text{2P}} \; \gamma_2 \; \Gamma_{12}\,]}

\begin{document}

\preprint{APS/123-QED}

\title{Deep learning parameter estimation and quantum control of single molecules 
}

\author{Juan M. Scarpetta$^{1, 2}$}
\author{Omar Calderón-Losada$^{1}$}
\author{Morten Hjorth-Jensen$^2$}
\author{John H. Reina$^{1, 2}$}

\affiliation{$^1$Department of Physics and Centre for Bioinformatics and Photonics---CIBioFi, 
Universidad del Valle, Cali 760032, Colombia}
\affiliation{$^2$Department of Physics and Center for Computing in Science Education, University of Oslo, Oslo N-0316, Norway}

\date{\today}

\begin{abstract}


Coherent control, a central concept in physics and chemistry, has sparked significant interest due to its ability to fine-tune interference effects in atoms and individual molecules for applications ranging from light-harvesting complexes to molecular qubits. However, precise characterization of the system's dissipative dynamics
is required for its implementation,  especially at high temperature. In a quantum control experiment, this means learning system-bath parameters and driving coupling strengths. Here, we demonstrate how to infer key physical parameters of a single molecule driven by spectrally modulated pulses at room temperature.  We develop and compare two computational approaches based on two-photon absorption photoluminescence signals: an optimization-based minimization scheme and a feed-forward neural network. 
The robustness of our approach highlights the importance of reliable parameter estimation in designing effective coherent control protocols. Our results have direct applications in ultrafast spectroscopy, quantum materials and technology.

\end{abstract}

\maketitle

\section{Introduction\label{sec:introduction}}

Coherent control is an experimental paradigm that has been extensively developed since the 1980s~\cite{Zewail_CohLaserSpectro_1980, Silberberg_CohControlSpectroscopy, Expemtal_CohLaserControl_physicochemicalProcc} to study and manipulate the ultrafast dynamics of quantum systems using ultrashort laser pulses in the picosecond and femtosecond regimes~\cite{CoherenceChem2024,Hildner_fsCohQControl_singleMol_RoomT, roomTQC2019,fs_LadderType_nanoWire, FemtoLaserControl_1992}. This strategy has also been implemented for developing  
molecular qubits in quantum information science~\cite{Molecgates2004,molecgate2018,QISchem2025,molecohQIS2023}. At these time scales it is possible to probe and influence phenomena such as electronic dephasing~\cite{Electronic_CohDephasing_Heide,ultrafastcoherence2021,gustin_mapping_2023}, vibrational motion~\cite{Brinks_ControlVibWavePacks_SingleMol, MasterinFemt_RamanSpectroscopy}, stimulated emission~\cite{StimulatedEmiss_SingleNanocrystals, StimulEmission_Microscopy} and charge/energy transfer processes~\cite{Cohert_ChargeTransfer_organicPV,photophysical2022,molecoptoel2018,Energy_Charge_Transfer_ElectronicMicroscopy,LongRange_ET_HildnerR}. A wide variety of physical platforms, including semiconductor quantum dots~\cite{Yan2023, QDot_CohControl_2016}, isolated atoms~\cite{Attosec_SingleAtoms_QCohControl}, molecular aggregates~\cite{MolecularAggregates_OptoElectronics} and conjugated polymers~\cite{piConjug_RelaxDecoherence, CohGeneration_ConjugatedPolymers,molecoptoel2018}, exhibit distinct transition energies and absorption cross sections that enable the implementation of coherent control protocols. A central challenge here is the identification of optimal pulse sequences that facilitate targeted transitions between system levels while accounting for the diversity of spectral and dynamical properties across these platforms. In particular, multiphoton absorption processes~\cite{Multiphoton_ControlCondPhses,FewPhoton_CoherentControl} are commonly exploited to create constructive or destructive interference between transition pathways~\cite{Hildner_2013_varPathways}, thereby enabling coherent manipulation of the system dynamics.

For organic molecules and aggregates, one accessible observable is photoluminescence (PL). Several studies have measured PL from organic systems at room temperature both at the single molecule level and in ensembles~\cite{MeLPPP_bulk_fsTPA,MeLPPP_spectrumVibrat}. However, experimental constraints complicate these measurements as many molecules undergo photobleaching on short time scales~\cite{PhotoBleaching_Complexes}, and the emission yield must be sufficiently high for detection~\cite{QuantumYield_SingleMol_Fluorsc}, which  limits the achievable detection times. Moreover, environmental effects are particularly significant  in ultrafast molecular experiments, as photophysical properties  at ambient conditions deviate from those at low temperature, resulting in substantially shorter dephasing times~\cite{ElectronicCoherence_RapidLoss}. 
Thus, single molecules must be considered open quantum systems that interact with the solvent and surrounding environment. Therefore, characterizing  these dissipative effects, for example via spectral density functions~\cite{SpinBoson_GenericModel, ExponentialDecomposition_CorrelationFunction, CoherentControl_TLS_nonMarkovian_ReinaEckel} or by estimating dephasing and relaxation rates~\cite{QDissipation_Weiss, Leggett_DisspTLS,JH_QDecoherence,Scarpetta_2025ML_twolevel,CoherentControl_TLS_nonMarkovian_ReinaEckel,QubitControlNoiseSpectroscopy,Quantum_Master_Equations} is essential, though not always feasible \textit{a priori}.
Despite that, many experiments have been successfully carried out on organic single molecules~\cite{Hildner_SingleMolsss_RoomTemp, CohControl_SingleMol_RoomTemp_Hilder_2025},  dyes, and molecular samples with large two-photon absorption cross sections  at room temperature~\cite{VillabonaOmar_TPCS_RhB_Zn, TwoPulsed_fluorophores_singleDNQDI} . These experiments have measured two-photon PL 
 emission using sequences of ultrashort laser pulses with spectral phase modulations~\cite{CohControl_SingleMol_RoomTemp_Hilder_2025, VisualizingWilma2019}, as well as delayed pulse schemes~\cite{TwoPhotonWilma2019, TwoPulsed_fluorophores_singleDNQDI}.

In this context, one molecule that has attracted much interest is the methyl-substituted ladder-type poly(para-phenylene) (MeLPPP)~\cite{MeLPPP_spectrumVibrat, MeLPPP_univPicture_Spectroscopy}, a $\pi$-conjugated polymeric molecule with favorable photophysical properties~\cite{MeLPPP_PL_Hildner_Spectroscopy}. MeLPPP has been shown to be photochemically stable~\cite{ExcitonDiff_Relaxat_MeLPPP}, exhibits a large two-photon absorption cross section~\cite{MeLPPP_2P_CrossSection_Hohenau},  and can be detected at the single-molecule level. These properties  make MeLPPP a strong candidate for room-temperature coherent control experiments~\cite{VisualizingWilma2019, TwoPhotonWilma2019}. However, as with all complex molecular systems, obtaining reliable information about the environment and the system–field coupling from PL measurements requires careful modeling of dissipative processes and an understanding of the relevant energy and temporal scales.

To address these challenges, research on quantum noise spectroscopy and open quantum dynamics has focused on automating the extraction of relevant environmental features and parameters in a robust, reliable, and interpretable manner~~\cite{ExtractInfo_QEnv_Correlations, ExtractInfo_from_ExpData}. Machine learning and deep learning techniques have been proposed and applied to infer physical parameters from dynamical observables~\cite{ML_Control_QuantumSystems, NN_Simulating_Quantum_Dynamics, OCC_ReinforcementLearning,Scarpetta_2025ML_twolevel}; for instance, in two-level systems~\cite{Scarpetta_2025ML_twolevel,Paternostro_1} and related models~\cite{Paternostro_2, Jbarr2025_ML_SDapproach}. These approaches aim to recover dissipative and driving parameters from experimentally accessible signals while minimizing the required prior knowledge about the system-environment coupling~\cite{ML_nonMark_Dynamics, ML_nonMark_noiseQD}.

An algorithm for PL measurements of the MeLPPP molecule, termed Quantum Dynamics Identification (QDI), has been reported in~\cite{VisualizingWilma2019}. 
QDI minimizes the residuals between the experimental data and the PL model and has been developed to estimate parameters associated with dissipative effects and system-environment coupling. However, algorithms of this type are sensitive to the chosen initial conditions and must be rigorously tested to determine their reliability and convergence behavior. 

In this work, we present two computational approaches for extracting key physical parameters, including dephasing and relaxation rates, system–field coupling, and energy levels, from a single $\pi$-conjugated MeLPPP molecule~\cite{Trebino_PulseShaping_2022, Weiner2010_PulseShapingTutorial} driven by spectrally modulated pulses with phase and amplitude shaping. 
We calculate the ultrafast dissipative dynamics and characterize how the control parameters influence the photoluminescence signal generated via two-photon absorption. 
 The first method is an optimization-based scheme that minimizes a defined loss function for PL, while the second method uses a feed-forward neural network trained to map PL traces onto the target parameters. We assess the robustness of both approaches under different pre-processing strategies and quantify the associated uncertainties in the inferred parameters. The experimental PL data used for final validation correspond to single-molecule MeLPPP measurements~\cite{CohControl_SingleMol_RoomTemp_Hilder_2025,VisualizingWilma2019,TwoPhotonWilma2019}.

This work is organized as follows. Section~\ref{sec:methods} provides the methods. Section~\ref{sec:theoretical_BG} introduces the theoretical background on the quantum dynamics of electronic populations in a three-level system and examines how they depend on modulated fields. Section~\ref{sec:num_results} presents numerical simulations of the MeLPPP system's dynamics. Section~\ref{sec:Methods} describes the generation of the PL database and outlines the optimization algorithms and neural network approaches. Section~\ref{sec:Results_Dissc} presents and discusses the results. Finally, Section~\ref{sec:Conclusions} summarizes the conclusions of this work.
\begin{figure}[h!]
\centering
\includegraphics[width=\linewidth]{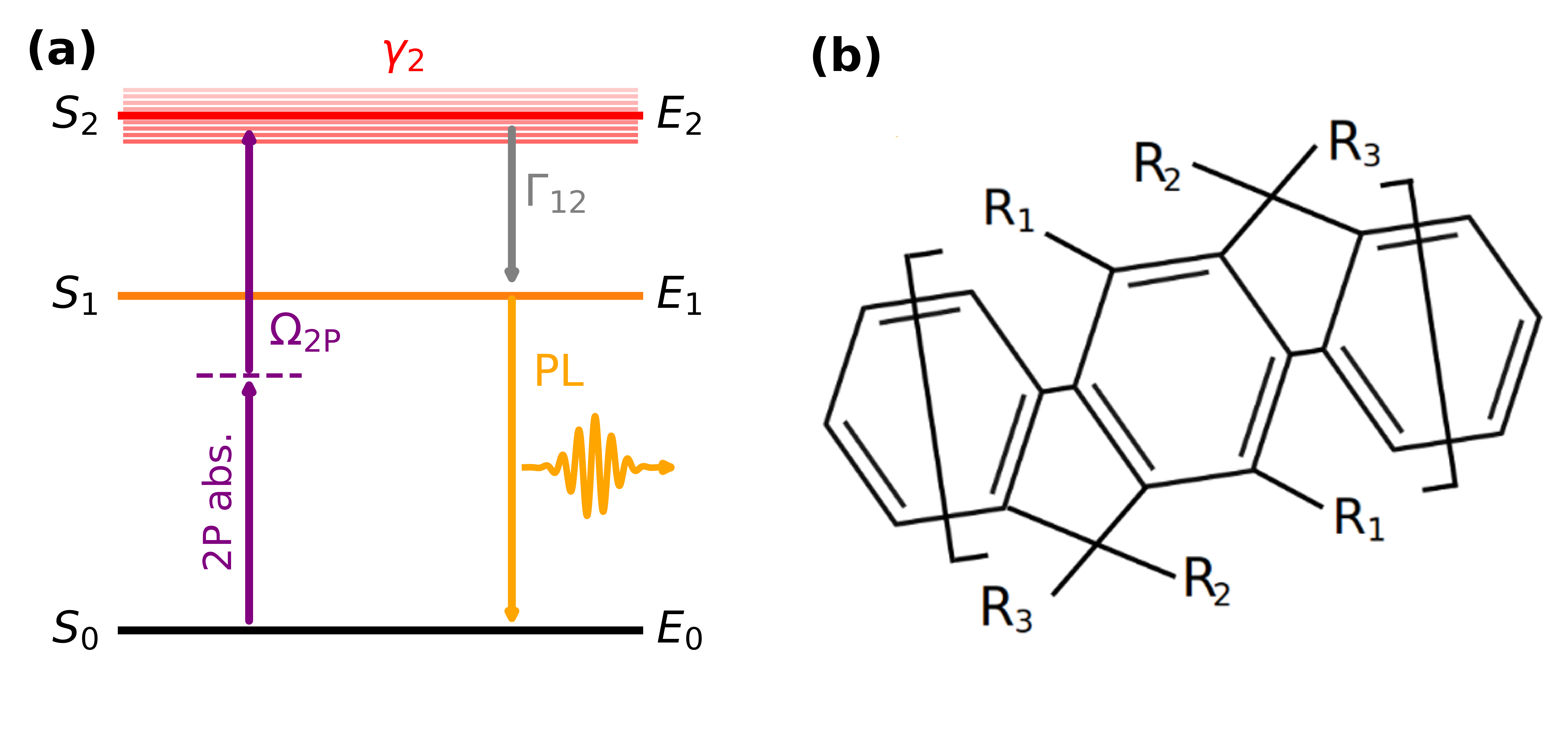}
  \caption{(a) Schematics of the individual molecules energy levels description. $E_i$ denotes the three-level energies of the molecule with corresponding two-photon absorption (2P abs.) transition  between $E_0$ and $E_2$ at Rabi frequency $\Omega_{\text{2P}}$ and $|S_i\rangle \equiv|i\rangle$ represents the electronic states. The relaxation processes between levels $E_2 \to E_1$ occur at a rate $\Gamma_{12}$, and dephasing in the singlet state $S_2$ at a rate $\gamma_2$. (b) Molecular structure of the MeLPPP (R1, n-hexyl; R2, methyl; R3, 1,4-decylphenyl).}
\label{fig:energy_levels}
\end{figure}


\section{Methods \label{sec:methods}}

\subsection{Theoretical framework\label{sec:theoretical_BG}}

\begin{figure*}[ht]
    \centering
     \begin{subfigure}[b]{0.34\linewidth}
         \centering
         \includegraphics[width=\linewidth]{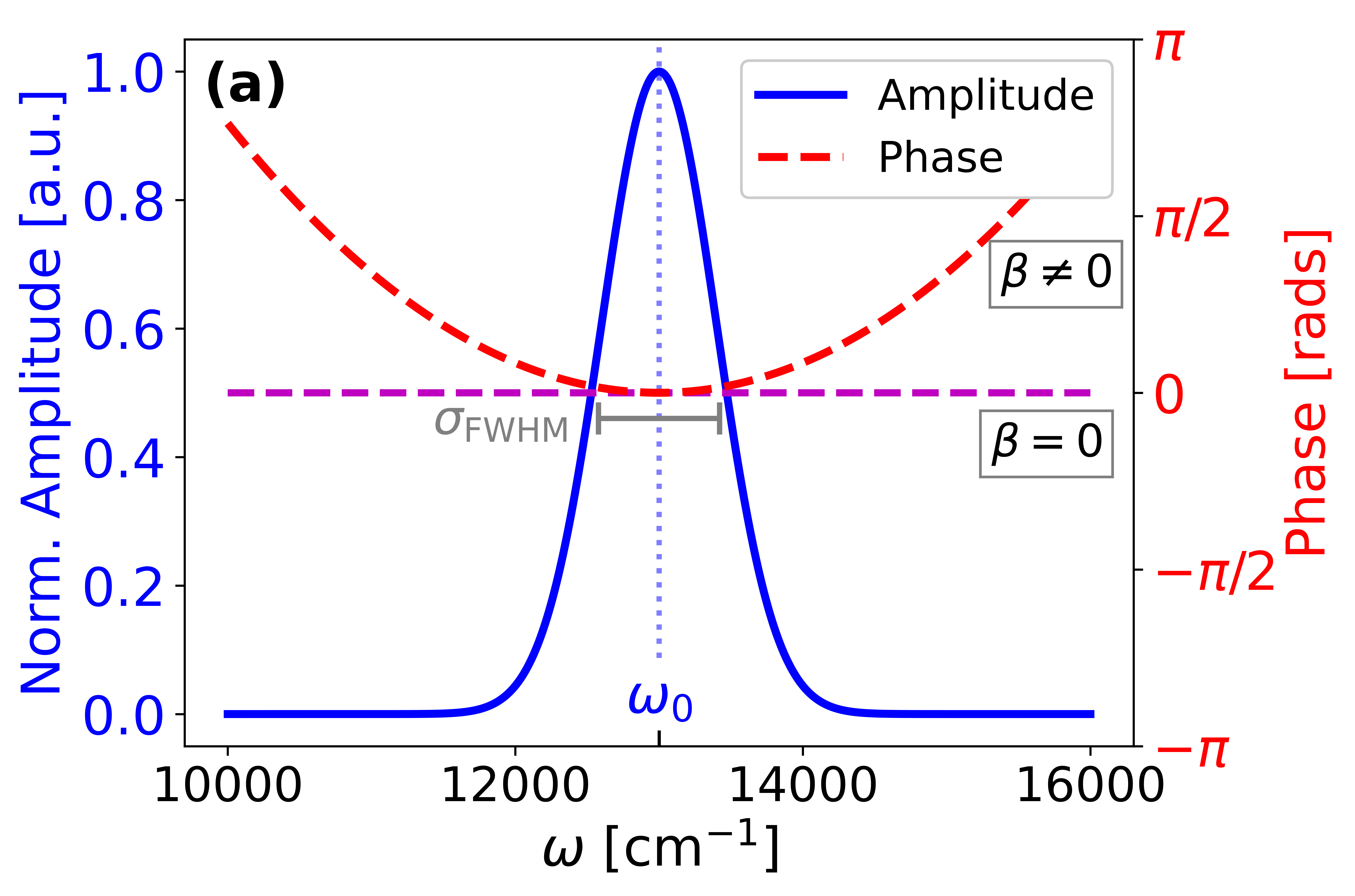}
     \end{subfigure}
     \hfill
     \begin{subfigure}[b]{0.31\linewidth}
         \centering
\includegraphics[width=\linewidth]{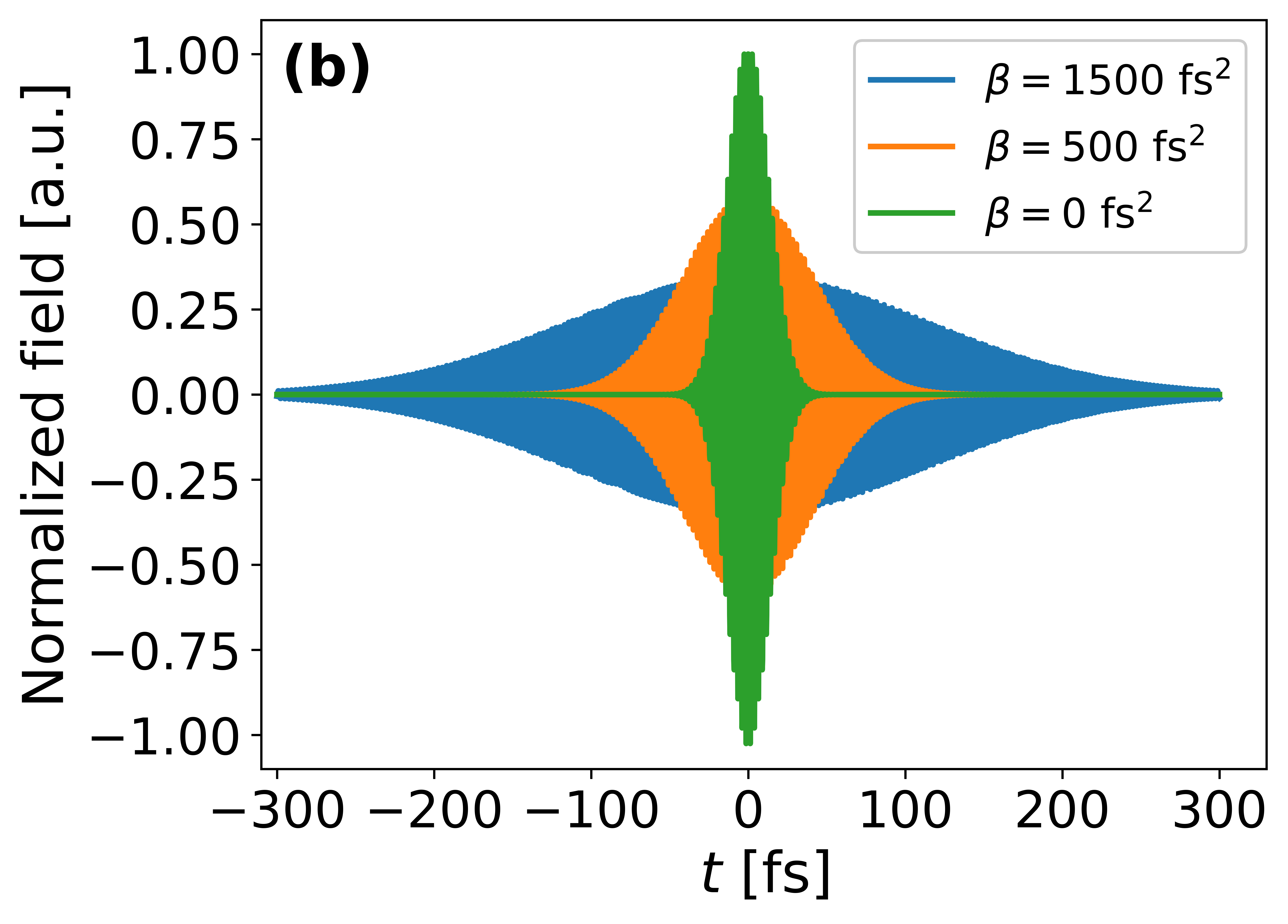}
     \end{subfigure}
     \hfill
     \begin{subfigure}[b]{0.31\linewidth}
         \centering
         \includegraphics[width=\linewidth]{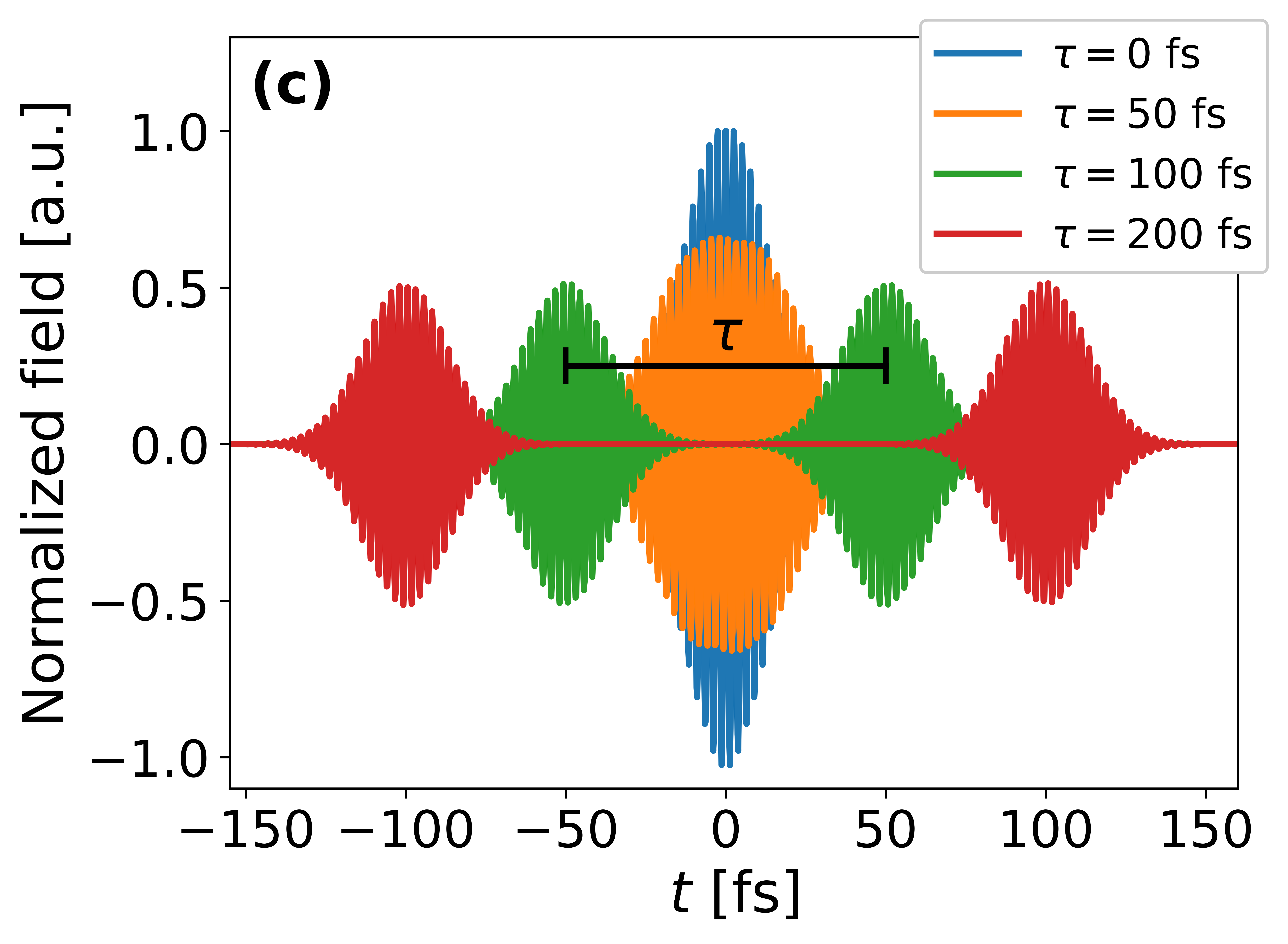}
     \end{subfigure}
        \caption{(a) Frequency domain profile of the field. The solid line represents the normalized spectrum $A(\omega)$ with central frequency $\omega_0= 12~987 \cm$ and spectral width $\sigma_\text{FWHM}=400 \cm$. The dashed lines represent the spectral phase $\varphi(\omega)$ for chirped $(\beta\neq 0)$ and unchirped $(\beta=0)$ fields. (b) Normalized temporal profile of the chirped field (see Eq.~\eqref{eq:phase_beta_mask}) for different $\beta$ values. (c) Normalized temporal profile of the  field with amplitude and phase modulation masks 
        (see Eq.~\eqref{eq:tau_mod_mask}) for different $\tau$ values.}
        \label{fig:Field_Profiles}
\end{figure*}

The molecular system can be described by a three level time-dependent Hamiltonian of the form
\begin{equation}
    \hat{H}(t) = \hat{H}_0 + f(t)\; \hat{H}_1,
\end{equation}
where $\hat{H}_0$ is diagonal with respect to its electronic energies $E_0$, $E_1$, and $E_2$. The function $f(t)$ represents the time-dependent driving corresponding to the laser-modulated pulses, and $\hat{H}_1$ accounts for the coupling between the states associated with the transitions. Specifically, the Hamiltonian of a non-linear two-photon absorption process between the energies $E_0$ and $E_2$~\cite{Buchleitner_TheoryCoherentControl, Silberberg_TP_fsLaser, Silberberg_TP_CohControl_BDC_light} is represented as
\begin{equation}
\label{eq:Ham}
    \hat{H}(t)=\begin{bmatrix} E_0 & 0 & -\Omega_{\text{2P}} ~ \text{Re} \lbrace E^2(t)\rbrace\\ 
0 & E_1 & 0\\ 
-\Omega_{\text{2P}} ~ \text{Re} \lbrace E^2(t)\rbrace & 0 & E_2 
\end{bmatrix},
\end{equation}
where $\Omega_{2\text{P}}$ denotes the two-photon Rabi frequency, taking into account the field intensity and the dipolar moment of the  $S_0 \to S_2$ transition (see Fig.~\ref{fig:energy_levels}(a)),  and $E(t)$ represents the normalized (complex) electric field of the chirped laser pulse. In the frequency domain, this electric field is expressed as $E(\omega) = A(\omega)e^{i\varphi(\omega)}$, comprising both spectral amplitude and phase components. In the time domain, the intensity field reads
\begin{equation}
    E(t)=\frac{1}{\sqrt{2\pi}} \int_{-\infty}^{\infty} A(\omega) M(\omega, \tau) \;e^{i\varphi(\omega, \tau, \beta)} \;e^{-i\omega t} \text{d}\omega,
    \label{eq:field}
\end{equation}
\noindent
where $A(\omega)$ represents the laser spectrum function (see, e.g.~Fig.~\ref{fig:Field_Profiles}a).
The parameters $\tau$ and $\beta$  control pulse shaping and  comprise both spectral phase modulation $\varphi(\omega, \beta, \tau)$, with quadratic chirp $\beta$, and amplitude mask modulation $M(\omega, \tau)$, with delay time $\tau$. 
These modulations affect the driving effects on the system dynamics and control the photoluminescence of the molecule~\cite{Buchleitner_TheoryCoherentControl}. The modulated field profiles are shown in Fig.~\ref{fig:Field_Profiles}, where panel (a) shows the pulse spectrum, while panels (b) and (c) depict the corresponding pulse broadening and splitting induced by the control parameters.

The dynamics of the electronic populations and coherences of an $N$-level system embedded in a dissipative environment and driven by the time-dependent Hamiltonian~Eq.~\eqref{eq:Ham}   can be described by the density operator  $\hat{\rho}$, as follows~\cite{mastereq2024}:
\begin{equation}
    \frac{d}{dt} \hat{\rho} (t) = \frac{-i}{\hbar} \left[\hat{H}(t), \hat{\rho}  \right] + \sum_{i \neq j=0}^{N-1} \mathcal{L}_{ij} (\hat{\rho}) + \sum_{i=0}^{N-1} \mathcal{D}_i (\hat{\rho}),
    \label{Eq:Lindblad_master_eqn}
\end{equation}
with superoperators
$\mathcal{L}_{ij} = \Gamma_{ij} \left( \rho_{ii}|j \rangle\langle j | - \frac{1}{2} \left\lbrace |i\rangle\langle i|, \hat{\rho}  \right\rbrace \right)$, and $\mathcal{D}_i =\frac{\gamma_i}{2}  \left( \rho_{ii}|i \rangle\langle i | - \frac{1}{2} \left\lbrace |i\rangle\langle i|, \hat{\rho}  \right\rbrace     \right)$. Here, $\rho_{ij}$, with $i,j = 0,1,2$, denotes the matrix elements of the density operator, and the computational basis $\left\lbrace \lvert 0 \rangle, \lvert 1 \rangle, \lvert 2 \rangle \right\rbrace$ is used.
$\mathcal{L}_{ij}$ and $\mathcal{D}_i$ 
describe the relaxation processes between two energy levels $E_j \to E_i$ at a rate $\Gamma_{ij}$, and dephasing in the state $S_k$ at a rate $\gamma_k$, respectively. Within this formalism, $\gamma_i$ and $\Gamma_{ij}$  completely encode the system-environment quantum dissipative effects.

\subsection{Numerical simulations\label{sec:num_results}}

For the three-level system depicted in Fig.~\ref{fig:energy_levels}(a), we consider the initial ground state $\hat{\rho}(0) = |0\rangle \langle 0|$. Over time, the system is driven by the pulse $E(t)$, resulting in a transient population transfer between states $S_0$ and $S_2$ via a two-photon absorption process.  The population in the virtual state $S_2$ then undergoes dephasing at a rate $\gamma_2$ and eventually relaxes to the $S_1$ level  at a rate $\Gamma_{12}$. At later times, compared to the pulse duration, populations of the three levels decay to their steady-state values, with the photoluminescent state $S_1$ becoming predominant.

For the MeLPPP molecule, the eigenenergies $E_0=0$ and $E_1=22\;000$ cm$^{-1}$ are retrieved from the one-photon emission characteristic spectrum~\cite{MeLPPP_spectrumVibrat, MeLPPP_univPicture_Spectroscopy}. The $S_2$ energy $E_2=25\;940$ cm$^{-1}$, relaxation rate $\Gamma_{12}=1/190$ fs$^{-1}$, and dephasing rate $\gamma_2=1/61$ fs$^{-1}$ are taken from the numerical results reported in~\cite{VisualizingWilma2019, TwoPhotonWilma2019}. For red-detuned light from a typical Titanium Sapphire laser, the Gaussian spectrum is centered at $\lambda_0=770$ nm ($12~987$ cm$^{-1}$) with a spectral 
full width at half maximum (FWHM) $\sigma_\text{FWHM}=24.3$ nm, corresponding to a Fourier-limited pulse of 40 fs.

\begin{figure}[h!]
\includegraphics[width=\linewidth]{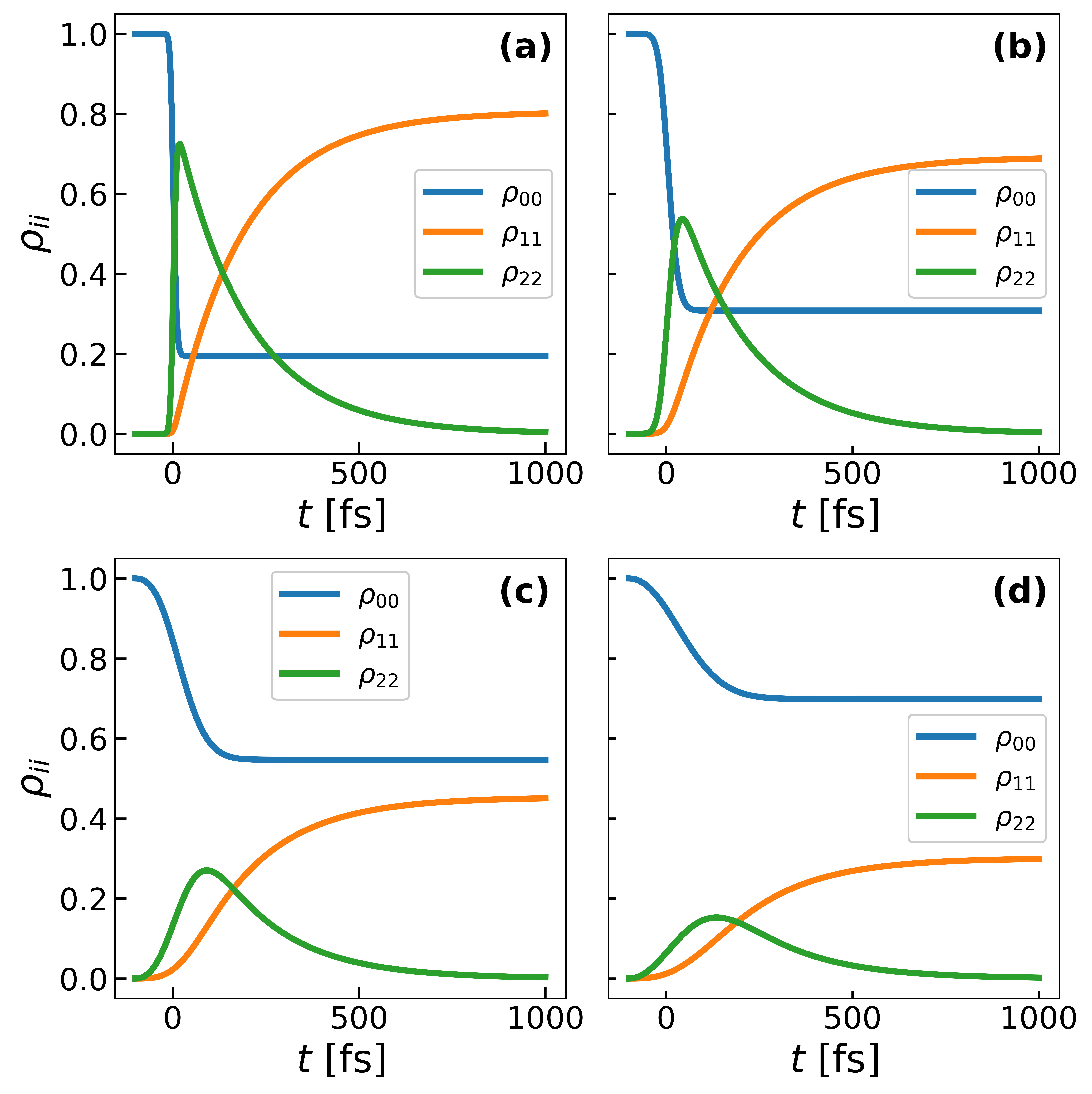}
\caption{Population dynamics of the MeLPPP molecule for $\Omega_{\text{2P}}=531\text{ cm}^{-1}$, $\gamma_2=1/61 \fs$, $\Gamma_{12}=1/190 \fs$ and   control parameter values: (a) $\beta=0$~fs$^2$,  (b) $\beta=500$~fs$^2$,  (c) $\beta=1500$~fs$^2$, and (d) $\beta=2500$~fs$^2$.}
\label{fig:beta_dynamics}
\end{figure}

In a first approach, we consider a modulation applied only to the spectral phase in quadratic form
\begin{equation}
    M(\omega, 0)=1 ,\quad\quad\;
\varphi(\omega, \beta) = \frac{1}{2}\beta(\omega - \omega_0)^2,
\label{eq:phase_beta_mask}
\end{equation}
with no modulation in the spectral amplitude. Figure~\ref{fig:beta_dynamics} shows the population evolution of the MeLPPP system for different values of $\beta$, demonstrating that the control parameter clearly reshapes the coupling and transient effects of the levels. 

At long times, the populations reach their steady values, allowing the evaluation of the expected PL associated with the corresponding quantum dynamics. This is shown in  Fig.~\ref{fig:beta_dynamics}, for small values of $\beta$ ($\beta < 1000~\mathrm{fs}^2$), the $S_1$ state is predominantly reached. For larger values, however, there is less excitation of $S_1$, resulting in a lower PL signal. Since the pulse $E(t)$ is centered at $t = 0~\mathrm{fs}$, the initial conditions are set at $t = -100~\mathrm{fs}$, and the steady-state values are taken at $t = 1~\mathrm{ps}$.

\begin{figure}[h!]
    \centering
\includegraphics[width=\linewidth]{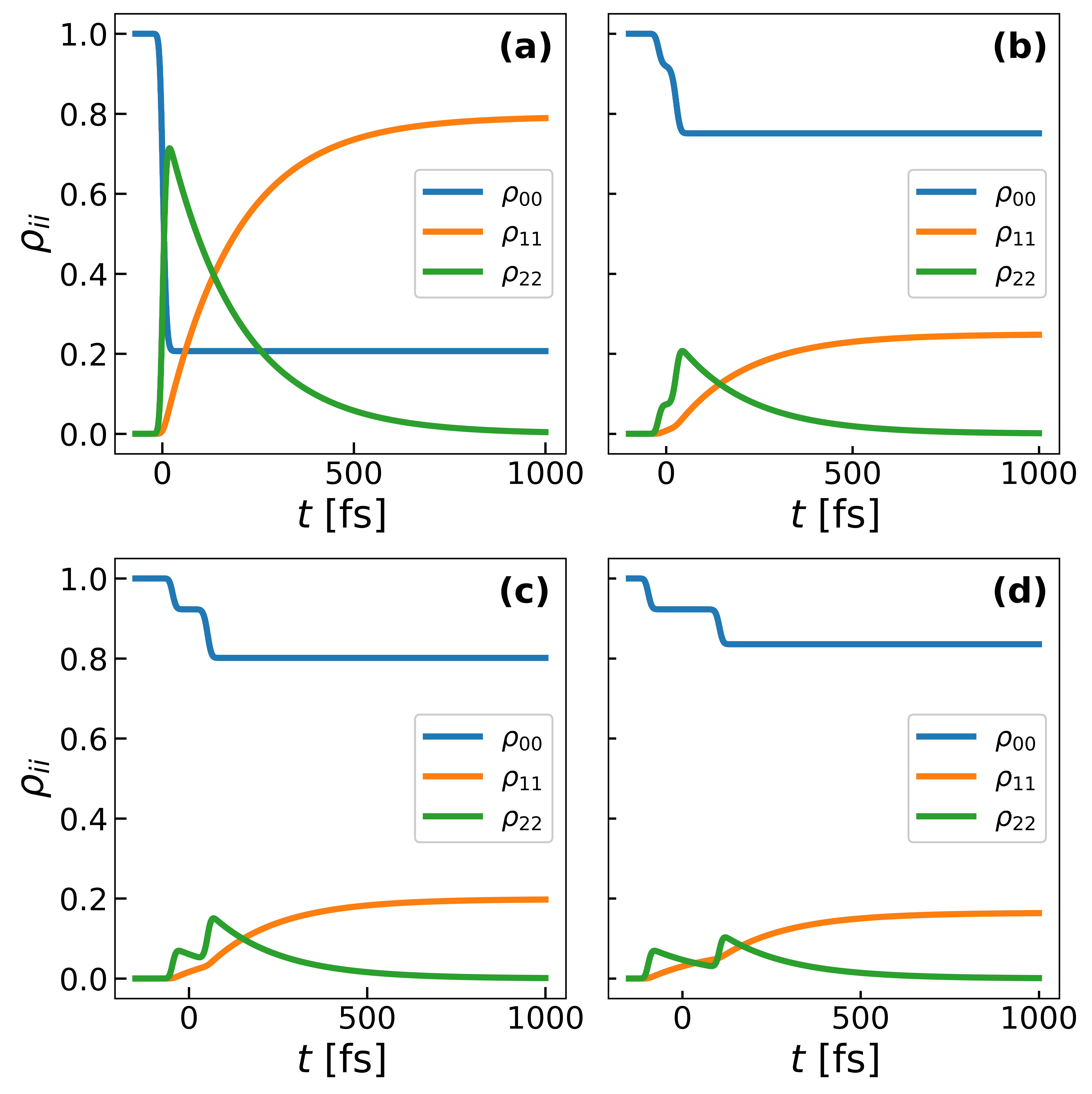}
    \caption{Ultrafast population dynamics of the MeLPPP molecule with values $\Omega_{\text{2P}}=531\text{ cm}^{-1}$, $\gamma_2=1/61 \fs$, $\Gamma_{12}=1/190 \fs$ and  different control parameter: (a) $\tau=0$ fs (b) $\tau=50$ fs (c) $\tau=100$ fs and (d) $\tau=200$ fs.}
    \label{fig:tau_dynamics}
\end{figure}

As a second approach, we consider spectral and phase modulation as follows:
\begin{align}
    M(\omega, \tau) & =  \left|\cos\left( \frac{1}{2}(\omega-\omega_0)\tau\right)\right| ,\\
\varphi(\omega, \tau) & =   \frac{\pi}{2} \text{sgn}\left[ \cos\left( \frac{1}{2}(\omega-\omega_0)\tau\right) \right],
\label{eq:tau_mod_mask}
\end{align}
where $\text{sgn}\, (\cdot)$ represents the sign function. Figure~\ref{fig:Field_Profiles} shows the modulation effects on the pulses. The results for  populations driven by these pulses are shown in Fig.~\ref{fig:tau_dynamics}.\\

For this second modulation, the population $\rho_{11}$ reaches its maximum at small values of $\tau$ ($\tau < 50~\mathrm{fs}$), corresponding to pulses that are split into two subpulses of half amplitude (see Fig.~\ref{fig:Field_Profiles}(c)). Consequently, this modulation offers only a limited degree of control over the PL intensity.

As can be seen in Figs.~\ref{fig:beta_dynamics} and~\ref{fig:tau_dynamics}, the control parameters $\beta$ and $\tau$ decisively determine the long-time behavior of the system. These parameters govern the final population of the steady state, particularly that of state $\rho_{11}$, the target level accessed via  the two-photon absorption process. The behavior of $\rho_{11}$ provides direct information about the photoluminescence of the molecular sample because the PL intensity is linearly related to this density matrix element~\cite{TwoPhotonWilma2019, VisualizingWilma2019}.
\begin{figure}[h!]
    \centering
\includegraphics[width=\linewidth]{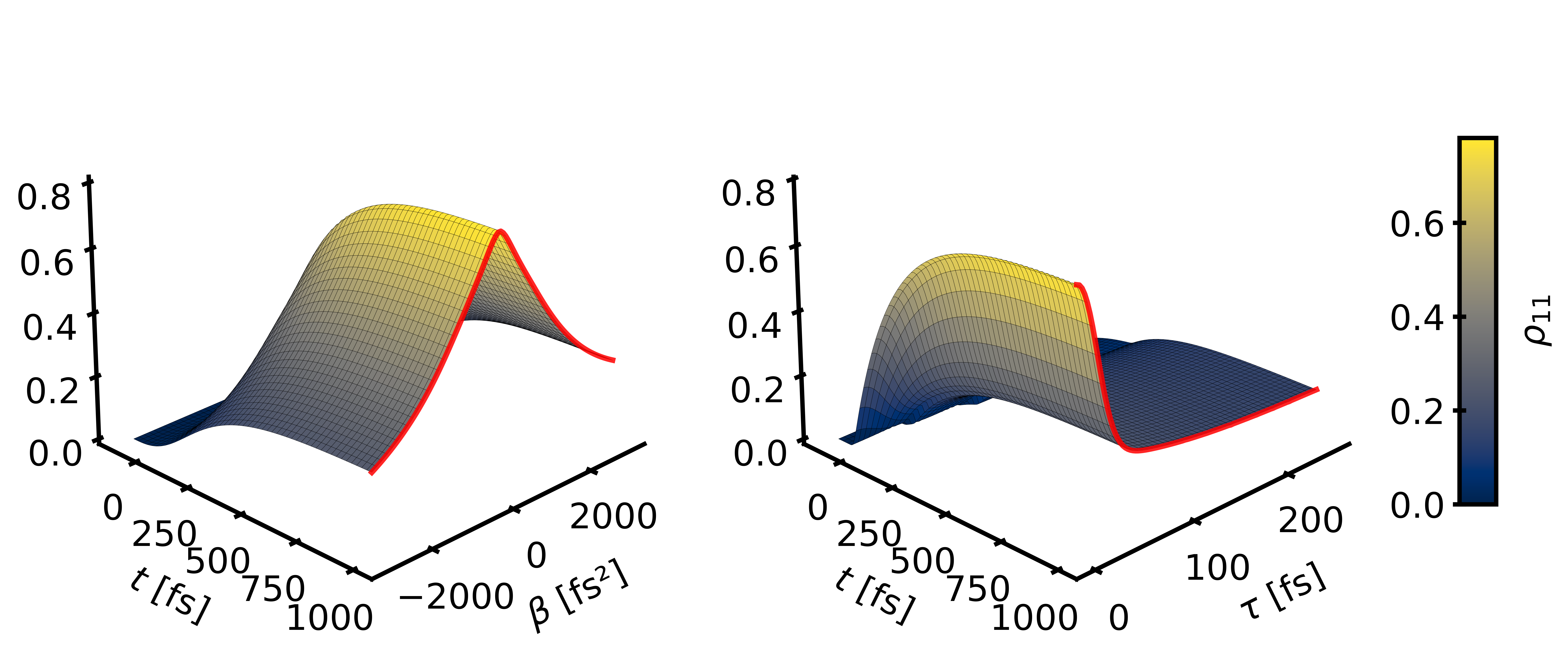}
\caption{Full map of the ultrafast dynamics of the population $\rho_{11}$ of the MeLPPP molecule as a function of time and the control parameters $\beta$ and $\tau$. At steady-state times ($\sim1~\text{ps}$), the obtained values reproduce the shape of the PL signal, as indicated by the red solid line. Maximum values of $\rho_{11}$ occur at $\beta = 0$ and $\tau = 0$, while deviations from these conditions lead to distinct PL intensities.}
\label{fig:surface_plot}
\end{figure}
\noindent

Figure~\ref{fig:surface_plot} shows the complete time evolution of $\rho_{11}$ as a function of the two control parameters, illustrating how the system's ultrafast dynamics and steady-state behavior are shaped by the coherent control conditions. This representation highlights the strong sensitivity of $\rho_{11}$ to phase dispersion and temporal delay parameters, providing direct insight into the interplay between field coherence, phase modulation and the molecular response.

These simulations reveal that the parameter $\tau$ leads to a pronounced maximum in the PL signal; however, the response rapidly decays as $\tau$ deviates from its optimal value. In contrast, the chirp parameter $\beta$ produces symmetric PL intensities around $\beta = 0$ and enables smoother, more controlled exploration of the parameter space. For this reason, the remainder of this work focuses on algorithms and protocols that rely on PL measurements as a function of the chirp parameter.

\subsection{Optimization Methods
\label{sec:Methods}}

In this section, we present two different numerical approaches for extracting relevant dissipative bath and driving coupling parameters from PL data. The first approach is based on numerically optimizing non-differentiable functions using standard simplex methods, such as the Nelder–Mead algorithm implemented in SciPy~\cite{2020SciPy}. The second approach employs a feed-forward neural network that takes the PL data as input.

\subsubsection{Parameter selection and constructed datasets\label{sec:Database}}
The dynamics and final populations are sensitive to specific parameters related to environmental dissipation and field coupling strength, which lead to appreciable effects on the stationary final state and consequently on the experimentally measured photoluminescence over the ensemble of molecules. As a result of collective emission, the steady-state value of the single-molecule population $\rho_{11}$ is related to the photoluminescence function $\text{PL}(\cdot)$ as~\cite{TwoPhotonWilma2019, VisualizingWilma2019}
\begin{equation}
\text{PL}(\mathcal{Q}) = A\; \rho_{11}(\mathcal{Q}) + B,  \label{eq:PL}
\end{equation}
where the constants $A$ and $B$ scale the experimentally measured PL signal to match the total emission intensity from all molecules within the laser spot and the dark-count background noise, respectively. Since no a priori information is available about the bath effects driving the dissipative dynamics or about the optimal field–system coupling, we selected the parameter set $\mathcal{Q} = [\, E_2 \;\, \Omega_{\text{2P}} \;  \gamma_2 \; \Gamma_{12}\,]$ as free variables to analyze the dependence of the ultrafast dynamics revealed in the PL signal. The remaining parameters were kept fixed, as they correspond to laser configurations and molecular properties that are not optically addressed by the excitation pulses~\cite{TwoPhotonWilma2019, MeLPPP_spectrumVibrat}.

To evaluate both algorithms, the experimental PL data reported in~\cite{VisualizingWilma2019} were used, where single-molecule MeLPPP dissolved in toluene (Sigma-Aldrich, 99.7\%) at a concentration of $10^{-9}$ M containing 5 mg$/$mL polystyrene was employed. This measured PL signal ranges from 169 cps to 705 cps using 55 different $\beta$ values. Therefore we set the scaling parameters to $A = 742.88$ cps and $B = 43.73$ cps.

Accordingly, for the first optimization algorithm a dataset $\mathcal{D}$ of 700 parameter sets was generated to represent the initial conditions of the parameters $\mathcal{Q}$ by randomly sweeping the values within their physical ranges. Two scenarios were considered. The first one includes the full parameter set $\mathcal{Q} = \Qset$, incorporating both the energy $E_2$ of the virtual state $S_2$ reached after two-photon excitation and the remaining parameters associated with driving and environmental decay processes. In the second scenario, the approximation $E_2 = 2\omega_0$ was applied, assuming that the two absorbed photons have an energy equal to the spectral peak, since the spectral FWHM is small compared with the central wavelength~\cite{LineShape_OptimalPL_Emission}. Under this approximation, the optimization problem is reduced to three free parameters. In both cases, these parameters were scanned in the same way. The two-photon Rabi frequency was swept over the range $[200 \cm, 800 \cm]$, while the dephasing rate values were varied within $[0.005 \fs, 0.05 \fs]$ and the decay rate values within $[0.002 \fs, 0.01 \fs]$. These values are consistent with the ultrafast femtosecond time scale of dephasing and decay processes~\cite{CouplingRegimes_MeLPPP_microcavities, ExcitonDiff_Relaxat_MeLPPP, VisualizingWilma2019}, corresponding to dephasing times from 20 fs up to 200 fs and relaxation times from 100 fs up to 500 fs. For the first scenario, $E_2$ was varied within the range $[22\,000 \cm, 30\,000 \cm]$, whereas in the second scenario, the energy was fixed at $E_2 = 2\omega_0 = 25\,940 \cm$.

For the NN based approach, a PL dataset was constructed consisting of $N = 10\,000$ training samples, each containing 55 features corresponding to intensity PL values for different $\beta$ values uniformly distributed within the range $[-3\,000 \;\text{fs}^2, 3\,000 \;\text{fs}^2]$. This dataset was generated by fixing $E_2 = 2\omega_0$ and sweeping the parameters $\mathcal{Q} = \QEset$ within the same intervals as before, and then computing the PL using Eq.~\eqref{eq:PL}. These parameters $\mathcal{Q}$ also represent the target quantities to be predicted and evaluated on the testing dataset.

Since the PL inputs and target variables span heterogeneous scales, ranging from tens to thousands of $\cm$ for the energy and Rabi frequency and from $10^{-2}$ to $10^{-3}$~$\fs$ for the decay rates, three scalers were considered for normalization and rescaling of the parameters: Standard Scaler, Robust Scaler and Robust Scaler with a Winsorization Preprocessing (WP)~\cite{Winsorization, NormalizationEffects_NN_Training}. These feature scaling techniques help to uniformly distribute the input and target values, facilitating the NN training by keeping the outputs on the order of unity, which matches the typical magnitude of their weights, biases and activation functions. Furthermore, the standard algorithms for numerical optimization also require consistent parameter scales to ensure proper convergence and computational efficiency~\cite{ImpactNormalization_ANN}.

\subsubsection{Optimization algorithm
\label{sec:OptimizationAlgo}
}
Using the set of experimental observations defined as $\lbrace(\beta, y_\beta) : \beta = \beta_1, \beta_2, \dots, \beta_n\rbrace$, where $y_\beta$ denotes the measured PL intensity corresponding to a specific value of the control parameter $\beta$, we define the mean squared error (MSE) metric as
\begin{equation}
\text{MSE} = \frac{1}{n} \sum_{{\beta}} \left[\, y_\beta - \text{PL}_\beta(\mathcal{Q}) \,\right]^2 ,
\label{eq:lossfunc}
\end{equation}
where $n$ represents the total number of experimental data points. In our case, we selected the experimental outcomes reported in Ref.~\cite{VisualizingWilma2019} as targets for parameter prediction. Additionally, we impose the constraint that the energy of the virtual state $S_2$, denoted by $E_2$, lies within an energy band close to $2\omega_0$, which corresponds to approximately twice the energy of the laser spectral peak as expected in a two-photon absorption process. To account for this physical constraint, we introduce a penalization term to the MSE, yielding the total loss function
\begin{equation}
\mathcal{L}(\mathcal{Q}) = \frac{1}{N} \sum_{{\beta}} \left[\, y_\beta - \text{PL}_\beta(\mathcal{Q}) \,\right]^2 + \lambda (E_2 - 2\omega_0)^2,
\label{eq:loss_fn_complete}
\end{equation}
where $\lambda$ is the hyperparameter that controls the strength of the constraint. The case $\lambda = 0$ corresponds to the first scenario, in which the approximation $E_2 = 2\omega_0$ is imposed and the optimization is reduced to three free parameters. Conversely, the case $\lambda \ll 1$ represents the full optimization problem, where $E_2$ is treated as an additional free variable. 

By minimizing the loss function in Eq.~\eqref{eq:loss_fn_complete}, we obtain the optimal set of parameters that best relate the experimental outcomes to the numerical simulations. Hence, the molecular and bath properties can be extracted from these measurements by solving the optimization task
\begin{equation}
\mathcal{Q}^* = \arg \min_{\mathcal{D}} \mathcal{L}(\mathcal{Q}) \,,
\label{eq:min_objective}
\end{equation}
where $\mathcal{Q}^*$ denotes the optimal parameter configuration. We employ the standard \texttt{minimize} function from \texttt{SciPy} using the Nelder–Mead algorithm~\cite{2020SciPy} to minimize the loss function in Eq.~\eqref{eq:loss_fn_complete} across the parameter set $\mathcal{Q}$. This choice is motivated by the fact that the problem involves the minimization of a black-box function with no analytical gradients and computationally expensive evaluations. A total of 700 initial conditions from $\mathcal{D}$ were tested in parallel to ensure convergence robustness. The optimization was carried out with a maximum of 500 iterations and a function tolerance (\texttt{ftol}) of $10^{-4}$, which provided a good balance between accuracy and computational efficiency.

\subsubsection{Feed forward neural network
\label{sec:NN_methods}
}

\begin{figure}[h!]
    \centering
\includegraphics[width=\linewidth]{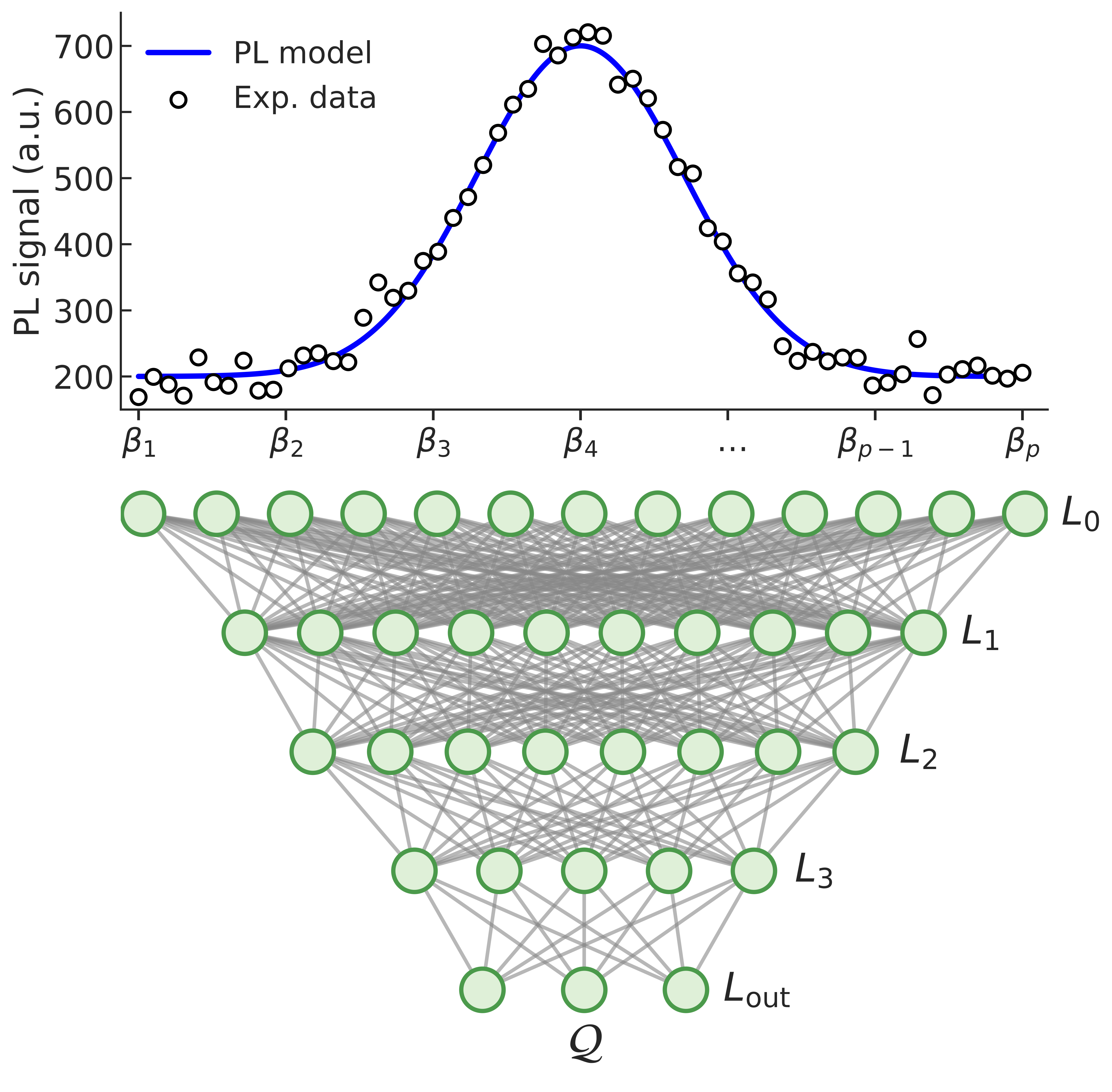}
    \caption{Sketch of the feed-forward neural network (FFNN) architecture used for the regression task. The input layer $L_0$ consists of $p = 55$ neurons corresponding to the PL components for different control parameters $\beta_k$. The hidden layers, with dimensions $[L_1, L_2, L_3, \dots]$, each employ the ReLU($\cdot$) activation function, while the output layer $L_{\text{out}}$ contains three neurons corresponding to the predicted parameters $\mathcal{Q}$.}
    \label{fig:FFNN_sketch}
\end{figure}

As an alternative approach, we implemented a neural network model that takes the PL values as input features and returns the predicted physical parameters. In this framework, we aim to find an optimal mapping function represented by a feed-forward neural network (FFNN) that learns the relationships and underlying patterns between the PL inputs and their associated parameters $\mathcal{Q}$, as illustrated in Fig.~\ref{fig:FFNN_sketch}. To obtain a highly accurate and sensitive predictor, we performed a systematic exploration of different neural network architectures and training strategies to evaluate their performance on unseen data.

The networks were implemented in \texttt{PyTorch}~\cite{Pytorch} to perform supervised learning for regression tasks, exploring ten different architectures from 1 to 10 hidden layers. Training was carried out by minimizing a mean-squared-error (MSE) cost function between the true and predicted values of $\mathcal{Q}$ using the PL dataset as input. Given the large amount of training data, the loss function was computed by dividing the dataset into mini-batches. Consequently, the total loss is expressed as
\begin{equation}
C(\mathcal{Q}, \tilde{\mathcal{Q}}) = \frac{1}{N}\sum_{k=1}^{N/\mathrm{Bs}} \sum_{i \in M_k} ||\mathcal{Q}_i - \tilde{\mathcal{Q}}_i||^2,
\label{eq:MSE_loss}
\end{equation}
where $\mathcal{Q}_i$ ($\tilde{\mathcal{Q}}_i$) denotes the target (predicted) values and $M_k$ is the $k$-th mini-batch with size $\mathrm{Bs} = |M_k|$. The training and validation sets contained $\frac{3}{5}N$ and $\frac{1}{5}N$ samples, respectively, where $N$ is the total number of data points, while the remaining data were used as a test set to assess the network predictive performance and generalization ability. Training was performed for 500 epochs, with batch sizes (Bs) ranging from 32 to 256 and learning rates between $10^{-1}$ and $10^{-6}$, ensuring proper convergence of the parameters and stability of the training process.

For the regression task, we used a total of $p = 55$ predictors corresponding to the PL values at different $\beta$ settings as input features to the network. The output layer was defined as a three-component vector representing the target parameters $\mathcal{Q}$, with no activation function applied. The number of hidden layers $[L_1, L_2, \dots, L_h]$ was varied from $h = 1$ to $h = 11$ to evaluate the model performance and determine the optimal architecture that minimized the validation loss. Each hidden layer employed the ReLU activation function, defined as $\text{ReLU}(x) = \max\lbrace x, 0\rbrace$, where the operation is applied element-wise over the vector components. The network is trained to determine the optimal weights and biases $\theta=\lbrace W_i, b_i\rbrace$ that minimize the cost function given in Eq.~\eqref{eq:MSE_loss}. For this task, the Adam optimizer~\cite{AutomaticDiff_Pytorch} was used, which evaluates the $n$-th gradient of the cost function $g^{(n)} = \nabla_\theta C\left(\theta^{(n)}\right)$ and, after random initialization, calculated iteratively the parameters $\theta^{(n)}$ until the stopping criterion is met. This approach allows for efficient and robust optimization of the network parameters, ensuring fast and stable convergence in few number of epochs.

\section{Results and discussion\label{sec:Results_Dissc}}

As a first step, a Monte Carlo sweep was performed over the entire parameter space $\mathcal{D}$ to use these values as initial conditions for the optimizer, testing in total 700 different values in order to analyze the parameters trending, convergence and statistical errors.

\begin{figure}[h!]
    \centering
\includegraphics[width=\linewidth]{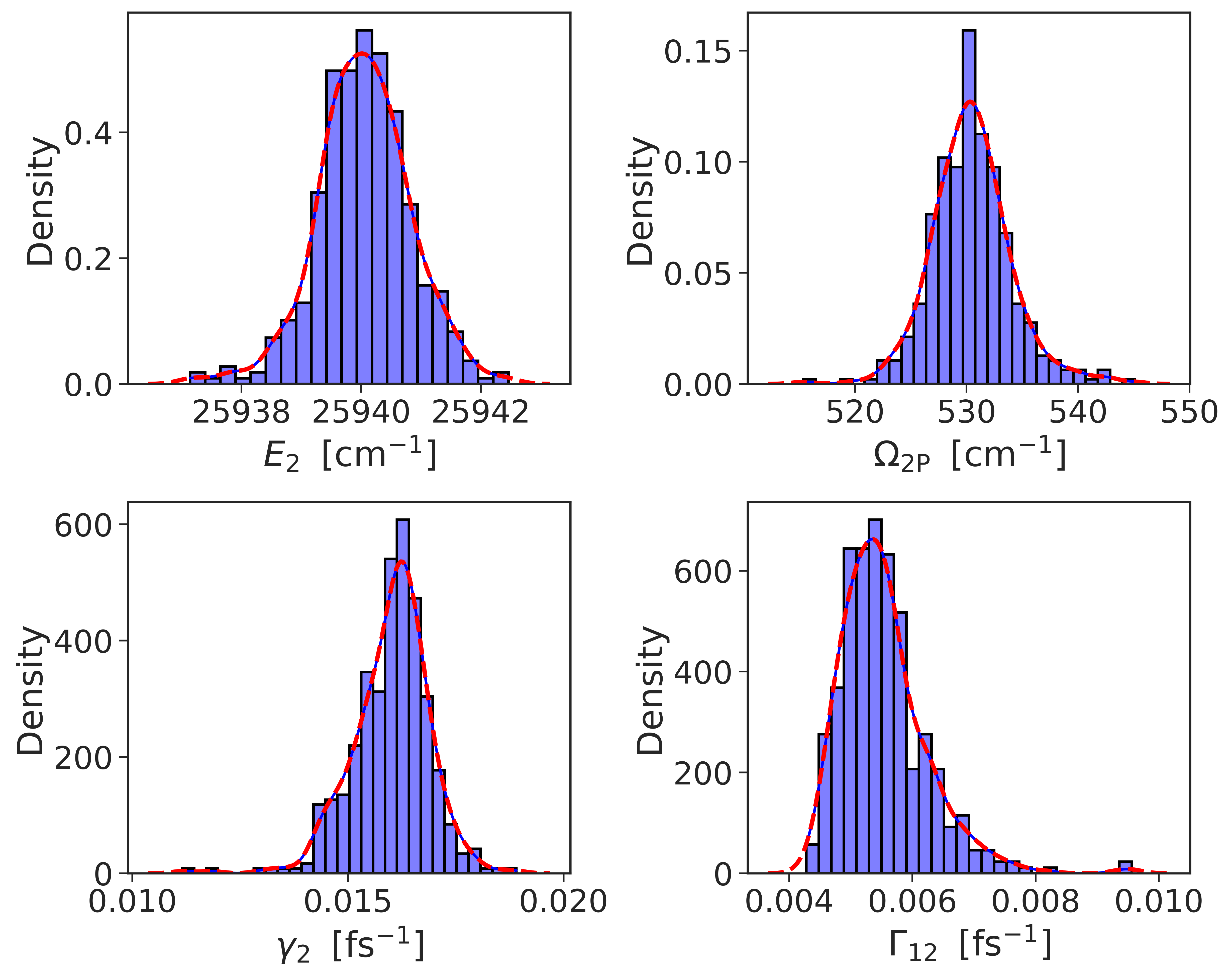}
    \caption{Histogram showing the distribution of the optimal parameter values $\Q^*$ obtained for all of the iterations of the optimization algorithm using $\lambda=10^{-2}$, \texttt{maxiter}=500 and \texttt{ftol}=$10^{-4}$.}
    \label{fig:histograms}
\end{figure}

In the first scenario, where all parameters were included, different values of the hyperparameter $\lambda$ were tested in order to analyze the convergence behavior as a function of this regularization strength. For large values of $\lambda \gg 1$, the optimization strongly enforced the constraint $E_2 \rightarrow 2\omega_0$, leading to a pronounced convergence toward this value but at the expense of a higher overall mean squared error (MSE). Conversely, for smaller values of $\lambda = 10^{-2}$ and $\lambda = 10^{-3}$, a more balanced behavior was achieved, in which the Ridge-like penalization~\cite{ML_probabilisticPersp_Murphy, Vapnik_2018} applied to the parameter $E_2$ remained significant while the total MSE was slightly affected. After collecting the final optimized parameters from all iterations, a histogram of the resulting values was constructed to visualize the distribution and identify emerging trends in the predicted parameters, as shown in Fig.~\ref{fig:histograms}. This analysis allowed us to assess the stability of the optimization outcomes and the sensitivity of the extracted parameters with respect to variations in the regularization strength $\lambda$.

From these histograms, the mean values obtained across all iterations were computed as the final results for each parameter. For the energy, a mean value of $\bar{E}_2 = 25\,940.04~\cm$ was obtained, together with a two-photon Rabi frequency of $\bar{\Omega}_{\text{2P}} = 530.45~\cm$, a dephasing rate of $\bar{\gamma}_2 = 0.016005~\fs$ and a decay rate of $\bar{\Gamma}_{12} = 0.0055336~\fs$. In contrast, for the second scenario, where $\lambda = 0$ and the energy was fixed at $E_2 = 25\,940~\text{cm}^{-1}$, the corresponding mean values were $\bar{\Omega}_{\text{2P}} = 533.04~\cm$, $\bar{\gamma}_2 = 0.016495~\fs$, and $\bar{\Gamma}_{12} = 0.0050602~\fs$. These results confirm the numerical consistency between both scenarios, with only minor deviations attributable to the imposed regularization constraint on $E_2$, which effectively reduces the parameter space from four to three variables.

\begin{figure}[h!]
    \centering
\includegraphics[width=\linewidth]{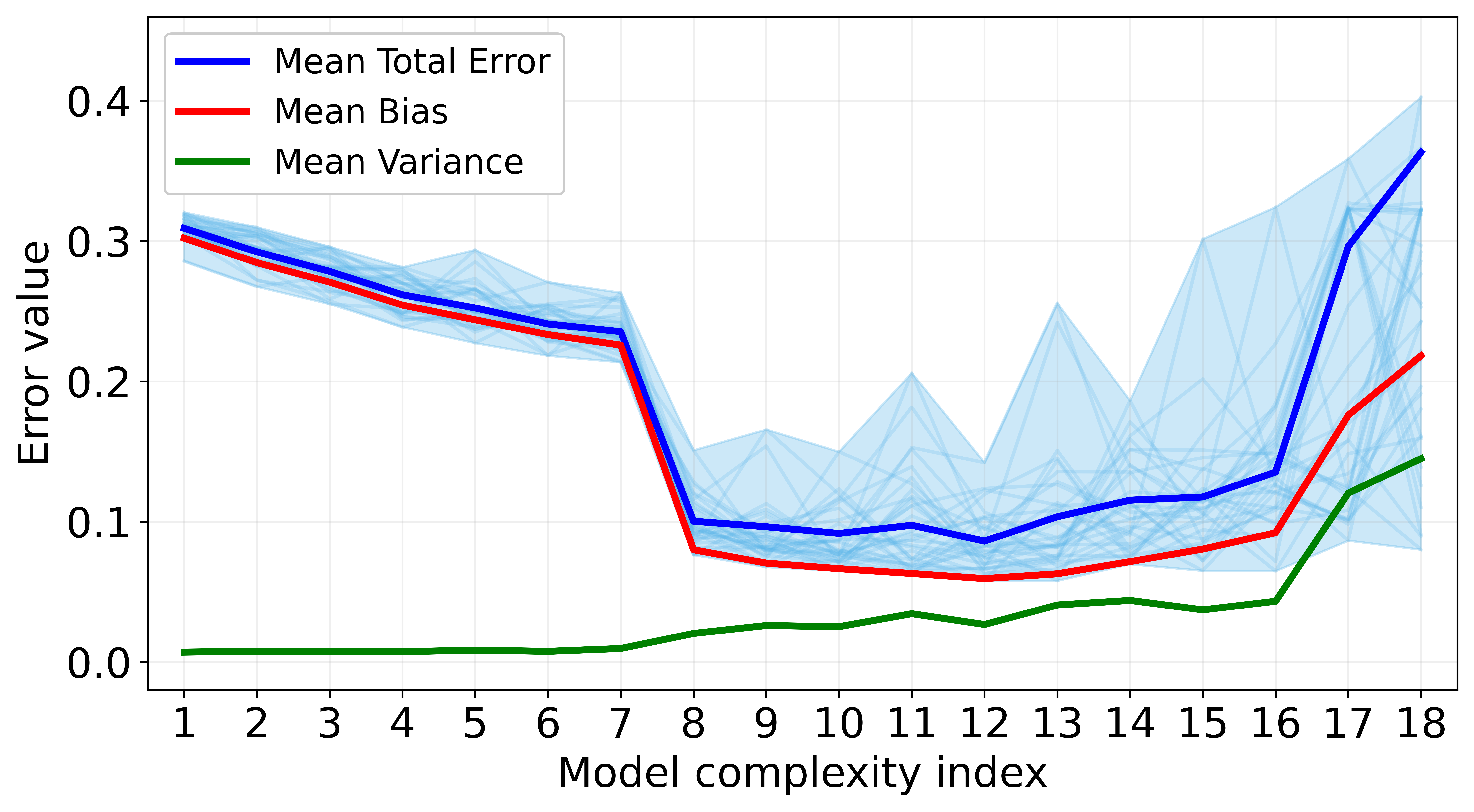}
    \caption{Bias–variance tradeoff curve showing the mean total error, mean bias squared and variance as a function of the model complexity  of the neural network. The complexity indices are listed in the Supplementary Information, Appendix \ref{app:Optim_Algorithm} Table~\ref{tab:BVTO_indexes}. The semi-transparent blue curves correspond to the ten new models generated through bagging~\cite{Goodfellow2016}, while the solid blue curve represents their average.}
    \label{fig:bias-var_tradeoff}
\end{figure}

For the neural network design, an initial sweep was performed over the possible architectures of the hidden layers to determine the optimal configuration that avoids overfitting in the training data while still capturing all relevant patterns and features of the input data to accurately predict the parameters. In Fig.~\ref{fig:bias-var_tradeoff} the Bias–Variance tradeoff curve~\cite{ML_probabilisticPersp_Murphy} was constructed using the validation set to compute the bias, variance and total MSE loss. For all training runs and for performing bagging across architectures, 300 epochs were used with a learning rate of $\delta = 10^{-3}$ and a batch size of $\text{Bs}=256$. 

From the figure, it can be observed that the best trade-off point is achieved at index 12 which corresponds to an optimal architecture consisting of three hidden layers of sizes $[128, 64, 32]$ (where the numbers refer to the number of nodes in each hidden layer), which was then used to perform hyperparameter tuning on the remaining training parameters to achieve fast and stable convergence within the epochs. This hyperparameter sweep was carried out considering a fixed number of 200 epochs, batch sizes of $\text{Bs}=\lbrace64, 128, 256, 512\rbrace$, and learning rates of $\delta=\lbrace 10^{-4}, 10^{-3}, 10^{-2}, 10^{-1}, 1\rbrace$, while also including and excluding dropout in the activation functions of the hidden layers. The lowest validation loss was achieved with a batch size of $\text{Bs}=64$, a learning rate of $\delta=10^{-4}$, and a dropout rate of $0.0$ (no dropout), see Appendix~\ref{app:NN_training}.

After training, the neural network was evaluated on the test set. For the three output parameter values, the performance metrics computed were: Root Mean Square Error (MSE), Mean Absolute Error (MAE), and the $R^2$ coefficient~\cite{ML_metrics}. Since three different scaling and normalization methods were applied to the data, these metrics were evaluated separately for each case. The results of the corresponding performance metrics are summarized in Table~\ref{tab:scaling_metrics}.
\begin{table}[h!]
\centering
\begin{tabular}{|c|c|c|c|c|}
\hline
\textbf{Method} & \textbf{Target} & \textbf{RMSE} & \textbf{MAE} & $\mathbf{R^2}$ \\
\hline
Standard Scaling & $\Omega_{\text{2P}}$ & 0.0283 & 0.0222 & 0.9992 \\
Standard Scaling & $\gamma_2$           & 0.0346 & 0.0228 & 0.9989 \\
Standard Scaling & $\Gamma_{12}$        & 0.1549 & 0.0850 & 0.9754 \\
\hline
Robust Scaling   & $\Omega_{\text{2P}}$ & 0.0318 & 0.0233 & 0.9990 \\
Robust Scaling   & $\gamma_2$           & 0.0365 & 0.0269 & 0.9987 \\
Robust Scaling   & $\Gamma_{12}$        & 0.1700 & 0.0947 & 0.9703 \\
\hline
Robust Scaling + WP & $\Omega_{\text{2P}}$ & 0.0720 & 0.0340 & 0.9949 \\
Robust Scaling + WP & $\gamma_2$           & 0.0997 & 0.0387 & 0.9903 \\
Robust Scaling + WP & $\Gamma_{12}$        & 0.2879 & 0.1277 & 0.9182 \\
\hline
\end{tabular}
\caption{Performance metrics of the neural network for three different scaling methods, computed on the test dataset.}
\label{tab:scaling_metrics}
\end{table}

\begin{figure*}[t!]
    \centering
\includegraphics[width=\textwidth]{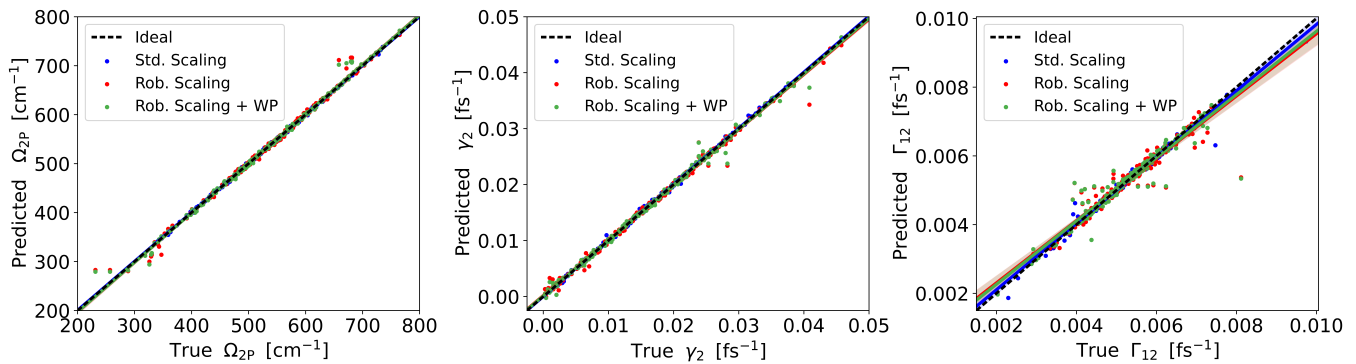}
    \caption{Predicted versus true values for the three target parameters $\Q = \QEset$ using the three different scaling methods calculated over the test dataset, highlighting deviations from the perfect one-to-one fit and including trend lines that illustrate the overall behavior and generalization of the predictions.}
    \label{fig:true_vs_predicted}
\end{figure*}

After evaluating the average performance metrics of the neural network on the entire test dataset, we proceeded to visualize the predictive accuracy of the model.  Figure~\ref{fig:true_vs_predicted} shows the predicted values in comparison with the true target values $\mathcal{Q}$, illustrating the overall agreement between the model outputs and the reference data, as well as the deviations from the ideal perfect fit. The trend lines of the scattered points indicate the general behavior of the predictions over larger value ranges.

Finally, after training and validating the optimal neural network architecture and quantifying its performance on a test set containing samples unseen by the model, we proceeded to evaluate its predictive capability on the real experimental PL data of the MeLPPP molecule reported in~\cite{VisualizingWilma2019}. This step provided a direct connection between the numerical model and the experimental observations, allowing us to assess the ability of the trained model to generalize beyond the simulated PL dataset. Since three neural networks were independently trained using different data scaling and normalization strategies, the predictions of the physical parameters $\mathcal{Q}$ were evaluated separately for each of them. This comparison enables us to identify how the input pre-processing affects the accuracy and robustness of the model when inferring the system parameters from experimental data. The results of this evaluation, together with the comparison between the optimization-based methods and the neural network predictions, are summarized in Table~\ref{tab:hildner_comparison}. In general terms, both algorithms performed well and successfully predicted the physical parameters associated with the molecular system, its environment, and the system–field coupling. However, depending on the specific target parameter, some methods outperformed others, as shown in Table~\ref{tab:scaling_metrics} and Table~\ref{tab:hildner_comparison}.

Regarding the first approach based on the optimization algorithm, despite its remarkable sensitivity to the choice of initial parameters during the minimization process, it was observed that when sampling across a wide range of initial conditions, the resulting estimates follow a consistent trend that agrees with the typical values reported in the literature~\cite{TwoPhotonWilma2019,VisualizingWilma2019}. Consequently, this method presents the drawback that the physical parameters $\Q$ cannot be reliably extracted from experimental PL data in a single optimization run. Instead, a broad sampling of initial conditions is required to reveal clear and stable parameter trends. Additionally, for the algorithm to operate properly, it is crucial to define appropriate physical constraints to ensure robust convergence. In turn, this requires prior knowledge or qualitative insights into the molecular system under coherent control, such as its relevant electronic states and lifetimes, or the energetic scales of the driving and dissipative processes, including temperature and solvent effects~\cite{ultrafastcoherence2021,roomTQC2019, QDissipation_Weiss, ultrafastcoherence2021,CoherentControl_TLS_nonMarkovian_ReinaEckel}. These considerations are essential for defining meaningful initial conditions and parameter boundaries, leading to results with a reasonable and physically interpretable meaning.

\begin{table*}[t!]
\centering
{\footnotesize
\begin{tabular}{|c|c|c|c|c|c|}
\hline
 & \multicolumn{4}{c|}{This work} & Hildner \textit{et al.} QDI~\cite{VisualizingWilma2019} \\
\hline
Target
& Optimization Algorithm
& NN with Std. Scaler
& NN with Rob. Scaler
& NN with Rob. Scaler + WP
&  \\
\hline
$\Omega_{\text{2P}}$
& 530.45 $\text{cm}^{-1}$
& 531.78 $\text{cm}^{-1}$
& 529.23 $\text{cm}^{-1}$
& 525.27 $\text{cm}^{-1}$
& 531 $\text{cm}^{-1}$ \\
\hline
$\gamma_2$
& 0.016005 $\text{fs}^{-1}$
& 0.014368 $\text{fs}^{-1}$
& 0.015966 $\text{fs}^{-1}$
& 0.016016 $\text{fs}^{-1}$
& $1/61\ \text{fs}^{-1}$ \\
\hline
$\Gamma_{12}$
& 0.0055336 $\text{fs}^{-1}$
& 0.0026379 $\text{fs}^{-1}$
& 0.0053537 $\text{fs}^{-1}$
& 0.0049061 $\text{fs}^{-1}$
& $1/190\ \text{fs}^{-1}$ \\
\hline
\end{tabular}
}
\caption{Comparison between parameters predicted by the methods in this work and experimental values obtained using the Quantum Dynamics Identification (QDI) protocol by Hildner \textit{et al.}~\cite{VisualizingWilma2019}. Results are shown for both optimization-based and neural-network approaches with different data pre-processing strategies.}
\label{tab:hildner_comparison}
\end{table*}

On the other hand, the neural network approach eliminates the need for performing multiple optimization runs to obtain reliable parameter estimates on real experimental data. Once properly trained and validated, the NN can evaluate any systematic set of experimental PL data and immediately extract the corresponding physical parameters. However, this advantage comes with the requirement of an extensive training and validation phase involving the exploration of different network architectures, feature-scaling methods and hyperparameter configurations. Moreover, the performance and generalization capability of the NN are inherently limited by the quality and representativeness of the training data~\cite{Hastie2009_ElementsStatLearning}, which, in this case, are generated from a photoluminescence model assuming collective emission as the sum of individual single-molecule contributions~\cite{MeLPPP_bulk_fsTPA}. Nevertheless, when the simulated PL data are produced under well-controlled experimental conditions (e.g., fixed emission intensities, pulse duration, temperature, and molecular parameters), the trained model can successfully predict physically meaningful values that closely agree with those reported experimentally~\cite{VisualizingWilma2019}.
In addition,  evaluation metrics obtained for  the test set and real PL data demonstrate that the trained model achieves strong predictive performance. Additionally,
it was observed that the feature scaling method applied to the input data can slightly influence the training dynamics and resulting predicted targets. This highlights the importance of selecting an appropriate normalization scheme during pre-processing.

\section{Conclusions\label{sec:Conclusions}}
In this work, we developed and validated robust algorithms that extract physically meaningful parameters associated with environmental interactions and driving couplings from experimentally accessible photoluminescence observables. We use the MeLPPP single molecule as a sample system. Our approach combines a description of the molecule’s ultrafast dynamics using a Lindblad master equation framework with a PL forward model mapping pulse, molecular, and environmental control parameters to the normalized PL intensity. This PL model formed the basis for developing the two complementary algorithms for parameter optimization and extraction using real experimental data.

Despite the well-known sensitivity of the inverse problem to initial conditions and the existence of multiple numerical solutions, the first approach, a direct inverse optimization algorithm, can reliably recover parameter estimates when characterized statistically. By running multiple independent optimizations, we obtained distributions that, while variable across individual runs, systematically converged to mean values consistent with previously reported parameters. We strengthened the cost function with additional terms that provide physical insight to create a more robust model and thoroughly tested the statistical behavior of the predictions, comparing its performance with that of similar algorithms. 

The second approach involves training a feed-forward neural network to predict dissipative and coupling parameters from PL data. During training, the network learns to identify correlations between the input PL traces and the dissipation and molecular parameters $\Q$. It adjusts its internal weights to construct a mapping between diverse PL signals and the underlying environmental parameters. This mapping enables the network to generalize well to unseen data and to produce physically reasonable parameter estimates when applied to real PL traces. Since statistical characterization and validation are embedded in the training stage, inference on  new molecular experimental data requires only one forward pass through the network, enabling fast, reliable parameter estimation.

Table~\ref{tab:hildner_comparison} summarizes the main findings and shows that both methods predict parameter values consistent with earlier reports using PL under spectral phase modulation. The two approaches offer complementary advantages. The inverse optimization method is straightforward and interpretable, though it requires repeated runs to assess uncertainty. In contrast, the neural network approach provides rapid single-shot inference, albeit with a more involved preprocessing, training, and architecture selection phase.

Overall, this work provides robust, validated tools for translating PL observables into quantitative physical parameters. This makes them practical for future coherent control experiments of single quantum objects and real-time experimental workflows. Future work will focus on extending the framework to broader classes of molecules and artificial atoms, incorporating experimental noise models during training, and integrating the methods into adaptive measurement schemes. 
Additionally, considering physical models beyond the standard Lindblad master equation could significantly expand the parameter space associated with the environment. Examples of such models include structured baths with temperature-dependent spectral densities, which offer a general, non-Markovian, broadband description of dissipative processes.

\section{Acknowledgments}   

We  thank  the financial support of the Colombian Ministry of Science (MINCIENCIAS), Contract No. BPIN 2022000100068, 
and the Norwegian Partnership Programme for Global Academic Cooperation (NORPART), Grant NORPART-2021/10436 (QTECNOS).

%


\begin{thebibliography}{86}%
\makeatletter
\providecommand \@ifxundefined [1]{%
 \@ifx{#1\undefined}
}%
\providecommand \@ifnum [1]{%
 \ifnum #1\expandafter \@firstoftwo
 \else \expandafter \@secondoftwo
 \fi
}%
\providecommand \@ifx [1]{%
 \ifx #1\expandafter \@firstoftwo
 \else \expandafter \@secondoftwo
 \fi
}%
\providecommand \natexlab [1]{#1}%
\providecommand \enquote  [1]{``#1''}%
\providecommand \bibnamefont  [1]{#1}%
\providecommand \bibfnamefont [1]{#1}%
\providecommand \citenamefont [1]{#1}%
\providecommand \href@noop [0]{\@secondoftwo}%
\providecommand \href [0]{\begingroup \@sanitize@url \@href}%
\providecommand \@href[1]{\@@startlink{#1}\@@href}%
\providecommand \@@href[1]{\endgroup#1\@@endlink}%
\providecommand \@sanitize@url [0]{\catcode `\\12\catcode `\$12\catcode
  `\&12\catcode `\#12\catcode `\^12\catcode `\_12\catcode `\%12\relax}%
\providecommand \@@startlink[1]{}%
\providecommand \@@endlink[0]{}%
\providecommand \url  [0]{\begingroup\@sanitize@url \@url }%
\providecommand \@url [1]{\endgroup\@href {#1}{\urlprefix }}%
\providecommand \urlprefix  [0]{URL }%
\providecommand \Eprint [0]{\href }%
\providecommand \doibase [0]{https://doi.org/}%
\providecommand \selectlanguage [0]{\@gobble}%
\providecommand \bibinfo  [0]{\@secondoftwo}%
\providecommand \bibfield  [0]{\@secondoftwo}%
\providecommand \translation [1]{[#1]}%
\providecommand \BibitemOpen [0]{}%
\providecommand \bibitemStop [0]{}%
\providecommand \bibitemNoStop [0]{.\EOS\space}%
\providecommand \EOS [0]{\spacefactor3000\relax}%
\providecommand \BibitemShut  [1]{\csname bibitem#1\endcsname}%
\let\auto@bib@innerbib\@empty
\bibitem [{\citenamefont {Zewail}(1980)}]{Zewail_CohLaserSpectro_1980}%
  \BibitemOpen
  \bibfield  {author} {\bibinfo {author} {\bibfnamefont {A.~H.}\ \bibnamefont
  {Zewail}},\ }\bibfield  {title} {\bibinfo {title} {Optical molecular
  dephasing: Principles of and probings by coherent laser spectroscopy},\
  }\href {https://doi.org/10.1021/ar50154a004} {\bibfield  {journal} {\bibinfo
  {journal} {Accounts of Chemical Research}\ }\textbf {\bibinfo {volume}
  {13}},\ \bibinfo {pages} {360} (\bibinfo {year} {1980})}\BibitemShut
  {NoStop}%
\bibitem [{\citenamefont
  {Silberberg}(2009)}]{Silberberg_CohControlSpectroscopy}%
  \BibitemOpen
  \bibfield  {author} {\bibinfo {author} {\bibfnamefont {Y.}~\bibnamefont
  {Silberberg}},\ }\bibfield  {title} {\bibinfo {title} {Quantum coherent
  control for nonlinear spectroscopy and microscopy},\ }\href
  {https://doi.org/10.1146/annurev.physchem.040808.090427} {\bibfield
  {journal} {\bibinfo  {journal} {Annual Review Of Physical Chemistry}\
  }\textbf {\bibinfo {volume} {60}},\ \bibinfo {pages} {277} (\bibinfo {year}
  {2009})}\BibitemShut {NoStop}%
\bibitem [{\citenamefont {Dantus}\ and\ \citenamefont
  {Lozovoy}(2004)}]{Expemtal_CohLaserControl_physicochemicalProcc}%
  \BibitemOpen
  \bibfield  {author} {\bibinfo {author} {\bibfnamefont {M.}~\bibnamefont
  {Dantus}}\ and\ \bibinfo {author} {\bibfnamefont {V.}~\bibnamefont
  {Lozovoy}},\ }\bibfield  {title} {\bibinfo {title} {Experimental coherent
  laser control of physicochemical processes},\ }\href
  {https://doi.org/10.1021/cr020668r} {\bibfield  {journal} {\bibinfo
  {journal} {Chemical Reviews}\ }\textbf {\bibinfo {volume} {104}},\ \bibinfo
  {pages} {1813} (\bibinfo {year} {2004})}\BibitemShut {NoStop}%
\bibitem [{\citenamefont {Schultz}\ \emph {et~al.}(2024)\citenamefont
  {Schultz}, \citenamefont {Yuly}, \citenamefont {Arsenault}, \citenamefont
  {Parker}, \citenamefont {Chowdhury}, \citenamefont {Dani}, \citenamefont
  {Kundu}, \citenamefont {Nuomin}, \citenamefont {Zhang}, \citenamefont
  {Valdiviezo}, \citenamefont {Zhang}, \citenamefont {Orcutt}, \citenamefont
  {Jang}, \citenamefont {Fleming}, \citenamefont {Makri}, \citenamefont
  {Ogilvie}, \citenamefont {Therien}, \citenamefont {Wasielewski},\ and\
  \citenamefont {Beratan}}]{CoherenceChem2024}%
  \BibitemOpen
  \bibfield  {author} {\bibinfo {author} {\bibfnamefont {J.~D.}\ \bibnamefont
  {Schultz}}, \bibinfo {author} {\bibfnamefont {J.~L.}\ \bibnamefont {Yuly}},
  \bibinfo {author} {\bibfnamefont {E.~A.}\ \bibnamefont {Arsenault}}, \bibinfo
  {author} {\bibfnamefont {K.}~\bibnamefont {Parker}}, \bibinfo {author}
  {\bibfnamefont {S.~N.}\ \bibnamefont {Chowdhury}}, \bibinfo {author}
  {\bibfnamefont {R.}~\bibnamefont {Dani}}, \bibinfo {author} {\bibfnamefont
  {S.}~\bibnamefont {Kundu}}, \bibinfo {author} {\bibfnamefont
  {H.}~\bibnamefont {Nuomin}}, \bibinfo {author} {\bibfnamefont
  {Z.}~\bibnamefont {Zhang}}, \bibinfo {author} {\bibfnamefont
  {J.}~\bibnamefont {Valdiviezo}}, \bibinfo {author} {\bibfnamefont
  {P.}~\bibnamefont {Zhang}}, \bibinfo {author} {\bibfnamefont
  {K.}~\bibnamefont {Orcutt}}, \bibinfo {author} {\bibfnamefont {S.~J.}\
  \bibnamefont {Jang}}, \bibinfo {author} {\bibfnamefont {G.~R.}\ \bibnamefont
  {Fleming}}, \bibinfo {author} {\bibfnamefont {N.}~\bibnamefont {Makri}},
  \bibinfo {author} {\bibfnamefont {J.~P.}\ \bibnamefont {Ogilvie}}, \bibinfo
  {author} {\bibfnamefont {M.~J.}\ \bibnamefont {Therien}}, \bibinfo {author}
  {\bibfnamefont {M.~R.}\ \bibnamefont {Wasielewski}},\ and\ \bibinfo {author}
  {\bibfnamefont {D.~N.}\ \bibnamefont {Beratan}},\ }\bibfield  {title}
  {\bibinfo {title} {Coherence in chemistry: Foundations and frontiers},\
  }\href@noop {} {\bibfield  {journal} {\bibinfo  {journal} {Chemical Reviews}\
  }\textbf {\bibinfo {volume} {124}},\ \bibinfo {pages} {11641} (\bibinfo
  {year} {2024})}\BibitemShut {NoStop}%
\bibitem [{\citenamefont {Hildner}\ \emph {et~al.}(2011)\citenamefont
  {Hildner}, \citenamefont {Brinks},\ and\ \citenamefont {van
  Hulst}}]{Hildner_fsCohQControl_singleMol_RoomT}%
  \BibitemOpen
  \bibfield  {author} {\bibinfo {author} {\bibfnamefont {R.}~\bibnamefont
  {Hildner}}, \bibinfo {author} {\bibfnamefont {D.}~\bibnamefont {Brinks}},\
  and\ \bibinfo {author} {\bibfnamefont {N.~F.}\ \bibnamefont {van Hulst}},\
  }\bibfield  {title} {\bibinfo {title} {Femtosecond coherence and quantum
  control of single molecules at room temperature},\ }\href
  {https://doi.org/10.1038/NPHYS1858} {\bibfield  {journal} {\bibinfo
  {journal} {Nature Physics}\ }\textbf {\bibinfo {volume} {7}},\ \bibinfo
  {pages} {172} (\bibinfo {year} {2011})}\BibitemShut {NoStop}%
\bibitem [{\citenamefont {Madrid-Úsuga}\ \emph {et~al.}(2019)\citenamefont
  {Madrid-Úsuga}, \citenamefont {Susa},\ and\ \citenamefont
  {Reina}}]{roomTQC2019}%
  \BibitemOpen
  \bibfield  {author} {\bibinfo {author} {\bibfnamefont {D.}~\bibnamefont
  {Madrid-Úsuga}}, \bibinfo {author} {\bibfnamefont {C.~E.}\ \bibnamefont
  {Susa}},\ and\ \bibinfo {author} {\bibfnamefont {J.~H.}\ \bibnamefont
  {Reina}},\ }\bibfield  {title} {\bibinfo {title} {Room temperature quantum
  coherence vs. electron transfer in a rhodanine derivative chromophore},\
  }\href {https://doi.org/10.1039/C9CP01398A} {\bibfield  {journal} {\bibinfo
  {journal} {Phys. Chem. Chem. Phys.}\ }\textbf {\bibinfo {volume} {21}},\
  \bibinfo {pages} {12640} (\bibinfo {year} {2019})}\BibitemShut {NoStop}%
\bibitem [{\citenamefont {Dai}\ and\ \citenamefont
  {Monkman}(2013)}]{fs_LadderType_nanoWire}%
  \BibitemOpen
  \bibfield  {author} {\bibinfo {author} {\bibfnamefont {D.~C.}\ \bibnamefont
  {Dai}}\ and\ \bibinfo {author} {\bibfnamefont {A.~P.}\ \bibnamefont
  {Monkman}},\ }\bibfield  {title} {\bibinfo {title} {Femtosecond hot-exciton
  emission in a ladder-type $\ensuremath{\pi}$-conjugated rigid-polymer
  nanowire},\ }\href {https://doi.org/10.1103/PhysRevB.87.045308} {\bibfield
  {journal} {\bibinfo  {journal} {Phys. Rev. B}\ }\textbf {\bibinfo {volume}
  {87}},\ \bibinfo {pages} {045308} (\bibinfo {year} {2013})}\BibitemShut
  {NoStop}%
\bibitem [{\citenamefont {Potter}\ \emph {et~al.}(1992)\citenamefont {Potter},
  \citenamefont {Herek}, \citenamefont {Pedersen}, \citenamefont {Liu},\ and\
  \citenamefont {Zewail}}]{FemtoLaserControl_1992}%
  \BibitemOpen
  \bibfield  {author} {\bibinfo {author} {\bibfnamefont {E.~D.}\ \bibnamefont
  {Potter}}, \bibinfo {author} {\bibfnamefont {J.~L.}\ \bibnamefont {Herek}},
  \bibinfo {author} {\bibfnamefont {S.}~\bibnamefont {Pedersen}}, \bibinfo
  {author} {\bibfnamefont {Q.}~\bibnamefont {Liu}},\ and\ \bibinfo {author}
  {\bibfnamefont {A.~H.}\ \bibnamefont {Zewail}},\ }\bibfield  {title}
  {\bibinfo {title} {Femtosecond laser control of a chemical reaction},\ }\href
  {https://doi.org/10.1038/355066a0} {\bibfield  {journal} {\bibinfo  {journal}
  {Nature}\ }\textbf {\bibinfo {volume} {355}},\ \bibinfo {pages} {66}
  (\bibinfo {year} {1992})}\BibitemShut {NoStop}%
\bibitem [{\citenamefont {Reina}\ \emph {et~al.}(2004)\citenamefont {Reina},
  \citenamefont {Beausoleil}, \citenamefont {Spiller},\ and\ \citenamefont
  {Munro}}]{Molecgates2004}%
  \BibitemOpen
  \bibfield  {author} {\bibinfo {author} {\bibfnamefont {J.~H.}\ \bibnamefont
  {Reina}}, \bibinfo {author} {\bibfnamefont {R.~G.}\ \bibnamefont
  {Beausoleil}}, \bibinfo {author} {\bibfnamefont {T.~P.}\ \bibnamefont
  {Spiller}},\ and\ \bibinfo {author} {\bibfnamefont {W.~J.}\ \bibnamefont
  {Munro}},\ }\bibfield  {title} {\bibinfo {title} {Radiative corrections and
  quantum gates in molecular systems},\ }\href
  {https://doi.org/10.1103/PhysRevLett.93.250501} {\bibfield  {journal}
  {\bibinfo  {journal} {Phys. Rev. Lett.}\ }\textbf {\bibinfo {volume} {93}},\
  \bibinfo {pages} {250501} (\bibinfo {year} {2004})}\BibitemShut {NoStop}%
\bibitem [{\citenamefont {Reina}\ \emph {et~al.}(2018)\citenamefont {Reina},
  \citenamefont {Susa},\ and\ \citenamefont {Hildner}}]{molecgate2018}%
  \BibitemOpen
  \bibfield  {author} {\bibinfo {author} {\bibfnamefont {J.~H.}\ \bibnamefont
  {Reina}}, \bibinfo {author} {\bibfnamefont {C.~E.}\ \bibnamefont {Susa}},\
  and\ \bibinfo {author} {\bibfnamefont {R.}~\bibnamefont {Hildner}},\
  }\bibfield  {title} {\bibinfo {title} {Conditional quantum dynamics and
  nonlocal states in dimeric and trimeric arrays of organic molecules},\ }\href
  {https://doi.org/10.1103/PhysRevA.97.063422} {\bibfield  {journal} {\bibinfo
  {journal} {Phys. Rev. A}\ }\textbf {\bibinfo {volume} {97}},\ \bibinfo
  {pages} {063422} (\bibinfo {year} {2018})}\BibitemShut {NoStop}%
\bibitem [{\citenamefont {Scholes}\ \emph {et~al.}(2025)\citenamefont
  {Scholes}, \citenamefont {Olaya-Castro}, \citenamefont {Mukamel},
  \citenamefont {Kirrander}, \citenamefont {Ni}, \citenamefont {Hedley},\ and\
  \citenamefont {Frank}}]{QISchem2025}%
  \BibitemOpen
  \bibfield  {author} {\bibinfo {author} {\bibfnamefont {G.~D.}\ \bibnamefont
  {Scholes}}, \bibinfo {author} {\bibfnamefont {A.}~\bibnamefont
  {Olaya-Castro}}, \bibinfo {author} {\bibfnamefont {S.}~\bibnamefont
  {Mukamel}}, \bibinfo {author} {\bibfnamefont {A.}~\bibnamefont {Kirrander}},
  \bibinfo {author} {\bibfnamefont {K.-K.}\ \bibnamefont {Ni}}, \bibinfo
  {author} {\bibfnamefont {G.~J.}\ \bibnamefont {Hedley}},\ and\ \bibinfo
  {author} {\bibfnamefont {N.~L.}\ \bibnamefont {Frank}},\ }\bibfield  {title}
  {\bibinfo {title} {The quantum information science challenge for chemistry},\
  }\href@noop {} {\bibfield  {journal} {\bibinfo  {journal} {The Journal of
  Physical Chemistry Letters}\ }\textbf {\bibinfo {volume} {16}},\ \bibinfo
  {pages} {1376} (\bibinfo {year} {2025})}\BibitemShut {NoStop}%
\bibitem [{\citenamefont {Wasielewski}(2023)}]{molecohQIS2023}%
  \BibitemOpen
  \bibfield  {author} {\bibinfo {author} {\bibfnamefont {M.~R.}\ \bibnamefont
  {Wasielewski}},\ }\bibfield  {title} {\bibinfo {title} {Light-driven spin
  chemistry for quantum information science},\ }\href
  {https://doi.org/10.1063/PT.3.5196} {\bibfield  {journal} {\bibinfo
  {journal} {Physics Today}\ }\textbf {\bibinfo {volume} {76}},\ \bibinfo
  {pages} {34} (\bibinfo {year} {2023})}\BibitemShut {NoStop}%
\bibitem [{\citenamefont {Heide}\ \emph {et~al.}(2021)\citenamefont {Heide},
  \citenamefont {Eckstein}, \citenamefont {Boolakee}, \citenamefont {Gerner},
  \citenamefont {Weber}, \citenamefont {Franco},\ and\ \citenamefont
  {Hommelhoff}}]{Electronic_CohDephasing_Heide}%
  \BibitemOpen
  \bibfield  {author} {\bibinfo {author} {\bibfnamefont {C.}~\bibnamefont
  {Heide}}, \bibinfo {author} {\bibfnamefont {T.}~\bibnamefont {Eckstein}},
  \bibinfo {author} {\bibfnamefont {T.}~\bibnamefont {Boolakee}}, \bibinfo
  {author} {\bibfnamefont {C.}~\bibnamefont {Gerner}}, \bibinfo {author}
  {\bibfnamefont {H.~B.}\ \bibnamefont {Weber}}, \bibinfo {author}
  {\bibfnamefont {I.}~\bibnamefont {Franco}},\ and\ \bibinfo {author}
  {\bibfnamefont {P.}~\bibnamefont {Hommelhoff}},\ }\bibfield  {title}
  {\bibinfo {title} {Electronic coherence and coherent dephasing in the optical
  control of electrons in graphene},\ }\href
  {https://doi.org/10.1021/acs.nanolett.1c02538} {\bibfield  {journal}
  {\bibinfo  {journal} {Nano Letters}\ }\textbf {\bibinfo {volume} {21}},\
  \bibinfo {pages} {9403} (\bibinfo {year} {2021})}\BibitemShut {NoStop}%
\bibitem [{\citenamefont {Madrid-Úsuga}\ and\ \citenamefont
  {Reina}(2021)}]{ultrafastcoherence2021}%
  \BibitemOpen
  \bibfield  {author} {\bibinfo {author} {\bibfnamefont {D.}~\bibnamefont
  {Madrid-Úsuga}}\ and\ \bibinfo {author} {\bibfnamefont {J.~H.}\ \bibnamefont
  {Reina}},\ }\bibfield  {title} {\bibinfo {title} {Molecular {Structure},
  {Quantum} {Coherence}, and {Solvent} {Effects} on the {Ultrafast} {Electron}
  {Transport} in {BODIPY}–{C60} {Derivatives}},\ }\href
  {https://doi.org/10.1021/acs.jpca.1c00603} {\bibfield  {journal} {\bibinfo
  {journal} {The Journal of Physical Chemistry A}\ }\textbf {\bibinfo {volume}
  {125}},\ \bibinfo {pages} {2518} (\bibinfo {year} {2021})}\BibitemShut
  {NoStop}%
\bibitem [{\citenamefont {Gustin}\ \emph {et~al.}(2023)\citenamefont {Gustin},
  \citenamefont {Kim}, \citenamefont {McCamant},\ and\ \citenamefont
  {Franco}}]{gustin_mapping_2023}%
  \BibitemOpen
  \bibfield  {author} {\bibinfo {author} {\bibfnamefont {I.}~\bibnamefont
  {Gustin}}, \bibinfo {author} {\bibfnamefont {C.~W.}\ \bibnamefont {Kim}},
  \bibinfo {author} {\bibfnamefont {D.~W.}\ \bibnamefont {McCamant}},\ and\
  \bibinfo {author} {\bibfnamefont {I.}~\bibnamefont {Franco}},\ }\bibfield
  {title} {\bibinfo {title} {Mapping electronic decoherence pathways in
  molecules},\ }\href {https://doi.org/10.1073/pnas.2309987120} {\bibfield
  {journal} {\bibinfo  {journal} {Proceedings of the National Academy of
  Sciences}\ }\textbf {\bibinfo {volume} {120}},\ \bibinfo {pages}
  {e2309987120} (\bibinfo {year} {2023})}\BibitemShut {NoStop}%
\bibitem [{\citenamefont {Brinks}\ \emph {et~al.}(2010)\citenamefont {Brinks},
  \citenamefont {Stefani}, \citenamefont {Kulzer}, \citenamefont {Hildner},
  \citenamefont {Taminiau}, \citenamefont {Avlasevich}, \citenamefont
  {Muellen},\ and\ \citenamefont
  {Van~Hulst}}]{Brinks_ControlVibWavePacks_SingleMol}%
  \BibitemOpen
  \bibfield  {author} {\bibinfo {author} {\bibfnamefont {D.}~\bibnamefont
  {Brinks}}, \bibinfo {author} {\bibfnamefont {F.~D.}\ \bibnamefont {Stefani}},
  \bibinfo {author} {\bibfnamefont {F.}~\bibnamefont {Kulzer}}, \bibinfo
  {author} {\bibfnamefont {R.}~\bibnamefont {Hildner}}, \bibinfo {author}
  {\bibfnamefont {T.~H.}\ \bibnamefont {Taminiau}}, \bibinfo {author}
  {\bibfnamefont {Y.}~\bibnamefont {Avlasevich}}, \bibinfo {author}
  {\bibfnamefont {K.}~\bibnamefont {Muellen}},\ and\ \bibinfo {author}
  {\bibfnamefont {N.~F.}\ \bibnamefont {Van~Hulst}},\ }\bibfield  {title}
  {\bibinfo {title} {Visualizing and controlling vibrational wave packets of
  single molecules},\ }\href {https://doi.org/10.1038/nature09110} {\bibfield
  {journal} {\bibinfo  {journal} {Nature}\ }\textbf {\bibinfo {volume} {465}},\
  \bibinfo {pages} {905} (\bibinfo {year} {2010})}\BibitemShut {NoStop}%
\bibitem [{\citenamefont {Lynch}\ \emph {et~al.}(2024)\citenamefont {Lynch},
  \citenamefont {Das}, \citenamefont {Alam}, \citenamefont {Rich},\ and\
  \citenamefont {Frontiera}}]{MasterinFemt_RamanSpectroscopy}%
  \BibitemOpen
  \bibfield  {author} {\bibinfo {author} {\bibfnamefont {P.~G.}\ \bibnamefont
  {Lynch}}, \bibinfo {author} {\bibfnamefont {A.}~\bibnamefont {Das}}, \bibinfo
  {author} {\bibfnamefont {S.}~\bibnamefont {Alam}}, \bibinfo {author}
  {\bibfnamefont {C.~C.}\ \bibnamefont {Rich}},\ and\ \bibinfo {author}
  {\bibfnamefont {R.~R.}\ \bibnamefont {Frontiera}},\ }\bibfield  {title}
  {\bibinfo {title} {Mastering femtosecond stimulated raman spectroscopy: A
  practical guide},\ }\href {https://doi.org/10.1021/acsphyschemau.3c00031}
  {\bibfield  {journal} {\bibinfo  {journal} {ACS Physical Chemistry Au}\
  }\textbf {\bibinfo {volume} {4}},\ \bibinfo {pages} {1} (\bibinfo {year}
  {2024})}\BibitemShut {NoStop}%
\bibitem [{\citenamefont {Piatkowski}\ \emph {et~al.}(2019)\citenamefont
  {Piatkowski}, \citenamefont {Accanto}, \citenamefont {Calbris}, \citenamefont
  {Christodoulou}, \citenamefont {Moreels},\ and\ \citenamefont {van
  Hulst}}]{StimulatedEmiss_SingleNanocrystals}%
  \BibitemOpen
  \bibfield  {author} {\bibinfo {author} {\bibfnamefont {L.}~\bibnamefont
  {Piatkowski}}, \bibinfo {author} {\bibfnamefont {N.}~\bibnamefont {Accanto}},
  \bibinfo {author} {\bibfnamefont {G.}~\bibnamefont {Calbris}}, \bibinfo
  {author} {\bibfnamefont {S.}~\bibnamefont {Christodoulou}}, \bibinfo {author}
  {\bibfnamefont {I.}~\bibnamefont {Moreels}},\ and\ \bibinfo {author}
  {\bibfnamefont {N.~F.}\ \bibnamefont {van Hulst}},\ }\bibfield  {title}
  {\bibinfo {title} {Ultrafast stimulated emission microscopy of single
  nanocrystals},\ }\href {https://doi.org/10.1126/science.aay1821} {\bibfield
  {journal} {\bibinfo  {journal} {Science}\ }\textbf {\bibinfo {volume}
  {366}},\ \bibinfo {pages} {1240} (\bibinfo {year} {2019})}\BibitemShut
  {NoStop}%
\bibitem [{\citenamefont {Min}\ \emph {et~al.}(2009)\citenamefont {Min},
  \citenamefont {Lu}, \citenamefont {Chong}, \citenamefont {Roy}, \citenamefont
  {Holtom},\ and\ \citenamefont {Xie}}]{StimulEmission_Microscopy}%
  \BibitemOpen
  \bibfield  {author} {\bibinfo {author} {\bibfnamefont {W.}~\bibnamefont
  {Min}}, \bibinfo {author} {\bibfnamefont {S.}~\bibnamefont {Lu}}, \bibinfo
  {author} {\bibfnamefont {S.}~\bibnamefont {Chong}}, \bibinfo {author}
  {\bibfnamefont {R.}~\bibnamefont {Roy}}, \bibinfo {author} {\bibfnamefont
  {G.~R.}\ \bibnamefont {Holtom}},\ and\ \bibinfo {author} {\bibfnamefont
  {X.~S.}\ \bibnamefont {Xie}},\ }\bibfield  {title} {\bibinfo {title} {Imaging
  chromophores with undetectable fluorescence by stimulated emission
  microscopy},\ }\href {https://doi.org/10.1038/nature08438} {\bibfield
  {journal} {\bibinfo  {journal} {Nature}\ }\textbf {\bibinfo {volume} {461}},\
  \bibinfo {pages} {1105} (\bibinfo {year} {2009})}\BibitemShut {NoStop}%
\bibitem [{\citenamefont {Falke}\ \emph {et~al.}(2014)\citenamefont {Falke},
  \citenamefont {Rozzi}, \citenamefont {Brida}, \citenamefont {Maiuri},
  \citenamefont {Amato}, \citenamefont {Sommer}, \citenamefont {De~Sio},
  \citenamefont {Rubio}, \citenamefont {Cerullo}, \citenamefont {Molinari},\
  and\ \citenamefont {Lienau}}]{Cohert_ChargeTransfer_organicPV}%
  \BibitemOpen
  \bibfield  {author} {\bibinfo {author} {\bibfnamefont {S.~M.}\ \bibnamefont
  {Falke}}, \bibinfo {author} {\bibfnamefont {C.~A.}\ \bibnamefont {Rozzi}},
  \bibinfo {author} {\bibfnamefont {D.}~\bibnamefont {Brida}}, \bibinfo
  {author} {\bibfnamefont {M.}~\bibnamefont {Maiuri}}, \bibinfo {author}
  {\bibfnamefont {M.}~\bibnamefont {Amato}}, \bibinfo {author} {\bibfnamefont
  {E.}~\bibnamefont {Sommer}}, \bibinfo {author} {\bibfnamefont
  {A.}~\bibnamefont {De~Sio}}, \bibinfo {author} {\bibfnamefont
  {A.}~\bibnamefont {Rubio}}, \bibinfo {author} {\bibfnamefont
  {G.}~\bibnamefont {Cerullo}}, \bibinfo {author} {\bibfnamefont
  {E.}~\bibnamefont {Molinari}},\ and\ \bibinfo {author} {\bibfnamefont
  {C.}~\bibnamefont {Lienau}},\ }\bibfield  {title} {\bibinfo {title} {Coherent
  ultrafast charge transfer in an organic photovoltaic blend},\ }\href
  {https://doi.org/10.1126/science.1249771} {\bibfield  {journal} {\bibinfo
  {journal} {Science}\ }\textbf {\bibinfo {volume} {344}},\ \bibinfo {pages}
  {1001} (\bibinfo {year} {2014})}\BibitemShut {NoStop}%
\bibitem [{\citenamefont {Madrid-Úsuga}\ \emph {et~al.}(2022)\citenamefont
  {Madrid-Úsuga}, \citenamefont {Ortiz},\ and\ \citenamefont
  {Reina}}]{photophysical2022}%
  \BibitemOpen
  \bibfield  {author} {\bibinfo {author} {\bibfnamefont {D.}~\bibnamefont
  {Madrid-Úsuga}}, \bibinfo {author} {\bibfnamefont {A.}~\bibnamefont
  {Ortiz}},\ and\ \bibinfo {author} {\bibfnamefont {J.~H.}\ \bibnamefont
  {Reina}},\ }\bibfield  {title} {\bibinfo {title} {Photophysical {Properties}
  of {BODIPY} {Derivatives} for the {Implementation} of {Organic} {Solar}
  {Cells}: {A} {Computational} {Approach}},\ }\href
  {https://doi.org/10.1021/acsomega.1c04598} {\bibfield  {journal} {\bibinfo
  {journal} {ACS Omega}\ }\textbf {\bibinfo {volume} {7}},\ \bibinfo {pages}
  {3963} (\bibinfo {year} {2022})}\BibitemShut {NoStop}%
\bibitem [{\citenamefont {Madrid-Úsuga}\ \emph {et~al.}(2018)\citenamefont
  {Madrid-Úsuga}, \citenamefont {Melo-Luna}, \citenamefont {Insuasty},
  \citenamefont {Ortiz},\ and\ \citenamefont {Reina}}]{molecoptoel2018}%
  \BibitemOpen
  \bibfield  {author} {\bibinfo {author} {\bibfnamefont {D.}~\bibnamefont
  {Madrid-Úsuga}}, \bibinfo {author} {\bibfnamefont {C.~A.}\ \bibnamefont
  {Melo-Luna}}, \bibinfo {author} {\bibfnamefont {A.}~\bibnamefont {Insuasty}},
  \bibinfo {author} {\bibfnamefont {A.}~\bibnamefont {Ortiz}},\ and\ \bibinfo
  {author} {\bibfnamefont {J.~H.}\ \bibnamefont {Reina}},\ }\bibfield  {title}
  {\bibinfo {title} {Optical and electronic properties of molecular systems
  derived from rhodanine},\ }\href {https://doi.org/10.1021/acs.jpca.8b08265}
  {\bibfield  {journal} {\bibinfo  {journal} {The Journal of Physical Chemistry
  A}\ }\textbf {\bibinfo {volume} {122}},\ \bibinfo {pages} {8469} (\bibinfo
  {year} {2018})}\BibitemShut {NoStop}%
\bibitem [{\citenamefont {Lewis}\ and\ \citenamefont
  {Ogilvie}(2012)}]{Energy_Charge_Transfer_ElectronicMicroscopy}%
  \BibitemOpen
  \bibfield  {author} {\bibinfo {author} {\bibfnamefont {K.~L.~M.}\
  \bibnamefont {Lewis}}\ and\ \bibinfo {author} {\bibfnamefont {J.~P.}\
  \bibnamefont {Ogilvie}},\ }\bibfield  {title} {\bibinfo {title} {Probing
  photosynthetic energy and charge transfer with two-dimensional electronic
  spectroscopy},\ }\href {https://doi.org/10.1021/jz201592v} {\bibfield
  {journal} {\bibinfo  {journal} {The Journal of Physical Chemistry Letters}\
  }\textbf {\bibinfo {volume} {3}},\ \bibinfo {pages} {503} (\bibinfo {year}
  {2012})}\BibitemShut {NoStop}%
\bibitem [{\citenamefont {Haedler}\ \emph {et~al.}(2015)\citenamefont
  {Haedler}, \citenamefont {Kreger}, \citenamefont {Issac}, \citenamefont
  {Wittmann}, \citenamefont {Kivala}, \citenamefont {Hammer}, \citenamefont
  {K{\"o}hler}, \citenamefont {Schmidt},\ and\ \citenamefont
  {Hildner}}]{LongRange_ET_HildnerR}%
  \BibitemOpen
  \bibfield  {author} {\bibinfo {author} {\bibfnamefont {A.~T.}\ \bibnamefont
  {Haedler}}, \bibinfo {author} {\bibfnamefont {K.}~\bibnamefont {Kreger}},
  \bibinfo {author} {\bibfnamefont {A.}~\bibnamefont {Issac}}, \bibinfo
  {author} {\bibfnamefont {B.}~\bibnamefont {Wittmann}}, \bibinfo {author}
  {\bibfnamefont {M.}~\bibnamefont {Kivala}}, \bibinfo {author} {\bibfnamefont
  {N.}~\bibnamefont {Hammer}}, \bibinfo {author} {\bibfnamefont
  {J.}~\bibnamefont {K{\"o}hler}}, \bibinfo {author} {\bibfnamefont {H.-W.}\
  \bibnamefont {Schmidt}},\ and\ \bibinfo {author} {\bibfnamefont
  {R.}~\bibnamefont {Hildner}},\ }\bibfield  {title} {\bibinfo {title}
  {Long-range energy transport in single supramolecular nanofibres at room
  temperature},\ }\href {https://doi.org/10.1038/nature14570} {\bibfield
  {journal} {\bibinfo  {journal} {Nature}\ }\textbf {\bibinfo {volume} {523}},\
  \bibinfo {pages} {196} (\bibinfo {year} {2015})}\BibitemShut {NoStop}%
\bibitem [{\citenamefont {Yan}\ \emph {et~al.}(2023)\citenamefont {Yan},
  \citenamefont {Chen}, \citenamefont {Zhang}, \citenamefont {Wang},
  \citenamefont {Babin}, \citenamefont {Wieck}, \citenamefont {Ludwig},
  \citenamefont {Meng}, \citenamefont {Hu}, \citenamefont {Duan}, \citenamefont
  {Chen}, \citenamefont {Fang}, \citenamefont {Cygorek}, \citenamefont {Lin},
  \citenamefont {Wang}, \citenamefont {Jin},\ and\ \citenamefont
  {Liu}}]{Yan2023}%
  \BibitemOpen
  \bibfield  {author} {\bibinfo {author} {\bibfnamefont {J.-Y.}\ \bibnamefont
  {Yan}}, \bibinfo {author} {\bibfnamefont {C.}~\bibnamefont {Chen}}, \bibinfo
  {author} {\bibfnamefont {X.-D.}\ \bibnamefont {Zhang}}, \bibinfo {author}
  {\bibfnamefont {Y.-T.}\ \bibnamefont {Wang}}, \bibinfo {author}
  {\bibfnamefont {H.-G.}\ \bibnamefont {Babin}}, \bibinfo {author}
  {\bibfnamefont {A.~D.}\ \bibnamefont {Wieck}}, \bibinfo {author}
  {\bibfnamefont {A.}~\bibnamefont {Ludwig}}, \bibinfo {author} {\bibfnamefont
  {Y.}~\bibnamefont {Meng}}, \bibinfo {author} {\bibfnamefont {X.}~\bibnamefont
  {Hu}}, \bibinfo {author} {\bibfnamefont {H.}~\bibnamefont {Duan}}, \bibinfo
  {author} {\bibfnamefont {W.}~\bibnamefont {Chen}}, \bibinfo {author}
  {\bibfnamefont {W.}~\bibnamefont {Fang}}, \bibinfo {author} {\bibfnamefont
  {M.}~\bibnamefont {Cygorek}}, \bibinfo {author} {\bibfnamefont
  {X.}~\bibnamefont {Lin}}, \bibinfo {author} {\bibfnamefont {D.-W.}\
  \bibnamefont {Wang}}, \bibinfo {author} {\bibfnamefont {C.-Y.}\ \bibnamefont
  {Jin}},\ and\ \bibinfo {author} {\bibfnamefont {F.}~\bibnamefont {Liu}},\
  }\bibfield  {title} {\bibinfo {title} {Coherent control of a high-orbital
  hole in a semiconductor quantum dot},\ }\href
  {https://doi.org/10.1038/s41565-023-01442-y} {\bibfield  {journal} {\bibinfo
  {journal} {Nature Nanotechnology}\ }\textbf {\bibinfo {volume} {18}},\
  \bibinfo {pages} {1139} (\bibinfo {year} {2023})}\BibitemShut {NoStop}%
\bibitem [{\citenamefont {Dory~Constantin}(2016)}]{QDot_CohControl_2016}%
  \BibitemOpen
  \bibfield  {author} {\bibinfo {author} {\bibfnamefont {M.}~\bibnamefont
  {Dory~Constantin}, \bibfnamefont {Kevin A.~Fischer}},\ }\bibfield  {title}
  {\bibinfo {title} {Complete coherent control of a quantum dot strongly
  coupled to a nanocavity},\ }\href@noop {} {\bibfield  {journal} {\bibinfo
  {journal} {Scientific Reports}\ }\textbf {\bibinfo {volume} {6}},\ \bibinfo
  {pages} {25172} (\bibinfo {year} {2016})}\BibitemShut {NoStop}%
\bibitem [{\citenamefont {Hikosaka}\ \emph {et~al.}(2019)\citenamefont
  {Hikosaka}, \citenamefont {Kaneyasu}, \citenamefont {Fujimoto}, \citenamefont
  {Iwayama},\ and\ \citenamefont {Katoh}}]{Attosec_SingleAtoms_QCohControl}%
  \BibitemOpen
  \bibfield  {author} {\bibinfo {author} {\bibfnamefont {Y.}~\bibnamefont
  {Hikosaka}}, \bibinfo {author} {\bibfnamefont {T.}~\bibnamefont {Kaneyasu}},
  \bibinfo {author} {\bibfnamefont {M.}~\bibnamefont {Fujimoto}}, \bibinfo
  {author} {\bibfnamefont {H.}~\bibnamefont {Iwayama}},\ and\ \bibinfo {author}
  {\bibfnamefont {M.}~\bibnamefont {Katoh}},\ }\bibfield  {title} {\bibinfo
  {title} {Coherent control in the extreme ultraviolet and attosecond regime by
  synchrotron radiation},\ }\href {https://doi.org/10.1038/s41467-019-12978-w}
  {\bibfield  {journal} {\bibinfo  {journal} {Nature Communications}\ }\textbf
  {\bibinfo {volume} {10}},\ \bibinfo {pages} {4988} (\bibinfo {year}
  {2019})}\BibitemShut {NoStop}%
\bibitem [{\citenamefont {Hernández}\ and\ \citenamefont
  {Crespo-Otero}(2023)}]{MolecularAggregates_OptoElectronics}%
  \BibitemOpen
  \bibfield  {author} {\bibinfo {author} {\bibfnamefont {F.~J.}\ \bibnamefont
  {Hernández}}\ and\ \bibinfo {author} {\bibfnamefont {R.}~\bibnamefont
  {Crespo-Otero}},\ }\bibfield  {title} {\bibinfo {title} {Modeling excited
  states of molecular organic aggregates for optoelectronics},\ }\href
  {https://doi.org/https://doi.org/10.1146/annurev-physchem-102822-100945}
  {\bibfield  {journal} {\bibinfo  {journal} {Annual Review of Physical
  Chemistry}\ }\textbf {\bibinfo {volume} {74}},\ \bibinfo {pages} {547}
  (\bibinfo {year} {2023})}\BibitemShut {NoStop}%
\bibitem [{\citenamefont {Mannouch}\ \emph {et~al.}(2018)\citenamefont
  {Mannouch}, \citenamefont {Barford},\ and\ \citenamefont
  {Al-Assam}}]{piConjug_RelaxDecoherence}%
  \BibitemOpen
  \bibfield  {author} {\bibinfo {author} {\bibfnamefont {J.~R.}\ \bibnamefont
  {Mannouch}}, \bibinfo {author} {\bibfnamefont {W.}~\bibnamefont {Barford}},\
  and\ \bibinfo {author} {\bibfnamefont {S.}~\bibnamefont {Al-Assam}},\
  }\bibfield  {title} {\bibinfo {title} {Ultra-fast relaxation, decoherence,
  and localization of photoexcited states in $\pi$-conjugated polymers},\
  }\href {https://doi.org/10.1063/1.5009393} {\bibfield  {journal} {\bibinfo
  {journal} {The Journal of Chemical Physics}\ }\textbf {\bibinfo {volume}
  {148}},\ \bibinfo {pages} {034901} (\bibinfo {year} {2018})}\BibitemShut
  {NoStop}%
\bibitem [{\citenamefont {De~Sio}\ \emph {et~al.}(2016)\citenamefont {De~Sio},
  \citenamefont {Troiani}, \citenamefont {Maiuri}, \citenamefont {Rehault},
  \citenamefont {Sommer}, \citenamefont {Lim}, \citenamefont {Huelga},
  \citenamefont {Plenio}, \citenamefont {Rozzi}, \citenamefont {Cerullo},
  \citenamefont {Molinari},\ and\ \citenamefont
  {Lienau}}]{CohGeneration_ConjugatedPolymers}%
  \BibitemOpen
  \bibfield  {author} {\bibinfo {author} {\bibfnamefont {A.}~\bibnamefont
  {De~Sio}}, \bibinfo {author} {\bibfnamefont {F.}~\bibnamefont {Troiani}},
  \bibinfo {author} {\bibfnamefont {M.}~\bibnamefont {Maiuri}}, \bibinfo
  {author} {\bibfnamefont {J.}~\bibnamefont {Rehault}}, \bibinfo {author}
  {\bibfnamefont {E.}~\bibnamefont {Sommer}}, \bibinfo {author} {\bibfnamefont
  {J.}~\bibnamefont {Lim}}, \bibinfo {author} {\bibfnamefont {S.~F.}\
  \bibnamefont {Huelga}}, \bibinfo {author} {\bibfnamefont {M.~B.}\
  \bibnamefont {Plenio}}, \bibinfo {author} {\bibfnamefont {C.~A.}\
  \bibnamefont {Rozzi}}, \bibinfo {author} {\bibfnamefont {G.}~\bibnamefont
  {Cerullo}}, \bibinfo {author} {\bibfnamefont {E.}~\bibnamefont {Molinari}},\
  and\ \bibinfo {author} {\bibfnamefont {C.}~\bibnamefont {Lienau}},\
  }\bibfield  {title} {\bibinfo {title} {Tracking the coherent generation of
  polaron pairs in conjugated polymers},\ }\bibfield  {journal} {\bibinfo
  {journal} {Nature Communications}\ }\textbf {\bibinfo {volume} {7}},\ \href
  {https://doi.org/10.1038/ncomms13742} {10.1038/ncomms13742} (\bibinfo {year}
  {2016})\BibitemShut {NoStop}%
\bibitem [{\citenamefont {Walowicz}\ \emph {et~al.}(2002)\citenamefont
  {Walowicz}, \citenamefont {Pastirk}, \citenamefont {Lozovoy},\ and\
  \citenamefont {Dantus}}]{Multiphoton_ControlCondPhses}%
  \BibitemOpen
  \bibfield  {author} {\bibinfo {author} {\bibfnamefont {K.~A.}\ \bibnamefont
  {Walowicz}}, \bibinfo {author} {\bibfnamefont {I.}~\bibnamefont {Pastirk}},
  \bibinfo {author} {\bibfnamefont {V.~V.}\ \bibnamefont {Lozovoy}},\ and\
  \bibinfo {author} {\bibfnamefont {M.}~\bibnamefont {Dantus}},\ }\bibfield
  {title} {\bibinfo {title} {Multiphoton intrapulse interference. 1. control of
  multiphoton processes in condensed phases},\ }\href
  {https://doi.org/10.1021/jp0258964} {\bibfield  {journal} {\bibinfo
  {journal} {The Journal of Physical Chemistry A}\ }\textbf {\bibinfo {volume}
  {106}},\ \bibinfo {pages} {9369} (\bibinfo {year} {2002})}\BibitemShut
  {NoStop}%
\bibitem [{\citenamefont {Chen}\ \emph {et~al.}(2010)\citenamefont {Chen},
  \citenamefont {Jaron-Becker},\ and\ \citenamefont
  {Becker}}]{FewPhoton_CoherentControl}%
  \BibitemOpen
  \bibfield  {author} {\bibinfo {author} {\bibfnamefont {S.}~\bibnamefont
  {Chen}}, \bibinfo {author} {\bibfnamefont {A.}~\bibnamefont {Jaron-Becker}},\
  and\ \bibinfo {author} {\bibfnamefont {A.}~\bibnamefont {Becker}},\
  }\bibfield  {title} {\bibinfo {title} {Time-dependent analysis of few-photon
  coherent control schemes},\ }\bibfield  {journal} {\bibinfo  {journal}
  {Physical Review A}\ }\textbf {\bibinfo {volume} {82}},\ \href
  {https://doi.org/10.1103/PhysRevA.82.013414} {10.1103/PhysRevA.82.013414}
  (\bibinfo {year} {2010})\BibitemShut {NoStop}%
\bibitem [{\citenamefont {Hildner}\ \emph {et~al.}(2013)\citenamefont
  {Hildner}, \citenamefont {Brinks}, \citenamefont {Nieder}, \citenamefont
  {Cogdell},\ and\ \citenamefont {van Hulst}}]{Hildner_2013_varPathways}%
  \BibitemOpen
  \bibfield  {author} {\bibinfo {author} {\bibfnamefont {R.}~\bibnamefont
  {Hildner}}, \bibinfo {author} {\bibfnamefont {D.}~\bibnamefont {Brinks}},
  \bibinfo {author} {\bibfnamefont {J.~B.}\ \bibnamefont {Nieder}}, \bibinfo
  {author} {\bibfnamefont {R.~J.}\ \bibnamefont {Cogdell}},\ and\ \bibinfo
  {author} {\bibfnamefont {N.~F.}\ \bibnamefont {van Hulst}},\ }\bibfield
  {title} {\bibinfo {title} {Quantum coherent energy transfer over varying
  pathways in single light-harvesting complexes},\ }\href
  {https://doi.org/10.1126/science.1235820} {\bibfield  {journal} {\bibinfo
  {journal} {Science}\ }\textbf {\bibinfo {volume} {340}},\ \bibinfo {pages}
  {1448} (\bibinfo {year} {2013})}\BibitemShut {NoStop}%
\bibitem [{\citenamefont {Harrison}\ \emph {et~al.}(1999)\citenamefont
  {Harrison}, \citenamefont {Urbasch}, \citenamefont {Mahrt}, \citenamefont
  {Giessen}, \citenamefont {B{\"a}{\ss}ler},\ and\ \citenamefont
  {Scherf}}]{MeLPPP_bulk_fsTPA}%
  \BibitemOpen
  \bibfield  {author} {\bibinfo {author} {\bibfnamefont {M.~G.}\ \bibnamefont
  {Harrison}}, \bibinfo {author} {\bibfnamefont {G.}~\bibnamefont {Urbasch}},
  \bibinfo {author} {\bibfnamefont {R.~F.}\ \bibnamefont {Mahrt}}, \bibinfo
  {author} {\bibfnamefont {H.}~\bibnamefont {Giessen}}, \bibinfo {author}
  {\bibfnamefont {H.}~\bibnamefont {B{\"a}{\ss}ler}},\ and\ \bibinfo {author}
  {\bibfnamefont {U.}~\bibnamefont {Scherf}},\ }\bibfield  {title} {\bibinfo
  {title} {Two-photon fluorescence and femtosecond two-photon absorption
  studies of melppp, a ladder-type poly(phenylene) with low intra-chain
  disorder},\ }\href {https://doi.org/10.1016/S0009-2614(99)00934-3} {\bibfield
   {journal} {\bibinfo  {journal} {Chem. Phys. Lett.}\ }\textbf {\bibinfo
  {volume} {313}},\ \bibinfo {pages} {755} (\bibinfo {year}
  {1999})}\BibitemShut {NoStop}%
\bibitem [{\citenamefont {M\"uller}\ \emph {et~al.}(2004)\citenamefont
  {M\"uller}, \citenamefont {Anni}, \citenamefont {Scherf}, \citenamefont
  {Lupton},\ and\ \citenamefont {Feldmann}}]{MeLPPP_spectrumVibrat}%
  \BibitemOpen
  \bibfield  {author} {\bibinfo {author} {\bibfnamefont {J.~G.}\ \bibnamefont
  {M\"uller}}, \bibinfo {author} {\bibfnamefont {M.}~\bibnamefont {Anni}},
  \bibinfo {author} {\bibfnamefont {U.}~\bibnamefont {Scherf}}, \bibinfo
  {author} {\bibfnamefont {J.~M.}\ \bibnamefont {Lupton}},\ and\ \bibinfo
  {author} {\bibfnamefont {J.}~\bibnamefont {Feldmann}},\ }\bibfield  {title}
  {\bibinfo {title} {Vibrational fluorescence spectroscopy of single conjugated
  polymer molecules},\ }\href {https://doi.org/10.1103/PhysRevB.70.035205}
  {\bibfield  {journal} {\bibinfo  {journal} {Phys. Rev. B}\ }\textbf {\bibinfo
  {volume} {70}},\ \bibinfo {pages} {035205} (\bibinfo {year}
  {2004})}\BibitemShut {NoStop}%
\bibitem [{\citenamefont {Bopp}\ \emph {et~al.}(1997)\citenamefont {Bopp},
  \citenamefont {Jia}, \citenamefont {Li}, \citenamefont {Cogdell},\ and\
  \citenamefont {Hochstrasser}}]{PhotoBleaching_Complexes}%
  \BibitemOpen
  \bibfield  {author} {\bibinfo {author} {\bibfnamefont {M.~A.}\ \bibnamefont
  {Bopp}}, \bibinfo {author} {\bibfnamefont {Y.}~\bibnamefont {Jia}}, \bibinfo
  {author} {\bibfnamefont {L.}~\bibnamefont {Li}}, \bibinfo {author}
  {\bibfnamefont {R.~J.}\ \bibnamefont {Cogdell}},\ and\ \bibinfo {author}
  {\bibfnamefont {R.~M.}\ \bibnamefont {Hochstrasser}},\ }\bibfield  {title}
  {\bibinfo {title} {Fluorescence and photobleaching dynamics of single
  light-harvesting complexes},\ }\href
  {https://doi.org/10.1073/pnas.94.20.10630} {\bibfield  {journal} {\bibinfo
  {journal} {Proceedings of the National Academy of Sciences}\ }\textbf
  {\bibinfo {volume} {94}},\ \bibinfo {pages} {10630} (\bibinfo {year}
  {1997})}\BibitemShut {NoStop}%
\bibitem [{\citenamefont {Lu}\ \emph {et~al.}(2020)\citenamefont {Lu},
  \citenamefont {Ye}, \citenamefont {Punj}, \citenamefont {Chiechi},\ and\
  \citenamefont {Orrit}}]{QuantumYield_SingleMol_Fluorsc}%
  \BibitemOpen
  \bibfield  {author} {\bibinfo {author} {\bibfnamefont {X.}~\bibnamefont
  {Lu}}, \bibinfo {author} {\bibfnamefont {G.}~\bibnamefont {Ye}}, \bibinfo
  {author} {\bibfnamefont {D.}~\bibnamefont {Punj}}, \bibinfo {author}
  {\bibfnamefont {R.~C.}\ \bibnamefont {Chiechi}},\ and\ \bibinfo {author}
  {\bibfnamefont {M.}~\bibnamefont {Orrit}},\ }\bibfield  {title} {\bibinfo
  {title} {Quantum yield limits for the detection of single-molecule
  fluorescence enhancement by a gold nanorod},\ }\href
  {https://doi.org/10.1021/acsphotonics.0c00803} {\bibfield  {journal}
  {\bibinfo  {journal} {ACS Photonics}\ }\textbf {\bibinfo {volume} {7}},\
  \bibinfo {pages} {2498} (\bibinfo {year} {2020})}\BibitemShut {NoStop}%
\bibitem [{\citenamefont {Akimov}\ and\ \citenamefont
  {Prezhdo}(2013)}]{ElectronicCoherence_RapidLoss}%
  \BibitemOpen
  \bibfield  {author} {\bibinfo {author} {\bibfnamefont {A.~V.}\ \bibnamefont
  {Akimov}}\ and\ \bibinfo {author} {\bibfnamefont {O.~V.}\ \bibnamefont
  {Prezhdo}},\ }\bibfield  {title} {\bibinfo {title} {Persistent electronic
  coherence despite rapid loss of electron--nuclear correlation},\ }\href
  {https://doi.org/10.1021/jz402035z} {\bibfield  {journal} {\bibinfo
  {journal} {The Journal of Physical Chemistry Letters}\ }\textbf {\bibinfo
  {volume} {4}},\ \bibinfo {pages} {3857} (\bibinfo {year} {2013})}\BibitemShut
  {NoStop}%
\bibitem [{\citenamefont {Wilhelm}\ \emph {et~al.}(2004)\citenamefont
  {Wilhelm}, \citenamefont {Kleff},\ and\ \citenamefont {von
  Delft}}]{SpinBoson_GenericModel}%
  \BibitemOpen
  \bibfield  {author} {\bibinfo {author} {\bibfnamefont {F.}~\bibnamefont
  {Wilhelm}}, \bibinfo {author} {\bibfnamefont {S.}~\bibnamefont {Kleff}},\
  and\ \bibinfo {author} {\bibfnamefont {J.}~\bibnamefont {von Delft}},\
  }\bibfield  {title} {\bibinfo {title} {The spin-boson model with a structured
  environment: a comparison of approaches},\ }\href
  {https://doi.org/10.1016/j.chemphys.2003.10.010} {\bibfield  {journal}
  {\bibinfo  {journal} {Chemical Physics}\ }\textbf {\bibinfo {volume} {296}},\
  \bibinfo {pages} {345–353} (\bibinfo {year} {2004})}\BibitemShut {NoStop}%
\bibitem [{\citenamefont {Takahashi}\ \emph {et~al.}(2024)\citenamefont
  {Takahashi}, \citenamefont {Rudge}, \citenamefont {Kaspar}, \citenamefont
  {Thoss},\ and\ \citenamefont
  {Borrelli}}]{ExponentialDecomposition_CorrelationFunction}%
  \BibitemOpen
  \bibfield  {author} {\bibinfo {author} {\bibfnamefont {H.}~\bibnamefont
  {Takahashi}}, \bibinfo {author} {\bibfnamefont {S.}~\bibnamefont {Rudge}},
  \bibinfo {author} {\bibfnamefont {C.}~\bibnamefont {Kaspar}}, \bibinfo
  {author} {\bibfnamefont {M.}~\bibnamefont {Thoss}},\ and\ \bibinfo {author}
  {\bibfnamefont {R.}~\bibnamefont {Borrelli}},\ }\bibfield  {title} {\bibinfo
  {title} {High accuracy exponential decomposition of bath correlation
  functions for arbitrary and structured spectral densities: Emerging
  methodologies and new approaches},\ }\href
  {https://doi.org/10.1063/5.0209348} {\bibfield  {journal} {\bibinfo
  {journal} {The Journal of Chemical Physics}\ }\textbf {\bibinfo {volume}
  {160}},\ \bibinfo {pages} {204105} (\bibinfo {year} {2024})}\BibitemShut
  {NoStop}%
\bibitem [{\citenamefont {Eckel}\ \emph {et~al.}(2009)\citenamefont {Eckel},
  \citenamefont {Reina},\ and\ \citenamefont
  {Thorwart}}]{CoherentControl_TLS_nonMarkovian_ReinaEckel}%
  \BibitemOpen
  \bibfield  {author} {\bibinfo {author} {\bibfnamefont {J.}~\bibnamefont
  {Eckel}}, \bibinfo {author} {\bibfnamefont {J.~H.}\ \bibnamefont {Reina}},\
  and\ \bibinfo {author} {\bibfnamefont {M.}~\bibnamefont {Thorwart}},\
  }\bibfield  {title} {\bibinfo {title} {Coherent control of an effective
  two-level system in a non-markovian biomolecular environment},\ }\href
  {https://doi.org/10.1088/1367-2630/11/8/085001} {\bibfield  {journal}
  {\bibinfo  {journal} {New Journal of Physics}\ }\textbf {\bibinfo {volume}
  {11}},\ \bibinfo {pages} {085001} (\bibinfo {year} {2009})}\BibitemShut
  {NoStop}%
\bibitem [{\citenamefont {Weiss}(2012)}]{QDissipation_Weiss}%
  \BibitemOpen
  \bibfield  {author} {\bibinfo {author} {\bibfnamefont {U.}~\bibnamefont
  {Weiss}},\ }\href {https://doi.org/10.1142/8334} {\emph {\bibinfo {title}
  {Quantum Dissipative Systems}}},\ \bibinfo {edition} {4th}\ ed.\ (\bibinfo
  {publisher} {World Scientific},\ \bibinfo {year} {2012})\ \Eprint
  {https://arxiv.org/abs/https://www.worldscientific.com/doi/pdf/10.1142/8334}
  {https://www.worldscientific.com/doi/pdf/10.1142/8334} \BibitemShut {NoStop}%
\bibitem [{\citenamefont {Leggett}\ \emph {et~al.}(1987)\citenamefont
  {Leggett}, \citenamefont {Chakravarty}, \citenamefont {Dorsey}, \citenamefont
  {Fisher}, \citenamefont {Garg},\ and\ \citenamefont
  {Zwerger}}]{Leggett_DisspTLS}%
  \BibitemOpen
  \bibfield  {author} {\bibinfo {author} {\bibfnamefont {A.~J.}\ \bibnamefont
  {Leggett}}, \bibinfo {author} {\bibfnamefont {S.}~\bibnamefont
  {Chakravarty}}, \bibinfo {author} {\bibfnamefont {A.~T.}\ \bibnamefont
  {Dorsey}}, \bibinfo {author} {\bibfnamefont {M.~P.~A.}\ \bibnamefont
  {Fisher}}, \bibinfo {author} {\bibfnamefont {A.}~\bibnamefont {Garg}},\ and\
  \bibinfo {author} {\bibfnamefont {W.}~\bibnamefont {Zwerger}},\ }\bibfield
  {title} {\bibinfo {title} {Dynamics of the dissipative two-state system},\
  }\href {https://doi.org/10.1103/RevModPhys.59.1} {\bibfield  {journal}
  {\bibinfo  {journal} {Rev. Mod. Phys.}\ }\textbf {\bibinfo {volume} {59}},\
  \bibinfo {pages} {1} (\bibinfo {year} {1987})}\BibitemShut {NoStop}%
\bibitem [{\citenamefont {Reina}\ \emph {et~al.}(2002)\citenamefont {Reina},
  \citenamefont {Quiroga},\ and\ \citenamefont {Johnson}}]{JH_QDecoherence}%
  \BibitemOpen
  \bibfield  {author} {\bibinfo {author} {\bibfnamefont {J.~H.}\ \bibnamefont
  {Reina}}, \bibinfo {author} {\bibfnamefont {L.}~\bibnamefont {Quiroga}},\
  and\ \bibinfo {author} {\bibfnamefont {N.~F.}\ \bibnamefont {Johnson}},\
  }\bibfield  {title} {\bibinfo {title} {Decoherence of quantum registers},\
  }\href {https://doi.org/10.1103/PhysRevA.65.032326} {\bibfield  {journal}
  {\bibinfo  {journal} {Phys. Rev. A}\ }\textbf {\bibinfo {volume} {65}},\
  \bibinfo {pages} {032326} (\bibinfo {year} {2002})}\BibitemShut {NoStop}%
\bibitem [{\citenamefont {Scarpetta}\ \emph {et~al.}(2025)\citenamefont
  {Scarpetta}, \citenamefont {Reina},\ and\ \citenamefont
  {Hjorth-Jensen}}]{Scarpetta_2025ML_twolevel}%
  \BibitemOpen
  \bibfield  {author} {\bibinfo {author} {\bibfnamefont {J.~M.}\ \bibnamefont
  {Scarpetta}}, \bibinfo {author} {\bibfnamefont {J.~H.}\ \bibnamefont
  {Reina}},\ and\ \bibinfo {author} {\bibfnamefont {M.}~\bibnamefont
  {Hjorth-Jensen}},\ }\bibfield  {title} {\bibinfo {title} {Machine learning
  non-markovian two-level quantum noise spectroscopy},\ }\href
  {https://doi.org/10.1103/2lzl-vpjd} {\bibfield  {journal} {\bibinfo
  {journal} {Phys. Rev. Res.}\ }\textbf {\bibinfo {volume} {7}},\ \bibinfo
  {pages} {043285} (\bibinfo {year} {2025})}\BibitemShut {NoStop}%
\bibitem [{\citenamefont {Maloney}\ \emph {et~al.}(2022)\citenamefont
  {Maloney}, \citenamefont {Oda}, \citenamefont {Quiroz}, \citenamefont
  {Clader},\ and\ \citenamefont {Norris}}]{QubitControlNoiseSpectroscopy}%
  \BibitemOpen
  \bibfield  {author} {\bibinfo {author} {\bibfnamefont {V.}~\bibnamefont
  {Maloney}}, \bibinfo {author} {\bibfnamefont {Y.}~\bibnamefont {Oda}},
  \bibinfo {author} {\bibfnamefont {G.}~\bibnamefont {Quiroz}}, \bibinfo
  {author} {\bibfnamefont {B.~D.}\ \bibnamefont {Clader}},\ and\ \bibinfo
  {author} {\bibfnamefont {L.~M.}\ \bibnamefont {Norris}},\ }\bibfield  {title}
  {\bibinfo {title} {Qubit control noise spectroscopy with optimal suppression
  of dephasing},\ }\href {https://doi.org/10.1103/PhysRevA.106.022425}
  {\bibfield  {journal} {\bibinfo  {journal} {Phys. Rev. A}\ }\textbf {\bibinfo
  {volume} {106}},\ \bibinfo {pages} {022425} (\bibinfo {year}
  {2022})}\BibitemShut {NoStop}%
\bibitem [{\citenamefont {Campaioli}\ \emph
  {et~al.}(2024{\natexlab{a}})\citenamefont {Campaioli}, \citenamefont {Cole},\
  and\ \citenamefont {Hapuarachchi}}]{Quantum_Master_Equations}%
  \BibitemOpen
  \bibfield  {author} {\bibinfo {author} {\bibfnamefont {F.}~\bibnamefont
  {Campaioli}}, \bibinfo {author} {\bibfnamefont {J.~H.}\ \bibnamefont
  {Cole}},\ and\ \bibinfo {author} {\bibfnamefont {H.}~\bibnamefont
  {Hapuarachchi}},\ }\bibfield  {title} {\bibinfo {title} {Quantum master
  equations: Tips and tricks for quantum optics, quantum computing, and
  beyond},\ }\href {https://doi.org/10.1103/PRXQuantum.5.020202} {\bibfield
  {journal} {\bibinfo  {journal} {PRX Quantum}\ }\textbf {\bibinfo {volume}
  {5}},\ \bibinfo {pages} {020202} (\bibinfo {year}
  {2024}{\natexlab{a}})}\BibitemShut {NoStop}%
\bibitem [{\citenamefont {Brinks}\ \emph {et~al.}(2011)\citenamefont {Brinks},
  \citenamefont {Hildner}, \citenamefont {Stefani},\ and\ \citenamefont {van
  Hulst}}]{Hildner_SingleMolsss_RoomTemp}%
  \BibitemOpen
  \bibfield  {author} {\bibinfo {author} {\bibfnamefont {D.}~\bibnamefont
  {Brinks}}, \bibinfo {author} {\bibfnamefont {R.}~\bibnamefont {Hildner}},
  \bibinfo {author} {\bibfnamefont {F.~D.}\ \bibnamefont {Stefani}},\ and\
  \bibinfo {author} {\bibfnamefont {N.~F.}\ \bibnamefont {van Hulst}},\
  }\bibfield  {title} {\bibinfo {title} {Coherent control of single molecules
  at room temperature},\ }\bibfield  {journal} {\bibinfo  {journal} {Faraday
  Discuss.}\ }\textbf {\bibinfo {volume} {153}},\ \href
  {https://doi.org/10.1039/C1FD00087J} {10.1039/C1FD00087J} (\bibinfo {year}
  {2011})\BibitemShut {NoStop}%
\bibitem [{\citenamefont {peng Xu}\ \emph {et~al.}(2025)\citenamefont {peng
  Xu}, \citenamefont {Herkenrath}, \citenamefont {Scherf},\ and\ \citenamefont
  {Hildner}}]{CohControl_SingleMol_RoomTemp_Hilder_2025}%
  \BibitemOpen
  \bibfield  {author} {\bibinfo {author} {\bibfnamefont {X.}~\bibnamefont {peng
  Xu}}, \bibinfo {author} {\bibfnamefont {T.~M.}\ \bibnamefont {Herkenrath}},
  \bibinfo {author} {\bibfnamefont {U.}~\bibnamefont {Scherf}},\ and\ \bibinfo
  {author} {\bibfnamefont {R.}~\bibnamefont {Hildner}},\ }\bibfield  {title}
  {\bibinfo {title} {Coherent control of single molecules via phase-shaped
  two-photon excitation at room temperature},\ }\href
  {https://doi.org/10.34133/ultrafastscience.0086} {\bibfield  {journal}
  {\bibinfo  {journal} {Ultrafast Science}\ }\textbf {\bibinfo {volume} {5}},\
  \bibinfo {pages} {0086} (\bibinfo {year} {2025})}\BibitemShut {NoStop}%
\bibitem [{\citenamefont {Villabona-Monsalve}\ \emph
  {et~al.}(2017)\citenamefont {Villabona-Monsalve}, \citenamefont
  {Calder{\'o}n-Losada}, \citenamefont {Nu{\~{n}}ez~Portela},\ and\
  \citenamefont {Valencia}}]{VillabonaOmar_TPCS_RhB_Zn}%
  \BibitemOpen
  \bibfield  {author} {\bibinfo {author} {\bibfnamefont {J.~P.}\ \bibnamefont
  {Villabona-Monsalve}}, \bibinfo {author} {\bibfnamefont {O.}~\bibnamefont
  {Calder{\'o}n-Losada}}, \bibinfo {author} {\bibfnamefont {M.}~\bibnamefont
  {Nu{\~{n}}ez~Portela}},\ and\ \bibinfo {author} {\bibfnamefont
  {A.}~\bibnamefont {Valencia}},\ }\bibfield  {title} {\bibinfo {title}
  {Entangled two photon absorption cross section on the 808 nm region for the
  common dyes zinc tetraphenylporphyrin and rhodamine b},\ }\href
  {https://doi.org/10.1021/acs.jpca.7b06450} {\bibfield  {journal} {\bibinfo
  {journal} {The Journal of Physical Chemistry A}\ }\textbf {\bibinfo {volume}
  {121}},\ \bibinfo {pages} {7869} (\bibinfo {year} {2017})}\BibitemShut
  {NoStop}%
\bibitem [{\citenamefont {Dall{'}Osto}\ \emph {et~al.}(2020)\citenamefont
  {Dall{'}Osto}, \citenamefont {Coccia}, \citenamefont {Guido},\ and\
  \citenamefont {Corni}}]{TwoPulsed_fluorophores_singleDNQDI}%
  \BibitemOpen
  \bibfield  {author} {\bibinfo {author} {\bibfnamefont {G.}~\bibnamefont
  {Dall{'}Osto}}, \bibinfo {author} {\bibfnamefont {E.}~\bibnamefont {Coccia}},
  \bibinfo {author} {\bibfnamefont {C.~A.}\ \bibnamefont {Guido}},\ and\
  \bibinfo {author} {\bibfnamefont {S.}~\bibnamefont {Corni}},\ }\bibfield
  {title} {\bibinfo {title} {Investigating ultrafast two-pulse experiments on
  single dnqdi fluorophores: a stochastic quantum approach},\ }\href
  {https://doi.org/10.1039/D0CP02557G} {\bibfield  {journal} {\bibinfo
  {journal} {Phys. Chem. Chem. Phys.}\ }\textbf {\bibinfo {volume} {22}},\
  \bibinfo {pages} {16734} (\bibinfo {year} {2020})}\BibitemShut {NoStop}%
\bibitem [{\citenamefont {Wilma}\ \emph {et~al.}(2018)\citenamefont {Wilma},
  \citenamefont {Shu}, \citenamefont {Scherf},\ and\ \citenamefont
  {Hildner}}]{VisualizingWilma2019}%
  \BibitemOpen
  \bibfield  {author} {\bibinfo {author} {\bibfnamefont {K.}~\bibnamefont
  {Wilma}}, \bibinfo {author} {\bibfnamefont {C.-C.}\ \bibnamefont {Shu}},
  \bibinfo {author} {\bibfnamefont {U.}~\bibnamefont {Scherf}},\ and\ \bibinfo
  {author} {\bibfnamefont {R.}~\bibnamefont {Hildner}},\ }\bibfield  {title}
  {\bibinfo {title} {Visualizing hidden ultrafast processes in individual
  molecules by single-pulse coherent control},\ }\href
  {https://doi.org/10.1021/jacs.8b08674} {\bibfield  {journal} {\bibinfo
  {journal} {Journal of the American Chemical Society}\ }\textbf {\bibinfo
  {volume} {140}},\ \bibinfo {pages} {15329} (\bibinfo {year}
  {2018})}\BibitemShut {NoStop}%
\bibitem [{\citenamefont {Wilma}\ \emph {et~al.}(2019)\citenamefont {Wilma},
  \citenamefont {Shu}, \citenamefont {Scherf},\ and\ \citenamefont
  {Hildner}}]{TwoPhotonWilma2019}%
  \BibitemOpen
  \bibfield  {author} {\bibinfo {author} {\bibfnamefont {K.}~\bibnamefont
  {Wilma}}, \bibinfo {author} {\bibfnamefont {C.-C.}\ \bibnamefont {Shu}},
  \bibinfo {author} {\bibfnamefont {U.}~\bibnamefont {Scherf}},\ and\ \bibinfo
  {author} {\bibfnamefont {R.}~\bibnamefont {Hildner}},\ }\bibfield  {title}
  {\bibinfo {title} {Two-photon induced ultrafast coherence decay of highly
  excited states in single molecules},\ }\href
  {https://doi.org/10.1088/1367-2630/ab115d} {\bibfield  {journal} {\bibinfo
  {journal} {New Journal of Physics}\ }\textbf {\bibinfo {volume} {21}},\
  \bibinfo {pages} {045001} (\bibinfo {year} {2019})}\BibitemShut {NoStop}%
\bibitem [{\citenamefont {Schindler}\ \emph {et~al.}(2004)\citenamefont
  {Schindler}, \citenamefont {Lupton}, \citenamefont {Feldmann},\ and\
  \citenamefont {Scherf}}]{MeLPPP_univPicture_Spectroscopy}%
  \BibitemOpen
  \bibfield  {author} {\bibinfo {author} {\bibfnamefont {F.}~\bibnamefont
  {Schindler}}, \bibinfo {author} {\bibfnamefont {J.~M.}\ \bibnamefont
  {Lupton}}, \bibinfo {author} {\bibfnamefont {J.}~\bibnamefont {Feldmann}},\
  and\ \bibinfo {author} {\bibfnamefont {U.}~\bibnamefont {Scherf}},\
  }\bibfield  {title} {\bibinfo {title} {A universal picture of chromophores in
  $\pi$-conjugated polymers derived from single-molecule spectroscopy},\ }\href
  {https://doi.org/10.1073/pnas.0403325101} {\bibfield  {journal} {\bibinfo
  {journal} {Proceedings of the National Academy of Sciences}\ }\textbf
  {\bibinfo {volume} {101}},\ \bibinfo {pages} {14695} (\bibinfo {year}
  {2004})}\BibitemShut {NoStop}%
\bibitem [{\citenamefont {Hildner}\ \emph {et~al.}(2007)\citenamefont
  {Hildner}, \citenamefont {Lemmer}, \citenamefont {Scherf},\ and\
  \citenamefont {Köhler}}]{MeLPPP_PL_Hildner_Spectroscopy}%
  \BibitemOpen
  \bibfield  {author} {\bibinfo {author} {\bibfnamefont {R.}~\bibnamefont
  {Hildner}}, \bibinfo {author} {\bibfnamefont {U.}~\bibnamefont {Lemmer}},
  \bibinfo {author} {\bibfnamefont {U.}~\bibnamefont {Scherf}},\ and\ \bibinfo
  {author} {\bibfnamefont {J.}~\bibnamefont {Köhler}},\ }\bibfield  {title}
  {\bibinfo {title} {Continuous-wave two-photon spectroscopy on a ladder-type
  conjugated polymer},\ }\href
  {https://doi.org/https://doi.org/10.1016/j.cplett.2007.10.011} {\bibfield
  {journal} {\bibinfo  {journal} {Chemical Physics Letters}\ }\textbf {\bibinfo
  {volume} {448}},\ \bibinfo {pages} {213} (\bibinfo {year}
  {2007})}\BibitemShut {NoStop}%
\bibitem [{\citenamefont {Gulbinas}\ \emph {et~al.}(2007)\citenamefont
  {Gulbinas}, \citenamefont {Minevičiūtė}, \citenamefont {Hertel},
  \citenamefont {Wellander}, \citenamefont {Yartsev},\ and\ \citenamefont
  {Sundström}}]{ExcitonDiff_Relaxat_MeLPPP}%
  \BibitemOpen
  \bibfield  {author} {\bibinfo {author} {\bibfnamefont {V.}~\bibnamefont
  {Gulbinas}}, \bibinfo {author} {\bibfnamefont {I.}~\bibnamefont
  {Minevičiūtė}}, \bibinfo {author} {\bibfnamefont {D.}~\bibnamefont
  {Hertel}}, \bibinfo {author} {\bibfnamefont {R.}~\bibnamefont {Wellander}},
  \bibinfo {author} {\bibfnamefont {A.}~\bibnamefont {Yartsev}},\ and\ \bibinfo
  {author} {\bibfnamefont {V.}~\bibnamefont {Sundström}},\ }\bibfield  {title}
  {\bibinfo {title} {Exciton diffusion and relaxation in methyl-substituted
  polyparaphenylene polymer films},\ }\href {https://doi.org/10.1063/1.2790901}
  {\bibfield  {journal} {\bibinfo  {journal} {The Journal of Chemical Physics}\
  }\textbf {\bibinfo {volume} {127}},\ \bibinfo {pages} {144907} (\bibinfo
  {year} {2007})}\BibitemShut {NoStop}%
\bibitem [{\citenamefont {Hohenau}\ \emph {et~al.}(2001)\citenamefont
  {Hohenau}, \citenamefont {Cagran}, \citenamefont {Kranzelbinder},
  \citenamefont {Scherf},\ and\ \citenamefont
  {Leising}}]{MeLPPP_2P_CrossSection_Hohenau}%
  \BibitemOpen
  \bibfield  {author} {\bibinfo {author} {\bibfnamefont {A.}~\bibnamefont
  {Hohenau}}, \bibinfo {author} {\bibfnamefont {C.}~\bibnamefont {Cagran}},
  \bibinfo {author} {\bibfnamefont {G.}~\bibnamefont {Kranzelbinder}}, \bibinfo
  {author} {\bibfnamefont {U.}~\bibnamefont {Scherf}},\ and\ \bibinfo {author}
  {\bibfnamefont {G.}~\bibnamefont {Leising}},\ }\bibfield  {title} {\bibinfo
  {title} {Efficient continuous-wave two-photon absorption in
  para-phenylene-type polymers},\ }\href@noop {} {\bibfield  {journal}
  {\bibinfo  {journal} {Advanced Materials}\ }\textbf {\bibinfo {volume}
  {13}},\ \bibinfo {pages} {1303} (\bibinfo {year} {2001})}\BibitemShut
  {NoStop}%
\bibitem [{\citenamefont {Reina}\ \emph {et~al.}(2014)\citenamefont {Reina},
  \citenamefont {Susa},\ and\ \citenamefont
  {Fanchini}}]{ExtractInfo_QEnv_Correlations}%
  \BibitemOpen
  \bibfield  {author} {\bibinfo {author} {\bibfnamefont {J.~H.}\ \bibnamefont
  {Reina}}, \bibinfo {author} {\bibfnamefont {C.~E.}\ \bibnamefont {Susa}},\
  and\ \bibinfo {author} {\bibfnamefont {F.~F.}\ \bibnamefont {Fanchini}},\
  }\bibfield  {title} {\bibinfo {title} {Extracting information from
  qubit-environment correlations},\ }\href {https://doi.org/10.1038/srep07443}
  {\bibfield  {journal} {\bibinfo  {journal} {Scientific Reports}\ }\textbf
  {\bibinfo {volume} {4}},\ \bibinfo {pages} {7443} (\bibinfo {year}
  {2014})}\BibitemShut {NoStop}%
\bibitem [{\citenamefont {Wallace}\ \emph {et~al.}(2025)\citenamefont
  {Wallace}, \citenamefont {Altmann}, \citenamefont {Gerardot}, \citenamefont
  {Gauger},\ and\ \citenamefont {Bonato}}]{ExtractInfo_from_ExpData}%
  \BibitemOpen
  \bibfield  {author} {\bibinfo {author} {\bibfnamefont {S.}~\bibnamefont
  {Wallace}}, \bibinfo {author} {\bibfnamefont {Y.}~\bibnamefont {Altmann}},
  \bibinfo {author} {\bibfnamefont {B.~D.}\ \bibnamefont {Gerardot}}, \bibinfo
  {author} {\bibfnamefont {E.~M.}\ \bibnamefont {Gauger}},\ and\ \bibinfo
  {author} {\bibfnamefont {C.}~\bibnamefont {Bonato}},\ }\bibfield  {title}
  {\bibinfo {title} {Learning the dynamics of markovian open quantum systems
  from experimental data},\ }\bibfield  {journal} {\bibinfo  {journal}
  {Physical Review Research}\ }\textbf {\bibinfo {volume} {7}},\ \href
  {https://doi.org/10.1103/m66m-lnjh} {10.1103/m66m-lnjh} (\bibinfo {year}
  {2025})\BibitemShut {NoStop}%
\bibitem [{\citenamefont {Ma}\ \emph {et~al.}(2025)\citenamefont {Ma},
  \citenamefont {Qi}, \citenamefont {Petersen}, \citenamefont {Wu},
  \citenamefont {Rabitz},\ and\ \citenamefont
  {Dong}}]{ML_Control_QuantumSystems}%
  \BibitemOpen
  \bibfield  {author} {\bibinfo {author} {\bibfnamefont {H.}~\bibnamefont
  {Ma}}, \bibinfo {author} {\bibfnamefont {B.}~\bibnamefont {Qi}}, \bibinfo
  {author} {\bibfnamefont {I.~R.}\ \bibnamefont {Petersen}}, \bibinfo {author}
  {\bibfnamefont {R.-B.}\ \bibnamefont {Wu}}, \bibinfo {author} {\bibfnamefont
  {H.}~\bibnamefont {Rabitz}},\ and\ \bibinfo {author} {\bibfnamefont
  {D.}~\bibnamefont {Dong}},\ }\bibfield  {title} {\bibinfo {title} {Machine
  learning for estimation and control of quantum systems},\ }\href
  {https://arxiv.org/abs/2503.03164} {\  (\bibinfo {year} {2025})},\ \Eprint
  {https://arxiv.org/abs/2503.03164} {arXiv:2503.03164 [quant-ph]} \BibitemShut
  {NoStop}%
\bibitem [{\citenamefont {Bandyopadhyay}\ \emph {et~al.}(2018)\citenamefont
  {Bandyopadhyay}, \citenamefont {Huang}, \citenamefont {Sun},\ and\
  \citenamefont {Zhao}}]{NN_Simulating_Quantum_Dynamics}%
  \BibitemOpen
  \bibfield  {author} {\bibinfo {author} {\bibfnamefont {S.}~\bibnamefont
  {Bandyopadhyay}}, \bibinfo {author} {\bibfnamefont {Z.}~\bibnamefont
  {Huang}}, \bibinfo {author} {\bibfnamefont {K.}~\bibnamefont {Sun}},\ and\
  \bibinfo {author} {\bibfnamefont {Y.}~\bibnamefont {Zhao}},\ }\bibfield
  {title} {\bibinfo {title} {Applications of neural networks to the simulation
  of dynamics of open quantum systems},\ }\href
  {https://doi.org/https://doi.org/10.1016/j.chemphys.2018.05.019} {\bibfield
  {journal} {\bibinfo  {journal} {Chemical Physics}\ }\textbf {\bibinfo
  {volume} {515}},\ \bibinfo {pages} {272} (\bibinfo {year} {2018})},\ \bibinfo
  {note} {ultrafast Photoinduced Processes in Polyatomic Molecules:Electronic
  Structure, Dynamics and Spectroscopy (Dedicated to Wolfgang Domcke on the
  occasion of his 70th birthday)}\BibitemShut {NoStop}%
\bibitem [{\citenamefont {Giannelli}\ \emph {et~al.}(2022)\citenamefont
  {Giannelli}, \citenamefont {Sgroi}, \citenamefont {Brown}, \citenamefont
  {Paraoanu}, \citenamefont {Paternostro}, \citenamefont {Paladino},\ and\
  \citenamefont {Falci}}]{OCC_ReinforcementLearning}%
  \BibitemOpen
  \bibfield  {author} {\bibinfo {author} {\bibfnamefont {L.}~\bibnamefont
  {Giannelli}}, \bibinfo {author} {\bibfnamefont {S.}~\bibnamefont {Sgroi}},
  \bibinfo {author} {\bibfnamefont {J.}~\bibnamefont {Brown}}, \bibinfo
  {author} {\bibfnamefont {G.~S.}\ \bibnamefont {Paraoanu}}, \bibinfo {author}
  {\bibfnamefont {M.}~\bibnamefont {Paternostro}}, \bibinfo {author}
  {\bibfnamefont {E.}~\bibnamefont {Paladino}},\ and\ \bibinfo {author}
  {\bibfnamefont {G.}~\bibnamefont {Falci}},\ }\bibfield  {title} {\bibinfo
  {title} {A tutorial on optimal control and reinforcement learning methods for
  quantum technologies},\ }\href
  {https://doi.org/10.1016/j.physleta.2022.128054} {\bibfield  {journal}
  {\bibinfo  {journal} {Physics Letters A}\ }\textbf {\bibinfo {volume}
  {434}},\ \bibinfo {pages} {128054} (\bibinfo {year} {2022})}\BibitemShut
  {NoStop}%
\bibitem [{\citenamefont {Barr}\ \emph {et~al.}(2024)\citenamefont {Barr},
  \citenamefont {Zicari}, \citenamefont {Ferraro},\ and\ \citenamefont
  {Paternostro}}]{Paternostro_1}%
  \BibitemOpen
  \bibfield  {author} {\bibinfo {author} {\bibfnamefont {J.}~\bibnamefont
  {Barr}}, \bibinfo {author} {\bibfnamefont {G.}~\bibnamefont {Zicari}},
  \bibinfo {author} {\bibfnamefont {A.}~\bibnamefont {Ferraro}},\ and\ \bibinfo
  {author} {\bibfnamefont {M.}~\bibnamefont {Paternostro}},\ }\bibfield
  {title} {\bibinfo {title} {Spectral density classification for environment
  spectroscopy},\ }\href {https://doi.org/10.1088/2632-2153/ad2cf1} {\bibfield
  {journal} {\bibinfo  {journal} {Machine Learning: Science and Technology}\
  }\textbf {\bibinfo {volume} {5}},\ \bibinfo {pages} {015043} (\bibinfo {year}
  {2024})}\BibitemShut {NoStop}%
\bibitem [{\citenamefont {Barr}\ \emph
  {et~al.}(2025{\natexlab{a}})\citenamefont {Barr}, \citenamefont {Ferraro},
  \citenamefont {Paternostro},\ and\ \citenamefont {Zicari}}]{Paternostro_2}%
  \BibitemOpen
  \bibfield  {author} {\bibinfo {author} {\bibfnamefont {J.}~\bibnamefont
  {Barr}}, \bibinfo {author} {\bibfnamefont {A.}~\bibnamefont {Ferraro}},
  \bibinfo {author} {\bibfnamefont {M.}~\bibnamefont {Paternostro}},\ and\
  \bibinfo {author} {\bibfnamefont {G.}~\bibnamefont {Zicari}},\ }\href
  {https://arxiv.org/abs/2501.07485} {\bibinfo {title} {Machine
  learning-enhanced characterisation of structured spectral densities:
  Leveraging the reaction coordinate mapping}} (\bibinfo {year}
  {2025}{\natexlab{a}}),\ \Eprint {https://arxiv.org/abs/2501.07485}
  {arXiv:2501.07485 [quant-ph]} \BibitemShut {NoStop}%
\bibitem [{\citenamefont {Barr}\ \emph
  {et~al.}(2025{\natexlab{b}})\citenamefont {Barr}, \citenamefont {Mukherjee},
  \citenamefont {Ferraro}, \citenamefont {Paternostro},\ and\ \citenamefont
  {Zicari}}]{Jbarr2025_ML_SDapproach}%
  \BibitemOpen
  \bibfield  {author} {\bibinfo {author} {\bibfnamefont {J.}~\bibnamefont
  {Barr}}, \bibinfo {author} {\bibfnamefont {S.}~\bibnamefont {Mukherjee}},
  \bibinfo {author} {\bibfnamefont {A.}~\bibnamefont {Ferraro}}, \bibinfo
  {author} {\bibfnamefont {M.}~\bibnamefont {Paternostro}},\ and\ \bibinfo
  {author} {\bibfnamefont {G.}~\bibnamefont {Zicari}},\ }\bibfield  {title}
  {\bibinfo {title} {A machine learning based approach to the identification of
  spectral densities in quantum open systems},\ }\href
  {https://arxiv.org/abs/2507.13730} {\  (\bibinfo {year}
  {2025}{\natexlab{b}})},\ \Eprint {https://arxiv.org/abs/2507.13730}
  {arXiv:2507.13730 [quant-ph]} \BibitemShut {NoStop}%
\bibitem [{\citenamefont {Luchnikov}\ \emph {et~al.}(2020)\citenamefont
  {Luchnikov}, \citenamefont {Vintskevich}, \citenamefont {Grigoriev},\ and\
  \citenamefont {Filippov}}]{ML_nonMark_Dynamics}%
  \BibitemOpen
  \bibfield  {author} {\bibinfo {author} {\bibfnamefont {I.}~\bibnamefont
  {Luchnikov}}, \bibinfo {author} {\bibfnamefont {S.}~\bibnamefont
  {Vintskevich}}, \bibinfo {author} {\bibfnamefont {D.}~\bibnamefont
  {Grigoriev}},\ and\ \bibinfo {author} {\bibfnamefont {S.}~\bibnamefont
  {Filippov}},\ }\bibfield  {title} {\bibinfo {title} {Machine learning
  non-markovian quantum dynamics},\ }\bibfield  {journal} {\bibinfo  {journal}
  {Physical Review Letters}\ }\textbf {\bibinfo {volume} {124}},\ \href
  {https://doi.org/10.1103/physrevlett.124.140502}
  {10.1103/physrevlett.124.140502} (\bibinfo {year} {2020})\BibitemShut
  {NoStop}%
\bibitem [{\citenamefont {Martina}\ \emph {et~al.}(2023)\citenamefont
  {Martina}, \citenamefont {Gherardini},\ and\ \citenamefont
  {Caruso}}]{ML_nonMark_noiseQD}%
  \BibitemOpen
  \bibfield  {author} {\bibinfo {author} {\bibfnamefont {S.}~\bibnamefont
  {Martina}}, \bibinfo {author} {\bibfnamefont {S.}~\bibnamefont
  {Gherardini}},\ and\ \bibinfo {author} {\bibfnamefont {F.}~\bibnamefont
  {Caruso}},\ }\bibfield  {title} {\bibinfo {title} {Machine learning
  classification of non-markovian noise disturbing quantum dynamics},\ }\href
  {https://doi.org/10.1088/1402-4896/acb39b} {\bibfield  {journal} {\bibinfo
  {journal} {Physica Scripta}\ }\textbf {\bibinfo {volume} {98}},\ \bibinfo
  {pages} {035104} (\bibinfo {year} {2023})}\BibitemShut {NoStop}%
\bibitem [{\citenamefont {Jafari}\ \emph {et~al.}(2022)\citenamefont {Jafari},
  \citenamefont {Khosravi},\ and\ \citenamefont
  {Trebino}}]{Trebino_PulseShaping_2022}%
  \BibitemOpen
  \bibfield  {author} {\bibinfo {author} {\bibfnamefont {R.}~\bibnamefont
  {Jafari}}, \bibinfo {author} {\bibfnamefont {S.~D.}\ \bibnamefont
  {Khosravi}},\ and\ \bibinfo {author} {\bibfnamefont {R.}~\bibnamefont
  {Trebino}},\ }\bibfield  {title} {\bibinfo {title} {Reliable determination of
  pulse-shape instability in trains of ultrashort laser pulses using
  frequency-resolved optical gating},\ }\href
  {https://doi.org/10.1038/s41598-022-25193-3} {\bibfield  {journal} {\bibinfo
  {journal} {Scientific Reports}\ }\textbf {\bibinfo {volume} {12}},\ \bibinfo
  {pages} {21006} (\bibinfo {year} {2022})}\BibitemShut {NoStop}%
\bibitem [{\citenamefont {Weiner}(2011)}]{Weiner2010_PulseShapingTutorial}%
  \BibitemOpen
  \bibfield  {author} {\bibinfo {author} {\bibfnamefont {A.~M.}\ \bibnamefont
  {Weiner}},\ }\bibfield  {title} {\bibinfo {title} {Ultrafast optical pulse
  shaping: A tutorial review},\ }\href
  {https://doi.org/https://doi.org/10.1016/j.optcom.2011.03.084} {\bibfield
  {journal} {\bibinfo  {journal} {Optics Communications}\ }\textbf {\bibinfo
  {volume} {284}},\ \bibinfo {pages} {3669} (\bibinfo {year} {2011})},\
  \bibinfo {note} {special Issue on Optical Pulse Shaping, Arbitrary Waveform
  Generation, and Pulse Characterization}\BibitemShut {NoStop}%
\bibitem [{\citenamefont {Schlawin}\ and\ \citenamefont
  {Buchleitner}(2017)}]{Buchleitner_TheoryCoherentControl}%
  \BibitemOpen
  \bibfield  {author} {\bibinfo {author} {\bibfnamefont {F.}~\bibnamefont
  {Schlawin}}\ and\ \bibinfo {author} {\bibfnamefont {A.}~\bibnamefont
  {Buchleitner}},\ }\bibfield  {title} {\bibinfo {title} {Theory of coherent
  control with quantum light},\ }\href
  {https://doi.org/10.1088/1367-2630/aa55ec} {\bibfield  {journal} {\bibinfo
  {journal} {New Journal of Physics}\ }\textbf {\bibinfo {volume} {19}},\
  \bibinfo {pages} {013009} (\bibinfo {year} {2017})}\BibitemShut {NoStop}%
\bibitem [{\citenamefont {{Meshulach}}\ and\ \citenamefont
  {{Silberberg}}(1998)}]{Silberberg_TP_fsLaser}%
  \BibitemOpen
  \bibfield  {author} {\bibinfo {author} {\bibfnamefont {D.}~\bibnamefont
  {{Meshulach}}}\ and\ \bibinfo {author} {\bibfnamefont {Y.}~\bibnamefont
  {{Silberberg}}},\ }\bibfield  {title} {\bibinfo {title} {{Coherent quantum
  control of two-photon transitions by a femtosecond laser pulse}},\ }\href
  {https://doi.org/10.1038/24329} {\bibfield  {journal} {\bibinfo  {journal}
  {\nat}\ }\textbf {\bibinfo {volume} {396}},\ \bibinfo {pages} {239} (\bibinfo
  {year} {1998})}\BibitemShut {NoStop}%
\bibitem [{\citenamefont {Dayan}\ \emph {et~al.}(2004)\citenamefont {Dayan},
  \citenamefont {Pe'er}, \citenamefont {Friesem},\ and\ \citenamefont
  {Silberberg}}]{Silberberg_TP_CohControl_BDC_light}%
  \BibitemOpen
  \bibfield  {author} {\bibinfo {author} {\bibfnamefont {B.}~\bibnamefont
  {Dayan}}, \bibinfo {author} {\bibfnamefont {A.}~\bibnamefont {Pe'er}},
  \bibinfo {author} {\bibfnamefont {A.~A.}\ \bibnamefont {Friesem}},\ and\
  \bibinfo {author} {\bibfnamefont {Y.}~\bibnamefont {Silberberg}},\ }\bibfield
   {title} {\bibinfo {title} {Two photon absorption and coherent control with
  broadband down-converted light},\ }\href
  {https://doi.org/10.1103/PhysRevLett.93.023005} {\bibfield  {journal}
  {\bibinfo  {journal} {Phys. Rev. Lett.}\ }\textbf {\bibinfo {volume} {93}},\
  \bibinfo {pages} {023005} (\bibinfo {year} {2004})}\BibitemShut {NoStop}%
\bibitem [{\citenamefont {Campaioli}\ \emph
  {et~al.}(2024{\natexlab{b}})\citenamefont {Campaioli}, \citenamefont {Cole},\
  and\ \citenamefont {Hapuarachchi}}]{mastereq2024}%
  \BibitemOpen
  \bibfield  {author} {\bibinfo {author} {\bibfnamefont {F.}~\bibnamefont
  {Campaioli}}, \bibinfo {author} {\bibfnamefont {J.~H.}\ \bibnamefont
  {Cole}},\ and\ \bibinfo {author} {\bibfnamefont {H.}~\bibnamefont
  {Hapuarachchi}},\ }\bibfield  {title} {\bibinfo {title} {Quantum {Master}
  {Equations}: {Tips} and {Tricks} for {Quantum} {Optics}, {Quantum}
  {Computing}, and {Beyond}},\ }\href
  {https://doi.org/10.1103/PRXQuantum.5.020202} {\bibfield  {journal} {\bibinfo
   {journal} {PRX Quantum}\ }\textbf {\bibinfo {volume} {5}},\ \bibinfo {pages}
  {020202} (\bibinfo {year} {2024}{\natexlab{b}})}\BibitemShut {NoStop}%
\bibitem [{\citenamefont {Virtanen}\ \emph {et~al.}(2020)\citenamefont
  {Virtanen}, \citenamefont {Gommers},\ and\ \citenamefont
  {Oliphant}}]{2020SciPy}%
  \BibitemOpen
  \bibfield  {author} {\bibinfo {author} {\bibfnamefont {P.}~\bibnamefont
  {Virtanen}}, \bibinfo {author} {\bibfnamefont {R.}~\bibnamefont {Gommers}},\
  and\ \bibinfo {author} {\bibfnamefont {e.~a.}\ \bibnamefont {Oliphant},
  \bibfnamefont {Travis~E.}},\ }\bibfield  {title} {\bibinfo {title} {Scipy
  1.0: fundamental algorithms for scientific computing in python},\ }\href
  {https://doi.org/10.1038/s41592-019-0686-2} {\bibfield  {journal} {\bibinfo
  {journal} {Nature Methods}\ }\textbf {\bibinfo {volume} {17}},\ \bibinfo
  {pages} {261–272} (\bibinfo {year} {2020})}\BibitemShut {NoStop}%
\bibitem [{\citenamefont {Beerepoot}\ \emph {et~al.}(2015)\citenamefont
  {Beerepoot}, \citenamefont {Friese}, \citenamefont {van Gisbergen},\ and\
  \citenamefont {Baerends}}]{LineShape_OptimalPL_Emission}%
  \BibitemOpen
  \bibfield  {author} {\bibinfo {author} {\bibfnamefont {M.~T.~P.}\
  \bibnamefont {Beerepoot}}, \bibinfo {author} {\bibfnamefont {D.~H.}\
  \bibnamefont {Friese}}, \bibinfo {author} {\bibfnamefont {S.~J.~A.}\
  \bibnamefont {van Gisbergen}},\ and\ \bibinfo {author} {\bibfnamefont
  {E.~J.}\ \bibnamefont {Baerends}},\ }\bibfield  {title} {\bibinfo {title}
  {Benchmarking two-photon absorption cross sections: performance of cc2 and
  cam-b3lyp},\ }\href {https://doi.org/10.1039/C5CP03241E} {\bibfield
  {journal} {\bibinfo  {journal} {Phys. Chem. Chem. Phys.}\ }\textbf {\bibinfo
  {volume} {17}},\ \bibinfo {pages} {19306} (\bibinfo {year}
  {2015})}\BibitemShut {NoStop}%
\bibitem [{\citenamefont {Sannikov}\ \emph {et~al.}(2025)\citenamefont
  {Sannikov}, \citenamefont {Urazova}, \citenamefont {Kolker}, \citenamefont
  {Averchenko}, \citenamefont {Ivanov}, \citenamefont {Putintsev},
  \citenamefont {Sahharova}, \citenamefont {Shlapakov}, \citenamefont
  {Ananikov},\ and\ \citenamefont
  {Lagoudakis}}]{CouplingRegimes_MeLPPP_microcavities}%
  \BibitemOpen
  \bibfield  {author} {\bibinfo {author} {\bibfnamefont {D.~A.}\ \bibnamefont
  {Sannikov}}, \bibinfo {author} {\bibfnamefont {N.~M.}\ \bibnamefont
  {Urazova}}, \bibinfo {author} {\bibfnamefont {M.~D.}\ \bibnamefont {Kolker}},
  \bibinfo {author} {\bibfnamefont {A.~V.}\ \bibnamefont {Averchenko}},
  \bibinfo {author} {\bibfnamefont {G.~D.}\ \bibnamefont {Ivanov}}, \bibinfo
  {author} {\bibfnamefont {A.~D.}\ \bibnamefont {Putintsev}}, \bibinfo {author}
  {\bibfnamefont {L.~T.}\ \bibnamefont {Sahharova}}, \bibinfo {author}
  {\bibfnamefont {N.~S.}\ \bibnamefont {Shlapakov}}, \bibinfo {author}
  {\bibfnamefont {V.~P.}\ \bibnamefont {Ananikov}},\ and\ \bibinfo {author}
  {\bibfnamefont {P.~G.}\ \bibnamefont {Lagoudakis}},\ }\bibfield  {title}
  {\bibinfo {title} {Strong- vs weak-coupling lasing in polymer-film
  microcavities},\ }\href {https://arxiv.org/abs/2510.13384} {\  (\bibinfo
  {year} {2025})},\ \Eprint {https://arxiv.org/abs/2510.13384}
  {arXiv:2510.13384 [physics.optics]} \BibitemShut {NoStop}%
\bibitem [{\citenamefont {Sharma}\ and\ \citenamefont
  {Chatterjee}(2021)}]{Winsorization}%
  \BibitemOpen
  \bibfield  {author} {\bibinfo {author} {\bibfnamefont {S.}~\bibnamefont
  {Sharma}}\ and\ \bibinfo {author} {\bibfnamefont {S.}~\bibnamefont
  {Chatterjee}},\ }\bibfield  {title} {\bibinfo {title} {Winsorization for
  robust bayesian neural networks},\ }\bibfield  {journal} {\bibinfo  {journal}
  {Entropy}\ }\textbf {\bibinfo {volume} {23}},\ \href
  {https://doi.org/10.3390/e23111546} {10.3390/e23111546} (\bibinfo {year}
  {2021})\BibitemShut {NoStop}%
\bibitem [{\citenamefont {Yu}\ and\ \citenamefont
  {Spiliopoulos}(2023)}]{NormalizationEffects_NN_Training}%
  \BibitemOpen
  \bibfield  {author} {\bibinfo {author} {\bibfnamefont {J.}~\bibnamefont
  {Yu}}\ and\ \bibinfo {author} {\bibfnamefont {K.}~\bibnamefont
  {Spiliopoulos}},\ }\bibfield  {title} {\bibinfo {title} {Normalization
  effects on deep neural networks},\ }\href
  {https://doi.org/10.3934/fods.2023004} {\bibfield  {journal} {\bibinfo
  {journal} {Foundations of Data Science}\ }\textbf {\bibinfo {volume} {5}},\
  \bibinfo {pages} {389} (\bibinfo {year} {2023})}\BibitemShut {NoStop}%
\bibitem [{\citenamefont {Kim}\ \emph {et~al.}(2025)\citenamefont {Kim},
  \citenamefont {Kim}, \citenamefont {Fu}, \citenamefont {Liu}, \citenamefont
  {Wang},\ and\ \citenamefont {Srebric}}]{ImpactNormalization_ANN}%
  \BibitemOpen
  \bibfield  {author} {\bibinfo {author} {\bibfnamefont {Y.-S.}\ \bibnamefont
  {Kim}}, \bibinfo {author} {\bibfnamefont {M.~K.}\ \bibnamefont {Kim}},
  \bibinfo {author} {\bibfnamefont {N.}~\bibnamefont {Fu}}, \bibinfo {author}
  {\bibfnamefont {J.}~\bibnamefont {Liu}}, \bibinfo {author} {\bibfnamefont
  {J.}~\bibnamefont {Wang}},\ and\ \bibinfo {author} {\bibfnamefont
  {J.}~\bibnamefont {Srebric}},\ }\bibfield  {title} {\bibinfo {title}
  {Investigating the impact of data normalization methods on predicting
  electricity consumption in a building using different artificial neural
  network models},\ }\href
  {https://doi.org/https://doi.org/10.1016/j.scs.2024.105570} {\bibfield
  {journal} {\bibinfo  {journal} {Sustainable Cities and Society}\ }\textbf
  {\bibinfo {volume} {118}},\ \bibinfo {pages} {105570} (\bibinfo {year}
  {2025})}\BibitemShut {NoStop}%
\bibitem [{\citenamefont {Paszke}\ \emph {et~al.}(2019)\citenamefont {Paszke},
  \citenamefont {Gross}, \citenamefont {Massa}, \citenamefont {Lerer},
  \citenamefont {Bradbury}, \citenamefont {Chanan}, \citenamefont {Killeen},
  \citenamefont {Lin}, \citenamefont {Gimelshein}, \citenamefont {Antiga},
  \citenamefont {Desmaison}, \citenamefont {Köpf}, \citenamefont {Yang},
  \citenamefont {DeVito}, \citenamefont {Raison}, \citenamefont {Tejani},
  \citenamefont {Chilamkurthy}, \citenamefont {Steiner}, \citenamefont {Fang},
  \citenamefont {Bai},\ and\ \citenamefont {Chintala}}]{Pytorch}%
  \BibitemOpen
  \bibfield  {author} {\bibinfo {author} {\bibfnamefont {A.}~\bibnamefont
  {Paszke}}, \bibinfo {author} {\bibfnamefont {S.}~\bibnamefont {Gross}},
  \bibinfo {author} {\bibfnamefont {F.}~\bibnamefont {Massa}}, \bibinfo
  {author} {\bibfnamefont {A.}~\bibnamefont {Lerer}}, \bibinfo {author}
  {\bibfnamefont {J.}~\bibnamefont {Bradbury}}, \bibinfo {author}
  {\bibfnamefont {G.}~\bibnamefont {Chanan}}, \bibinfo {author} {\bibfnamefont
  {T.}~\bibnamefont {Killeen}}, \bibinfo {author} {\bibfnamefont
  {Z.}~\bibnamefont {Lin}}, \bibinfo {author} {\bibfnamefont {N.}~\bibnamefont
  {Gimelshein}}, \bibinfo {author} {\bibfnamefont {L.}~\bibnamefont {Antiga}},
  \bibinfo {author} {\bibfnamefont {A.}~\bibnamefont {Desmaison}}, \bibinfo
  {author} {\bibfnamefont {A.}~\bibnamefont {Köpf}}, \bibinfo {author}
  {\bibfnamefont {E.}~\bibnamefont {Yang}}, \bibinfo {author} {\bibfnamefont
  {Z.}~\bibnamefont {DeVito}}, \bibinfo {author} {\bibfnamefont
  {M.}~\bibnamefont {Raison}}, \bibinfo {author} {\bibfnamefont
  {A.}~\bibnamefont {Tejani}}, \bibinfo {author} {\bibfnamefont
  {S.}~\bibnamefont {Chilamkurthy}}, \bibinfo {author} {\bibfnamefont
  {B.}~\bibnamefont {Steiner}}, \bibinfo {author} {\bibfnamefont
  {L.}~\bibnamefont {Fang}}, \bibinfo {author} {\bibfnamefont {J.}~\bibnamefont
  {Bai}},\ and\ \bibinfo {author} {\bibfnamefont {S.}~\bibnamefont
  {Chintala}},\ }\bibfield  {title} {\bibinfo {title} {Pytorch: An imperative
  style, high-performance deep learning library},\ }\href
  {https://arxiv.org/abs/1912.01703} {\  (\bibinfo {year} {2019})},\ \Eprint
  {https://arxiv.org/abs/1912.01703} {arXiv:1912.01703 [cs.LG]} \BibitemShut
  {NoStop}%
\bibitem [{\citenamefont {Paszke}\ \emph {et~al.}(2017)\citenamefont {Paszke},
  \citenamefont {Gross}, \citenamefont {Chintala}, \citenamefont {Chanan},
  \citenamefont {Yang}, \citenamefont {DeVito}, \citenamefont {Lin},
  \citenamefont {Desmaison}, \citenamefont {Antiga},\ and\ \citenamefont
  {Lerer}}]{AutomaticDiff_Pytorch}%
  \BibitemOpen
  \bibfield  {author} {\bibinfo {author} {\bibfnamefont {A.}~\bibnamefont
  {Paszke}}, \bibinfo {author} {\bibfnamefont {S.}~\bibnamefont {Gross}},
  \bibinfo {author} {\bibfnamefont {S.}~\bibnamefont {Chintala}}, \bibinfo
  {author} {\bibfnamefont {G.}~\bibnamefont {Chanan}}, \bibinfo {author}
  {\bibfnamefont {E.}~\bibnamefont {Yang}}, \bibinfo {author} {\bibfnamefont
  {Z.}~\bibnamefont {DeVito}}, \bibinfo {author} {\bibfnamefont
  {Z.}~\bibnamefont {Lin}}, \bibinfo {author} {\bibfnamefont {A.}~\bibnamefont
  {Desmaison}}, \bibinfo {author} {\bibfnamefont {L.}~\bibnamefont {Antiga}},\
  and\ \bibinfo {author} {\bibfnamefont {A.}~\bibnamefont {Lerer}},\ }\bibfield
   {title} {\bibinfo {title} {Automatic differentiation in pytorch},\ }in\
  \href@noop {} {\emph {\bibinfo {booktitle} {NIPS-W}}}\ (\bibinfo {year}
  {2017})\BibitemShut {NoStop}%
\bibitem [{\citenamefont {Murphy}(2013)}]{ML_probabilisticPersp_Murphy}%
  \BibitemOpen
  \bibfield  {author} {\bibinfo {author} {\bibfnamefont {K.~P.}\ \bibnamefont
  {Murphy}},\ }\href
  {https://www.amazon.com/Machine-Learning-Probabilistic-Perspective-Computation/dp/0262018020/ref=sr_1_2?ie=UTF8&qid=1336857747&sr=8-2}
  {\emph {\bibinfo {title} {Machine learning: A probabilistic perspective}}}\
  (\bibinfo  {publisher} {MIT Press},\ \bibinfo {address} {Cambridge, Mass.
  [u.a.]},\ \bibinfo {year} {2013})\BibitemShut {NoStop}%
\bibitem [{\citenamefont {Vapnik}(2018)}]{Vapnik_2018}%
  \BibitemOpen
  \bibfield  {author} {\bibinfo {author} {\bibfnamefont {V.~N.}\ \bibnamefont
  {Vapnik}},\ }\href {https://link.springer.com/book/10.1007/978-1-4757-3264-1}
  {\emph {\bibinfo {title} {The nature of statistical learning theory}}}\
  (\bibinfo  {publisher} {Springer New York},\ \bibinfo {year}
  {2018})\BibitemShut {NoStop}%
\bibitem [{\citenamefont {Goodfellow}\ \emph {et~al.}(2016)\citenamefont
  {Goodfellow}, \citenamefont {Bengio},\ and\ \citenamefont
  {Courville}}]{Goodfellow2016}%
  \BibitemOpen
  \bibfield  {author} {\bibinfo {author} {\bibfnamefont {I.}~\bibnamefont
  {Goodfellow}}, \bibinfo {author} {\bibfnamefont {Y.}~\bibnamefont {Bengio}},\
  and\ \bibinfo {author} {\bibfnamefont {A.}~\bibnamefont {Courville}},\
  }\href@noop {} {\emph {\bibinfo {title} {Deep Learning}}}\ (\bibinfo
  {publisher} {The MIT Press, Cambridge, Massachusetts},\ \bibinfo {year}
  {2016})\BibitemShut {NoStop}%
\bibitem [{\citenamefont {Rainio}\ \emph {et~al.}(2024)\citenamefont {Rainio},
  \citenamefont {Teuho},\ and\ \citenamefont {Kl{\'e}n}}]{ML_metrics}%
  \BibitemOpen
  \bibfield  {author} {\bibinfo {author} {\bibfnamefont {O.}~\bibnamefont
  {Rainio}}, \bibinfo {author} {\bibfnamefont {J.}~\bibnamefont {Teuho}},\ and\
  \bibinfo {author} {\bibfnamefont {R.}~\bibnamefont {Kl{\'e}n}},\ }\bibfield
  {title} {\bibinfo {title} {Evaluation metrics and statistical tests for
  machine learning},\ }\href {https://doi.org/10.1038/s41598-024-56706-x}
  {\bibfield  {journal} {\bibinfo  {journal} {Scientific Reports}\ }\textbf
  {\bibinfo {volume} {14}},\ \bibinfo {pages} {6086} (\bibinfo {year}
  {2024})}\BibitemShut {NoStop}%
\bibitem [{\citenamefont {Hastie}\ \emph {et~al.}(2009)\citenamefont {Hastie},
  \citenamefont {Tibshirani},\ and\ \citenamefont
  {Friedman}}]{Hastie2009_ElementsStatLearning}%
  \BibitemOpen
  \bibfield  {author} {\bibinfo {author} {\bibfnamefont {T.}~\bibnamefont
  {Hastie}}, \bibinfo {author} {\bibfnamefont {R.}~\bibnamefont {Tibshirani}},\
  and\ \bibinfo {author} {\bibfnamefont {J.}~\bibnamefont {Friedman}},\
  }\href@noop {} {\emph {\bibinfo {title} {The Elements of Statistical
  Learning: Data Mining, Inference, and Prediction}}},\ \bibinfo {edition}
  {2nd}\ ed.\ (\bibinfo  {publisher} {Springer},\ \bibinfo {address} {New York,
  NY},\ \bibinfo {year} {2009})\BibitemShut {NoStop}%
\end{thebibliography}


\clearpage
\onecolumngrid
\section*{Supplementary Information}
\appendix
\section{Neural network training\label{app:NN_training}}
\noindent
\begin{minipage}[h]{0.40\textwidth}
In Sec.~\ref{sec:NN_methods}, we described the training strategy employed for the neural network. As a preliminary step, exploratory tests were conducted to identify the optimal hyperparameters for the subsequent training and validation procedures. These tests focused on determining an appropriate network architecture with enough hidden layers to provide sufficient capacity to prevent underfitting and adequately capture the relevant structures relating the input PL data to the output parameters $\Q$, while simultaneously avoiding excessive complexity that could lead to overfitting. To perform this analysis, we used a learning rate of $\delta = 10^{-3}$, a batch size of 256 and no dropout, and we trained each model for up to 100 epochs with early stopping using a patience of 30 epochs. This procedure was used to train and record the validation error for each architecture considered, spanning from one to eleven hidden layers. The validation MSE was further decomposed into its bias-squared and variance contributions. In order to obtain robust estimates, the total validation error for each architecture was bagged over 10 independent validation curves. The results are summarized in Fig.~\ref{fig:bias-var_tradeoff} and the definitions of the corresponding indices are provided in Table~\ref{tab:BVTO_indexes}.\\
\end{minipage}
\hfill
\begin{minipage}[h]{0.56\textwidth}
\renewcommand{\arraystretch}{1.2}
\setlength{\tabcolsep}{6pt} 
\begin{tabular}{cl}
\textbf{ID} & {\centering\textbf{Model Sizes}} \\
\hline
1 & [8] \\
2 & [12] \\
3 & [16] \\
4 & [20] \\
5 & [24] \\
6 & [28] \\
7 & [32] \\
8 & [64, 32] \\
9 & [76, 48] \\
10 & [88, 60] \\
11 & [128, 64] \\
12 & [128, 64, 32] \\
13 & [256, 128, 64, 32] \\
14 & [512, 256, 128, 64, 32, 16] \\
15 & [1024, 512, 256, 128, 64, 32, 16] \\
16 & [2048, 1024, 512, 256, 128, 64, 32, 16] \\
17 & [4096, 2048, 1024, 512, 256, 128, 64, 32, 16, 4] \\
18 & [8192, 4096, 2048, 1024, 512, 256, 128, 64, 32, 16, 4] \\
\end{tabular}
\captionof{table}{Representation of the model complexity indices {\bf ID} shown in Fig.~\ref{fig:bias-var_tradeoff} mapping them to the corresponding neural network hidden layer architectures (model complexity). The numbers in the parentheses refer to number of nodes in each layer. The deepest network has eleven hidden layers and is given by {\bf ID} 18.}
\label{tab:BVTO_indexes}
\end{minipage}

\begin{figure}[h!]
    \centering
    \includegraphics[width=0.8\linewidth]{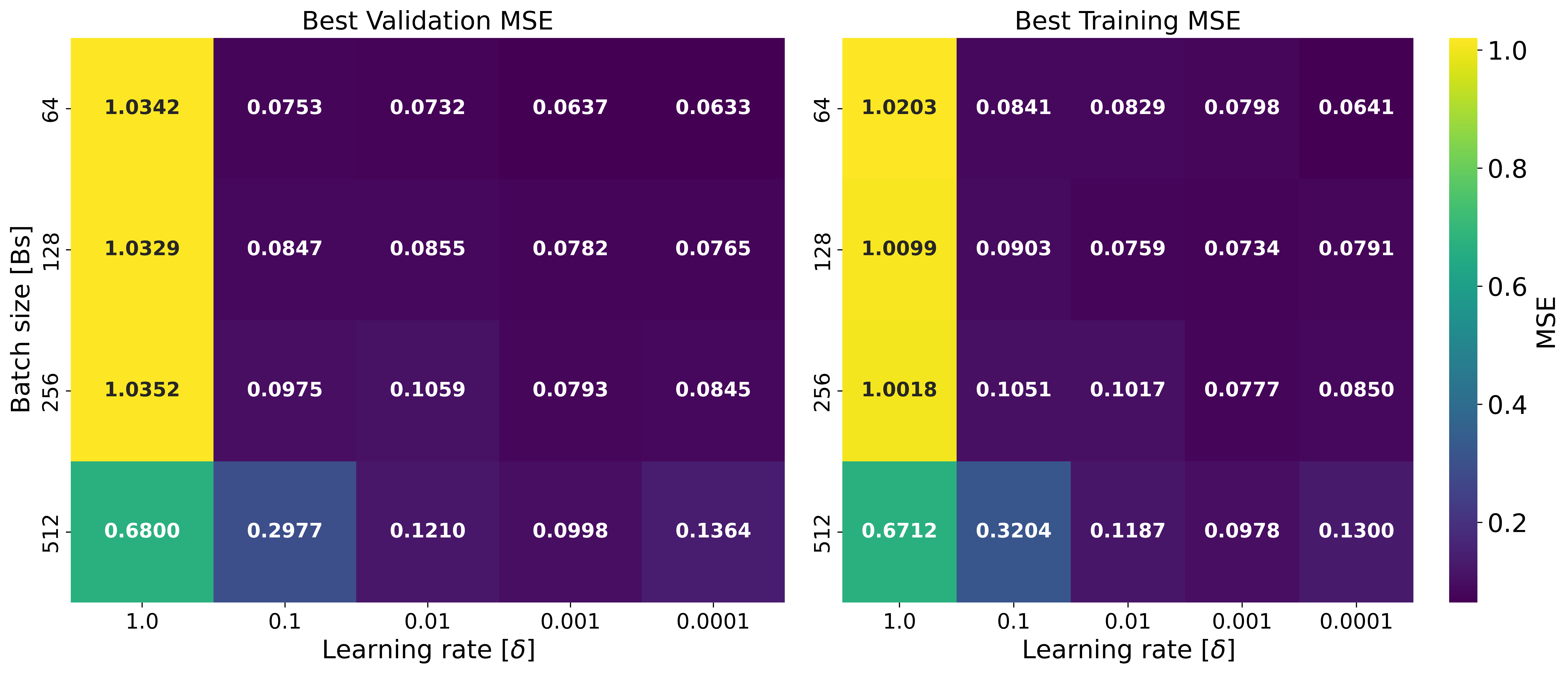}
    \caption{Heatmaps of the best MSE achieved across the hyperparameter grid. The left panel shows the best validation MSE, and the right panel shows the best training MSE. It is seen that the minimum MSE within 200 epochs is achieved for $\delta=10^{-4}$ and $\text{Bs}=64$.}
    \label{fig:val_train_MSE_heatmap}
\end{figure}

Once the optimal network hidden layers architecture of [128,64,32] was determined, the remaining hyperparameters were optimized, keeping this architecture fixed. We set fixed training conditions of 200 epochs and a patience of 30 epochs and performed a grid search over batch sizes $\text{Bs}=\lbrace 64,128,256,512\rbrace$ and learning rates $\delta=\lbrace 1, 10^{-1}, 10^{-2}, 10^{-3}, 10^{-4}\rbrace$. The resulting training and validation errors are presented in Fig.~\ref{fig:val_train_MSE_heatmap}.

\begin{figure}[h!]
    \centering
    \includegraphics[width=\linewidth]{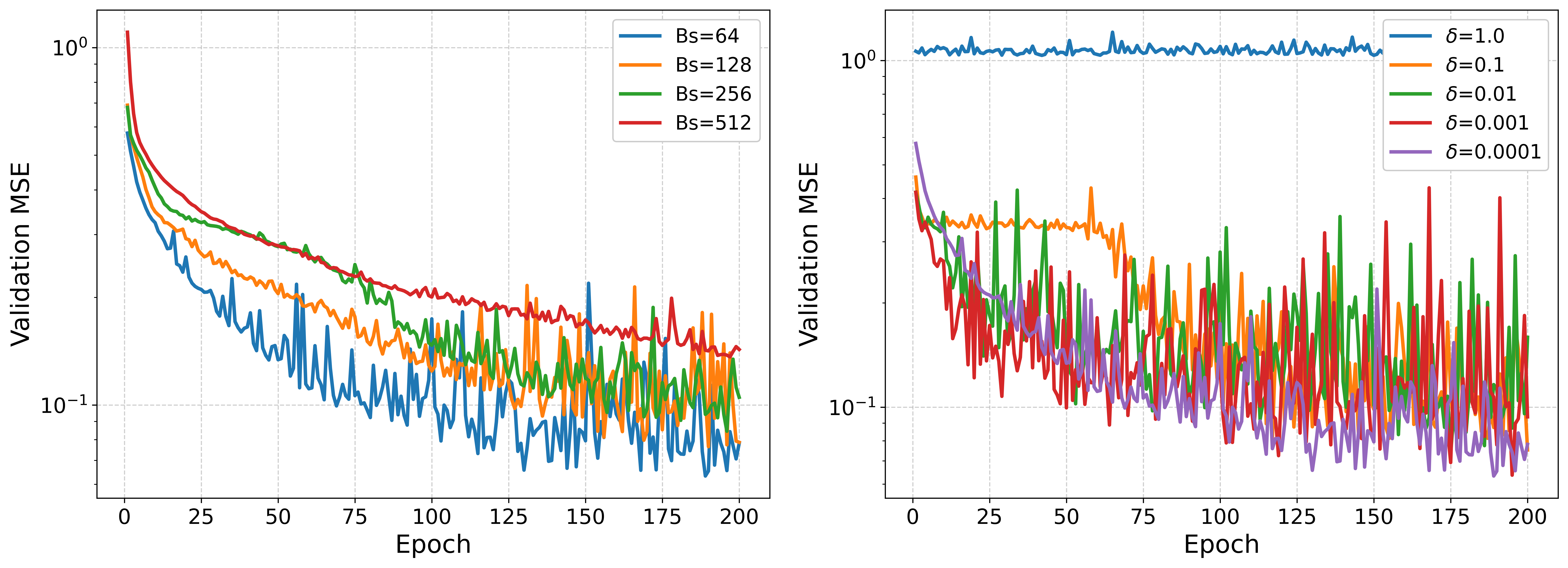}
    \caption{Validation MSE as a function of training epochs. The left panel shows curves for a fixed learning rate with varying batch sizes, while the right panel shows curves for a fixed batch size with varying learning rates. It is shown that a batch size $\text{Bs} = 64$ outperforms larger batch sizes, while learning rates on the order of $10^{-3}$ or lower exhibit similar convergence behavior.}
    \label{fig:MSE_vs_epochs}
\end{figure}

From Fig.~\ref{fig:val_train_MSE_heatmap}, it is clear that a simultaneous minimum in both the training and validation MSE occurs for the configuration with a learning rate $\delta=10^{-4}$ and a batch size $\text{Bs}=64$. These values were therefore adopted as the optimal hyperparameters for training the network. In Fig.~\ref{fig:MSE_vs_epochs}, these profiles are further illustrated by showing their evolution over the course of the training epochs.

\begin{figure}[h!]
    \centering
    \includegraphics[width=0.9\linewidth]{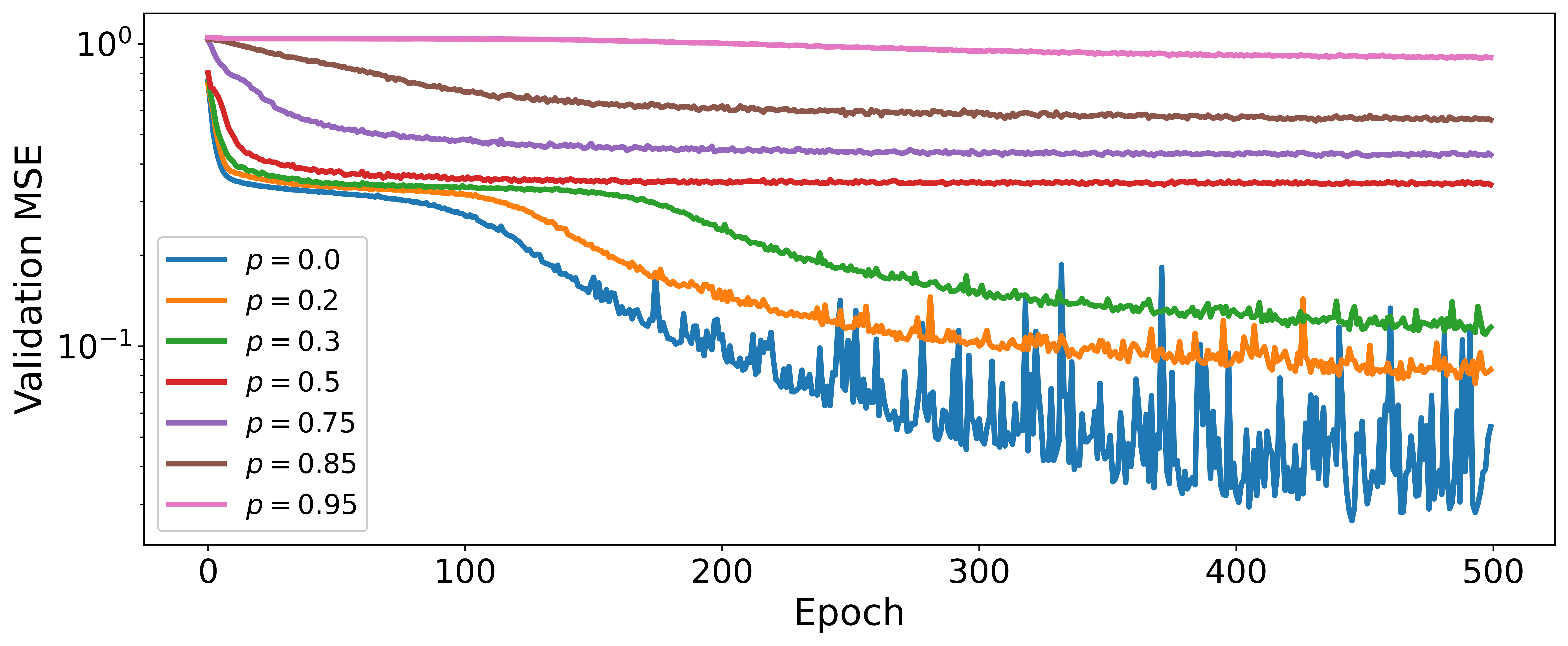}
    \caption{Test training for different dropout rates $p$, using a learning rate of $\delta=10^{-3}$, a batch size of 64 and a total of 500 fixed epochs without a scheduler. It is clear that the fastest convergence and lowest validation error are obtained for $p=0$, that is, with no dropout.}
    \label{fig:dropout_analysis}
\end{figure}

After identifying these hyperparameters, we analyzed the effect of including dropout in the hidden layers to determine the optimal dropout rate for training. The results are shown in Fig.~\ref{fig:dropout_analysis} as a function of different dropout probabilities $p$.
\newpage
\section{Optimization Algorithm\label{app:Optim_Algorithm}}

\begin{figure}[h!]
\begin{minipage}{0.48\textwidth}
\centering
\includegraphics[width=0.9\linewidth]{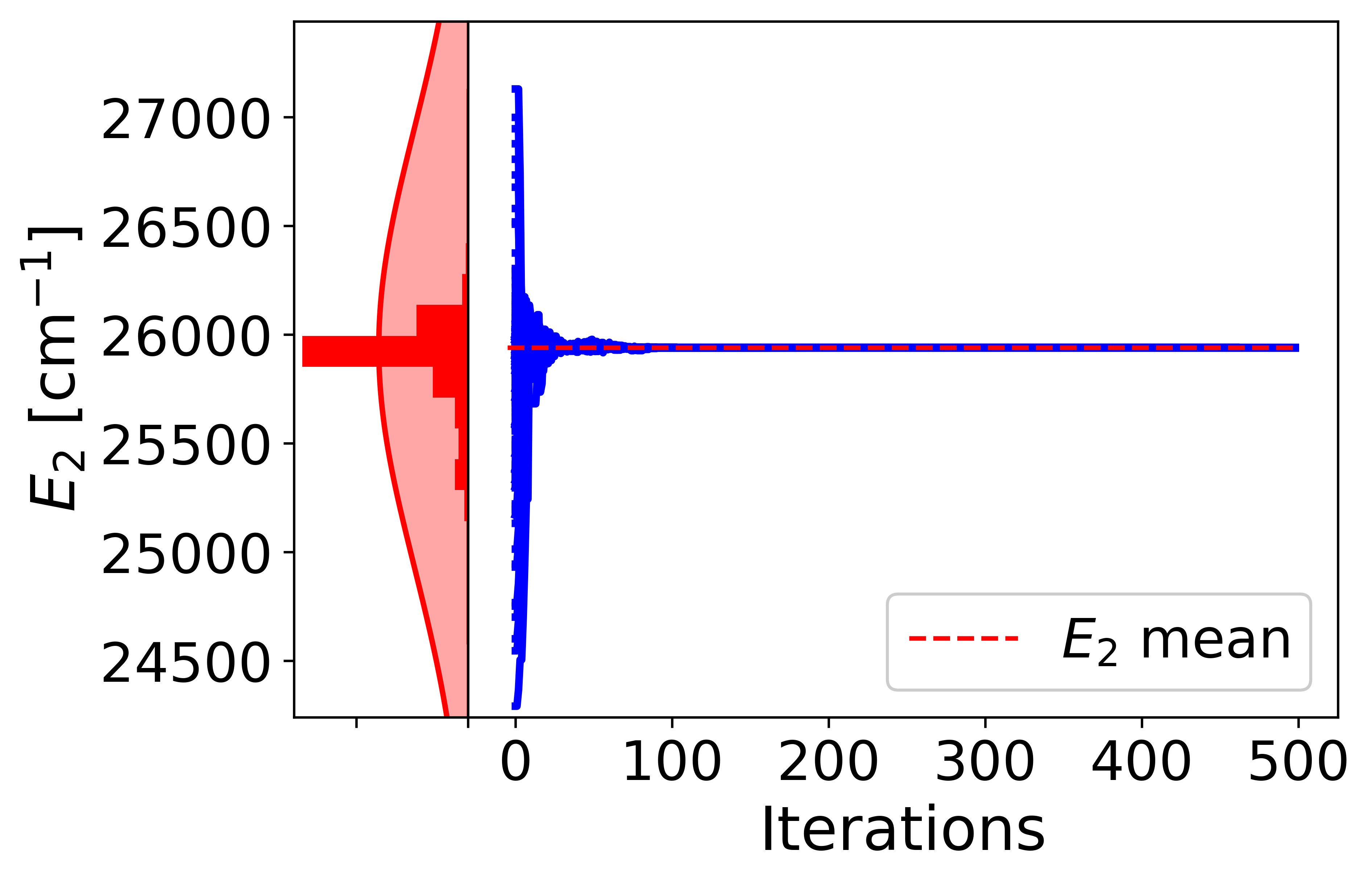}
\end{minipage}
\begin{minipage}{0.48\textwidth}
\centering
\includegraphics[width=0.9\linewidth]{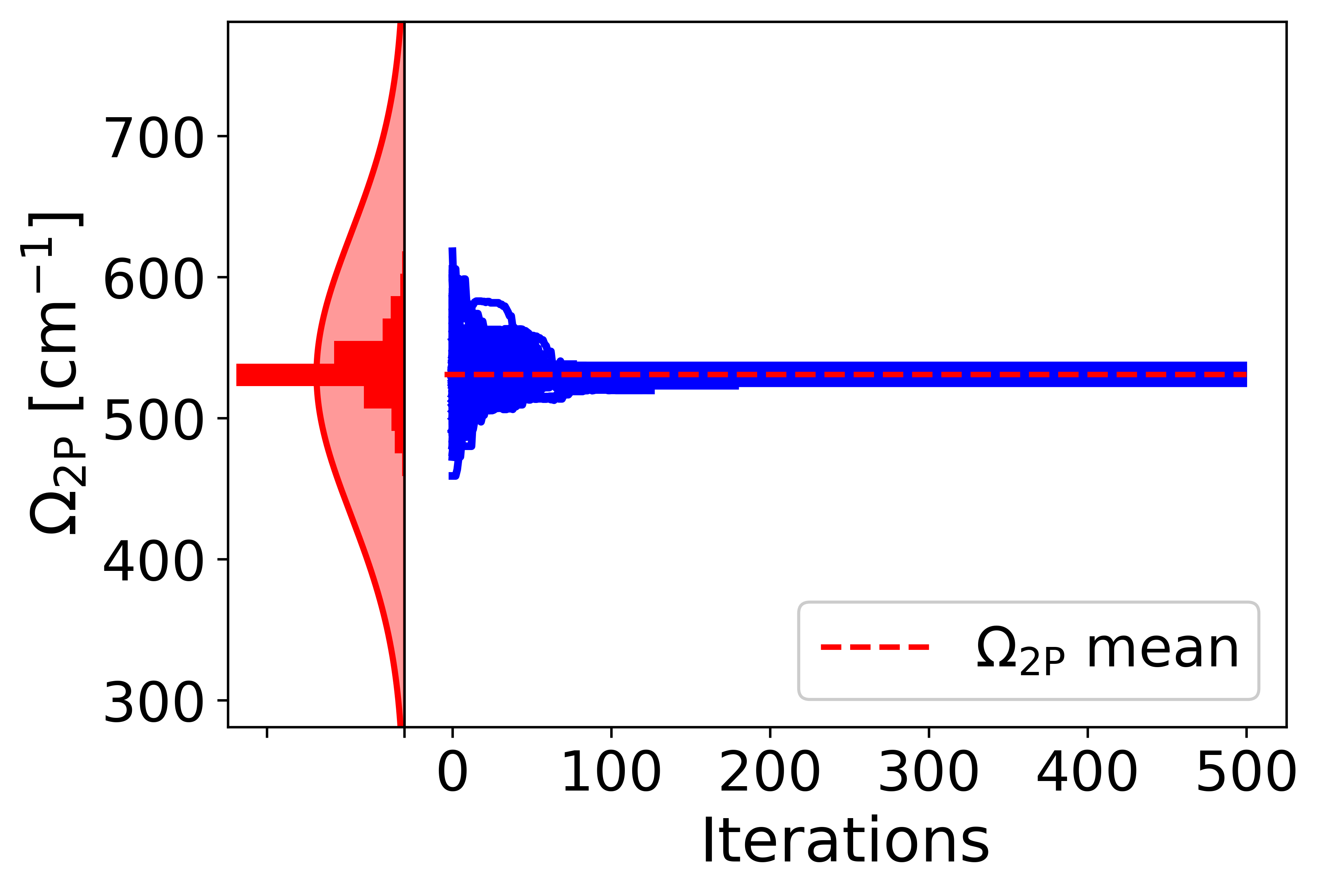}
\end{minipage}

\begin{minipage}{0.48\textwidth}
\centering
\includegraphics[width=0.9\linewidth]{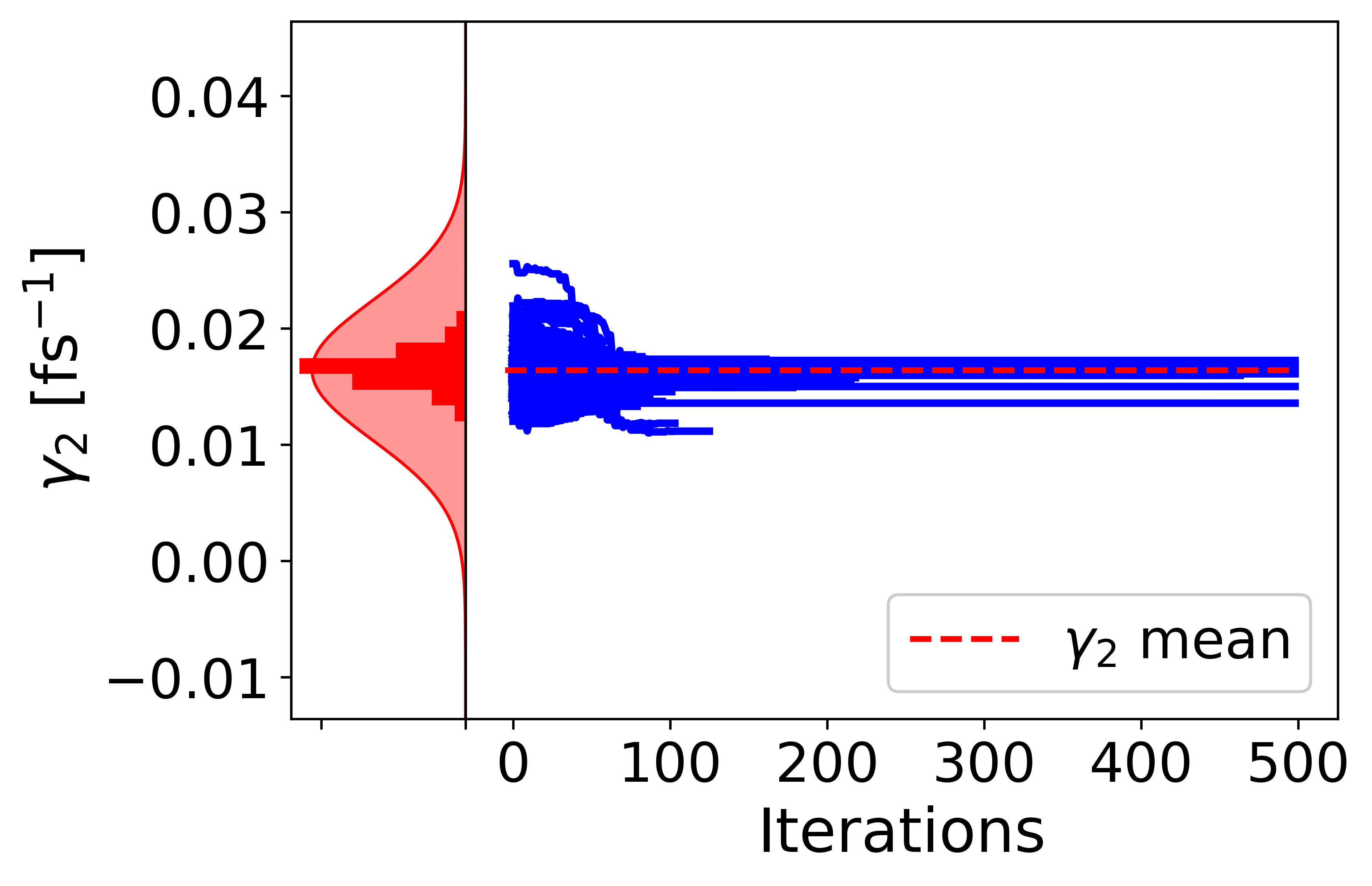}
\end{minipage}
\begin{minipage}{0.48\textwidth}
\centering
\includegraphics[width=0.9\linewidth]{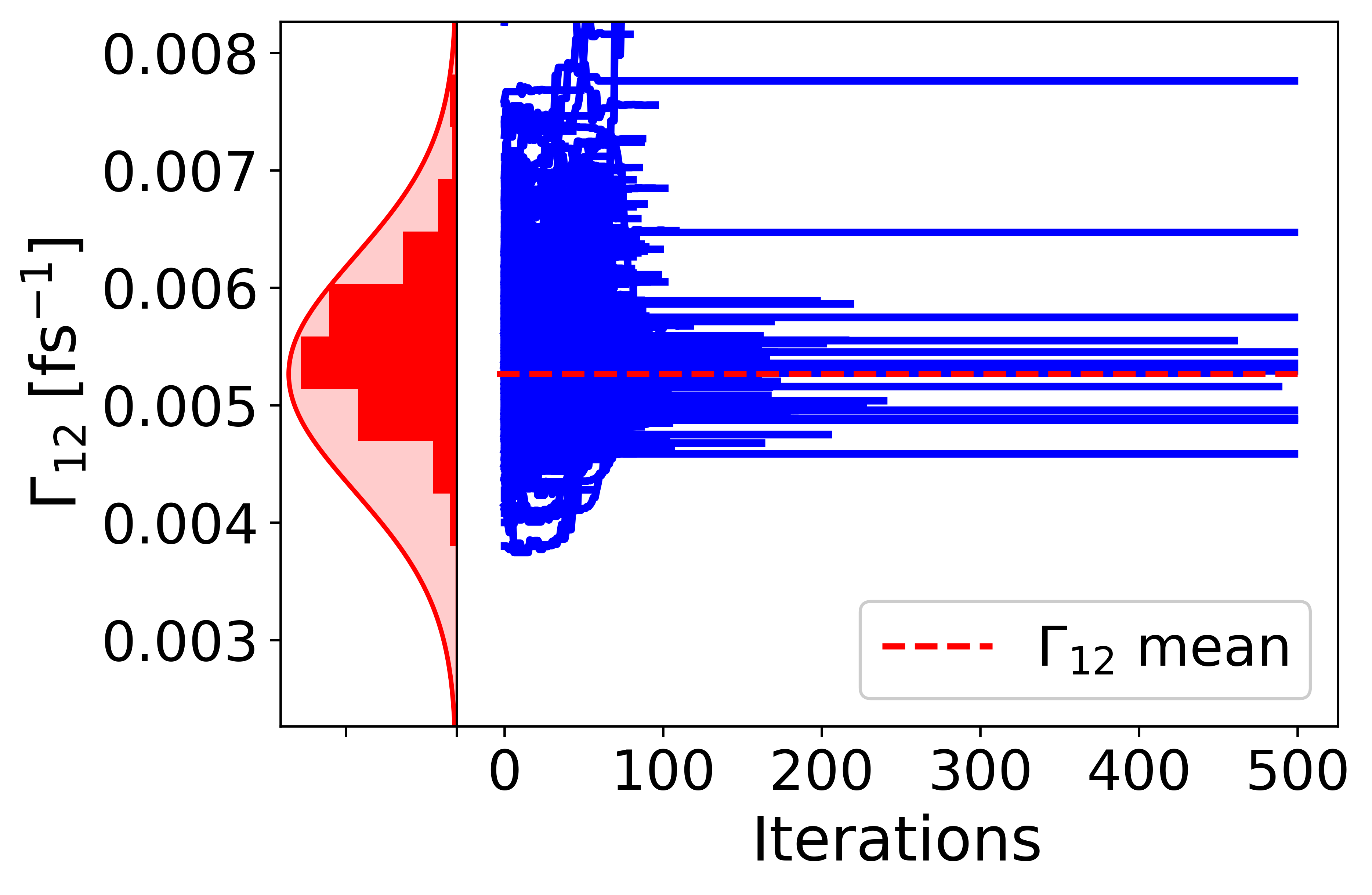}
\end{minipage}
\caption{Evolution of the target parameters across iterations of the optimization algorithm using $\lambda=10^{-2}$ in the scenario with four free parameters.}
\label{fig:histograms_plus_evolutions}
\end{figure}

\begin{figure}[h!]
    \centering
    \includegraphics[width=0.8\linewidth]{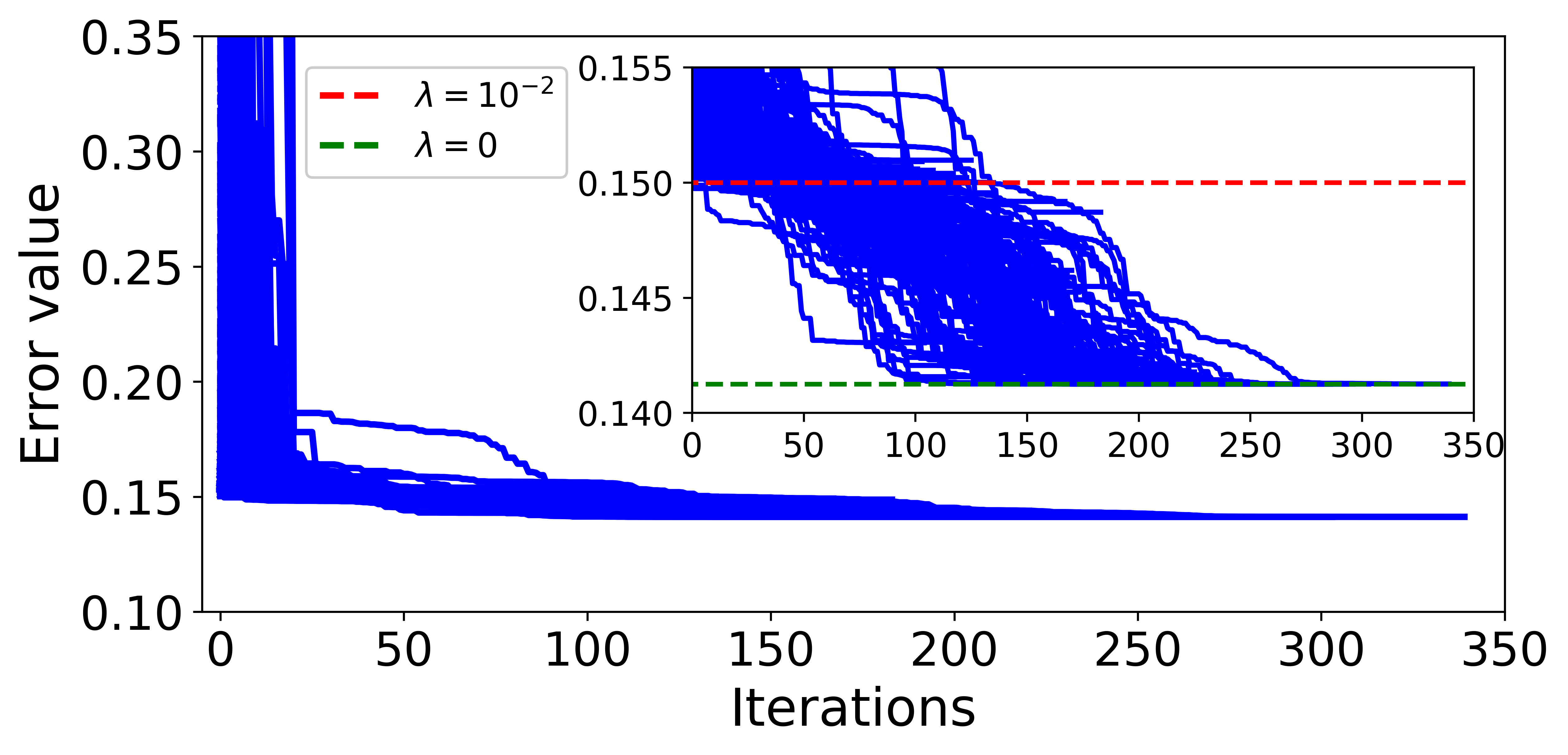}
    \caption{Evolution of the mean error between the theoretical PL prediction and the experimental data as a function of the iteration number. After approximately 100 iterations, the error decreases rapidly and reaches convergence. The dashed red and green lines highlight the final error values for the cases $\lambda=10^{-2}$ and $\lambda = 0$, respectively.}
    \label{fig:error_plots_evolution_algorithm}
\end{figure}

For the optimization algorithm described in Sec.~\ref{sec:OptimizationAlgo}, once all iterations across all initial conditions were completed, the histograms shown in Fig.~\ref{fig:histograms} were obtained. Figures~\ref{fig:histograms_plus_evolutions} and \ref{fig:error_plots_evolution_algorithm} provide additional details illustrating how these histograms are generated as the iterations progress and how the error is gradually minimized toward its final value.

\end{document}